\documentclass[aps,prd,nofootinbib,showkeys,preprint,floatfix]{revtex4-1}

\usepackage{amsmath}
\usepackage{amssymb}
\usepackage{graphicx}
\usepackage{rotating}
\usepackage{color} 
\usepackage{multirow} 
 \usepackage[T1]{fontenc}

\def\rpv{$R_p \hspace{-1em}/\;\:\hspace{0.2em}$}
\newcommand{\AddrAHEP}{
  {\it AHEP Group, Instituto de F\'{\i}sica Corpuscular --
    C.S.I.C./Universitat de Val{\`e}ncia \\
    Edificio de Institutos de Paterna, Apartado 22085,
  E--46071 Val{\`e}ncia, Spain}}

\newcommand{\AddrLisb}{%
 Departamento de F\'\i sica and CFTP, Instituto Superior T\'ecnico\\
          Av. Rovisco Pais 1, 1049-001 Lisboa, Portugal }

\newcommand{\AddrWur}{%
Institut f\"ur Theoretische Physik und Astronomie, 
Universit\"at W\"urzburg\\
Am Hubland, 
97074 Wuerzburg}

\def\gsim{\raise0.3ex\hbox{$\;>$\kern-0.75em\raise-1.1ex\hbox{$\sim\;$}}}
\def\lsim{\raise0.3ex\hbox{$\;<$\kern-0.75em\raise-1.1ex\hbox{$\sim\;$}}}

\begin{document}

\preprint{CFTP/10-014}  
\preprint{IFIC/10-43}  

\title{LHC and lepton flavour violation phenomenology of a left-right extension of the MSSM}

\author{J. N. Esteves}\email{joaomest@cftp.ist.utl.pt}
\author{J. C. Romao}\email{jorge.romao@ist.utl.pt}\affiliation{\AddrLisb}
\author{M. Hirsch} \email{mahirsch@ific.uv.es}
\author{A. Vicente}\email{Avelino.Vicente@ific.uv.es}\affiliation{\AddrAHEP}
\author{W. Porod} \email{porod@physik.uni-wuerzburg.de}
\author{F. Staub}\email{florian.staub@physik.uni-wuerzburg.de}
\affiliation{\AddrWur}



\begin{abstract}
We study the phenomenology of a supersymmetric left-right model, 
assuming minimal supergravity boundary conditions. 
Both left-right and (B-L) symmetries are broken at an energy scale 
close to, but significantly below the GUT scale. Neutrino data is 
explained via a seesaw mechanism. We calculate the RGEs for 
superpotential and soft parameters complete at 2-loop order. At 
low energies lepton flavour violation (LFV) and small, but potentially 
measurable mass splittings in the charged scalar lepton sector 
appear, due to the RGE running. Different from the supersymmetric 
``pure seesaw'' models, both, LFV and slepton mass splittings, occur 
not only in the left- but also in the right slepton sector. Especially, 
 ratios of LFV slepton decays, such as 
Br(${\tilde\tau}_R \to \mu \chi^0_1$)/Br(${\tilde\tau}_L \to \mu \chi^0_1$)
are sensitive to the ratio of (B-L) and left-right symmetry breaking scales. 
Also the model predicts a polarization asymmetry of the outgoing positrons 
in the decay $\mu^+ \to e^+ \gamma$, ${\cal A} \sim [0,1]$, which differs from 
the pure seesaw ``prediction'' ${\cal A}=1$. Observation of any of these 
signals allows to distinguish this model from any of the three standard, 
pure (mSugra) seesaw setups.

\end{abstract}

\maketitle


\section{Introduction}

The most popular explanation for the observed smallness of 
neutrino masses is certainly the seesaw mechanism 
\cite{Minkowski:1977sc,MohSen,Schechter:1980gr,Cheng:1980qt}.
Literally hundreds of theoretical papers based on ``the seesaw'' 
have been published since the discovery of neutrino oscillations 
\cite{Fukuda:1998mi}. 
The seesaw can be implemented at tree-level in exactly three 
realizations \cite{Ma:1998dn}: exchange of a fermionic singlet, 
a.k.a. the right-handed neutrino (type-I) \cite{Minkowski:1977sc,MohSen}; 
of a scalar triplet (type-II) \cite{MohSen,Schechter:1980gr,Cheng:1980qt}; 
or of a fermionic triplet (type-III) \cite{Foot:1988aq}. In any 
of these ``seesaw mechanisms'' neutrino masses are given by 
$m_\nu \sim v^2 /\Lambda$, where $v$ is the Higgs vacuum expectation value
(vev) and $\Lambda$ 
the scale of the seesaw. For coefficients ${\cal O}(1)$ and 
$\Lambda \sim (10^{14}-10^{15})$ GeV one finds neutrinos with sub-eV 
masses, just as experimental data demands. Unfortunately, attractive 
as this idea might appear from the theoretical point of view, this 
estimate also implies that ``the seesaw'' will never be {\em directly} 
tested. 

This situation might change slightly, if supersymmetry (SUSY) is found
at the LHC, essentially because scalar leptons provide potentially
additional information about seesaw parameters. Assuming SUSY gets
broken at a high energy scale, the seesaw parameters leave their
imprint on the soft parameters in the Renormalization Group Equation
(RGE) running. Then, at least in principle, indirect tests of the
seesaw become possible\footnote{In the general minimal supersymmetric
  extension of the standard model (MSSM) {\em all} soft terms are free
  parameters, fixed at the electroweak scale and nothing can be
  learned about the high energy world.}. Indeed, this has been pointed
out already in \cite{Borzumati:1986qx}, where it was shown that lepton
flavour violating (LFV) off-diagonal mass terms for sleptons are
automatically generated in seesaw (type-I), even if SUSY breaking is
completely flavour blind at the GUT scale as in minimal supergravity
(mSugra)\footnote{It might be technically more correct to call this
  setup the ``constrained MSSM'' (CMSSM). We will stick to the
  terminology mSugra.}.

Motivated by the above arguments, many authors have then studied LFV 
in SUSY models. For the seesaw type-I, low energy LFV decays such as 
$l_i \to l_j \gamma$ and $l_i \to 3 l_j$ have been calculated in 
\cite{Hisano:1995nq,Hisano:1995cp,Ellis:2002fe,Deppisch:2002vz,Petcov:2003zb,
Arganda:2005ji,Petcov:2005yh,Antusch:2006vw,Deppisch:2004fa,Hirsch:2008dy}; 
$\mu-e$ conversion in nuclei has been studied in 
\cite{Arganda:2007jw,Deppisch:2005zm}. The type-II seesaw has 
received much less attention, although it has actually fewer 
free parameters than type-I. The latter implies that ratios of LFV 
decays of leptons can actually be predicted as a function of neutrino 
angles in mSugra, as has been shown in \cite{Rossi:2002zb,Hirsch:2008gh}.
Finally, for completeness we mention that LFV in SUSY seesaw type-III 
has been studied in \cite{OurTypeIII}.

Measurements at colliders, once SUSY is discovered, can provide
additional information. LFV decays of left sleptons within mSugra have
been studied for type-I in \cite{Hisano:1998wn} and for type-II in
\cite{Hirsch:2008gh,Esteves:2009vg}.  Precise mass measurements might
also show indirect effects of the seesaw
\cite{Blair:2002pg,Freitas:2005et,Deppisch:2007xu}. Most prominently,
type-II and type-III seesaw contain non-singlet superfields, so gauge
couplings run differently from pure MSSM. One then expects that
sparticle spectra show a characteristic ``deformation'' with respect
to mSugra predictions. From different combinations of masses one can
form ``invariants'', i.e. numbers which to leading order depend only
on the seesaw scale \cite{Buckley:2006nv}, although there are
important corrections at 2-loop \cite{Hirsch:2008gh,OurTypeIII}, which
have to be included before any quantitative analysis can be done.
Experimentally interesting is also that at the LHC the mass splitting
between selectrons and smuons may be constrained down to ${\cal
O}(10^{-4})$ for 30 $fb^{-1}$ of integrated luminosity
\cite{Allanach:2008ib}.  In mSugra, one expects this splitting to be
unmeasurably tiny, whereas in mSugra plus seesaw significantly
different masses can be generated, as has been shown for type-I in
\cite{Abada:2010kj}.

Interestingly, in pure seesaw models with flavour blind SUSY boundary 
conditions all of the effects discussed above show up only in the 
left slepton sector. Naturally one expects that in a supersymmetric 
model with an intermediate left-right symmetric stage, also the 
right sleptons should contain some indirect information about the 
high energy parameters. This simple observation forms the main 
motivation for the current paper. Before entering in the details 
of our calculation, let us first briefly discuss left-right 
symmetric models. 

Quite a large number of different left-right (LR) symmetric models 
have been discussed in the literature. Originally LR models were 
introduced to explain the observed left-handedness of the weak interaction 
as a consequence of symmetry breaking 
\cite{Pati:1974yy,Mohapatra:1974gc,Senjanovic:1975rk}. 
However, LR models offer other advantages as well. First, the 
particle content of LR models contains automatically the right-handed 
neutrino and thus the ingredients for generating a (type-I) 
seesaw mechanism \footnote{Breaking the LR symmetry with triplets 
can generate also a type-II \cite{MohSen}.}. Second, the gauge group 
$SU(3)_c \times SU(2)_L \times SU(2)_R \times U(1)_{B-L}$ is one 
of the possible chains through which $SO(10)$ 
\cite{Georgi:1974my,Fritzsch:1974nn} can be broken to the standard 
model gauge group \footnote{Not all $SO(10)$ breaking chains contain 
a seesaw. Neither does $SU(5)$. It is, of course, straightforward to 
{\em add} a seesaw to $SU(5)$.}. In addition, it has been shown that 
they provide technical solutions to the SUSY CP and strong CP problems 
\cite{Mohapatra:1996vg} and they give an understanding of the $U(1)$ 
charges of the standard model fermions. Interesting only for the 
supersymmetric versions of LR models, (B-L) is gauged and thus, 
potentially, the low energy theory conserves R-parity \cite{Martin:1992mq}. 

This last argument requires possibly some elaboration. R-parity, defined 
as $R_P = (-1)^{3(B-L)+2s}$ (where $B$ and $L$ stand for baryon and 
lepton numbers and $s$ for the spin of the particle), is imposed in 
the MSSM to avoid dangerous baryon and lepton number violating 
operators. However, the origin of $R_P$ is not explained within the 
MSSM. In early LR models $SU(2)_R$ doublets were used to break the gauge
symmetry. The non-supersymmetric model proposed in references
\cite{Mohapatra:1974gc,Senjanovic:1975rk} introduced two additional 
scalar doublets $\chi_L$ and $\chi_R$, where $\chi_L \equiv \chi_L(1,2,1,1)$ 
and $\chi_R \equiv \chi_R(1,1,2,-1)$ under $SU(3)_c \times SU(2)_L \times 
SU(2)_R \times U(1)_{B-L}$. Parity conservation 
implies that both, $\chi_L$ and $\chi_R$, are needed. When the neutral 
component of $\chi_R$ gets a vev, $\langle \chi_R^0 \rangle \neq 0$, the 
gauge symmetry is broken down to the SM gauge group. However, $\chi_R$ 
is odd under $U(1)_{B-L}$ and thus, in the SUSY versions of this 
setup, $R_P$ is broken at the same time \footnote{This could be solved 
by imposing additional discrete symmetries on the model that forbid 
the dangerous \rpv operators \cite{Malinsky:2005bi}, but this cannot 
be regarded as automatic R-parity conservation.}. A possible solution to 
this problem is to break the gauge symmetry by $SU(2)_R$ fields with 
even charge under $U(1)_{B-L}$, i.e. by triplets. For a SUSY LR model, 
this was in fact proposed in reference \cite{Cvetic:1983su}, where four 
triplets were added to the MSSM spectrum: $\Delta(1,3,1,2)$, 
$\Delta^c(1,1,3,-2)$, $\bar{\Delta}(1,3,1,-2)$ and $\bar{\Delta}^c(1,1,3,2)$. 
Breaking the symmetry by the vev of $\Delta^c$ produces at the same 
time a right-handed neutrino mass via the operator $L^c \Delta^c L^c$, 
leading to a type-I seesaw mechanism. Depending on whether or not 
$\Delta$ gets a vev, also a type-II seesaw can be generated 
\cite{Akhmedov:2006de}.

However, whether R-parity is conserved in this setup is not clear. 
The reason is that the minimum of the potential might prefer a solution 
in which also the right-handed scalar neutrino gets a vev, thus breaking 
$R_P$, as has been claimed to be the case in \cite{Kuchimanchi:1993jg}. 
Later \cite{Babu:2008ep} calculated some 1-loop corrections to the 
scalar potential, concluding that $R_P$ conserving minima can be found. 
However, this contradicts the earlier claim \cite{Kuchimanchi:1993jg} 
that 1-loop corrections can not eliminate the dangerous \rpv 
minima. Aulakh and collaborators \cite{Aulakh:1997ba,Aulakh:1997fq}, 
on the other hand, showed that by the addition of two more triplets, 
$\Omega(1,3,1,0)$ and $\Omega^c(1,1,3,0)$, with zero lepton number 
one can achieve LR breaking with conserved $R_P$ guaranteed already 
at tree-level. Lacking a general proof that the model \cite{Cvetic:1983su} 
conserves $R_P$ we will follow \cite{Aulakh:1997ba,Aulakh:1997fq} as 
the setup for our numerical calculations.

Finally, for completeness we mention the existence of left-right models with
R-parity violation. For example, if the left-right symmetry is broken
with the vevs of right-handed sneutrinos R-parity gets broken as well
and the resulting phenomenology is
totally different, as shown in \cite{Hayashi:1984rd,FileviezPerez:2008sx}.

Compared to the long list of papers about indirect tests of the seesaw, 
surprisingly little work on the ``low-energy'' phenomenology of SUSY LR 
models has been done. One loop RGEs for two left-right SUSY models 
have been calculated in \cite{Setzer:2005hg}. These two models are (with 
one additional singlet): (a) breaking LR by doublets a la 
\cite{Mohapatra:1974gc,Senjanovic:1975rk} and (b) by triplets following 
\cite{Cvetic:1983su}, but no numerical work at all was done in this paper. 
The possibility that right sleptons might have flavour violating decays in 
the left-right symmetric SUSY model of \cite{Cvetic:1983su} was mentioned 
in \cite{Chao:2007ye}. A systematic study of all the possible signals 
discussed above for the seesaw case is lacking and to our knowledge 
there is no publication of any calculation of these signals for the model 
of \cite{Aulakh:1997ba,Aulakh:1997fq}.
(For completeness we would like to mention that in GUTs based on SU(5) one 
can have the situation the LFV occurs {\em only} in the right slepton 
sector, as pointed out in \cite{Barbieri:1994pv}. However, this 
model \cite{Barbieri:1994pv} is in a different class from all the models 
discussed above, since it does not contain non-zero neutrino masses.)

The rest of this paper is organized as follows. In the next section we 
define the model \cite{Aulakh:1997ba,Aulakh:1997fq} and discuss its 
particle content and main features at each symmetry breaking scale. 
We have calculated the RGEs for each step complete at the 2-loop level 
following the general description by \cite{Martin:1993zk} using the 
Mathematica package SARAH \cite{Staub:2008uz,Staub:2009bi,Staub:2010jh}. 
A summary is given in the appendix, the complete set of equations 
and the SARAH model files can be found at \cite{SarahWeb}. 
Neutrino masses can be fitted to experimental data via a type-I seesaw 
mechanism and we discuss different ways to implement the fit. We then 
turn to the numerical results. The output of SARAH has been passed to 
the program package SPheno \cite{Porod:2003um} for numerical evaluation. 
We calculate the SUSY spectra and LFV slepton decays, such as 
${\tilde\tau}_{L/R} \to \mu \tilde{\chi}^0_1$ and 
${\tilde\tau}_{L/R} \to e \tilde{\chi}^0_1$ and 
$\tilde{\chi}^0_2 \to e \mu \tilde{\chi}^0_1$, 
as well as low-energy decays $l_i \to l_j \gamma$ for some sample 
points as a function of the 
LR and (B-L) scales. Potentially measurable signals are found in both, 
left and right slepton sectors, if (a) the seesaw scale is above 
(very roughly) $10^{13}$ GeV and (b) if the scale of LR breaking is 
significantly below the GUT scale. Since we find sizable LFV soft 
masses in both slepton sectors, also the polarization in 
$\mu \to e \gamma$ is different from the pure seesaw expectation. 
We then close with a short summary and outlook.

\section{Left-right supersymmetric model}

In this section we define the model, its particle content and give 
a description of the different symmetry breaking steps. The fit 
to neutrino masses and its connection to LFV violation in the 
slepton sector is discussed in some detail, to prepare for 
the numerical results given in the next section. We summarize briefly 
the free parameters of the theory.

The model essentially follows \cite{Aulakh:1997ba,Aulakh:1997fq}. 
We have not attempted to find a GUT completion. We will, however, 
assume that gauge couplings and soft SUSY parameters can be unified, 
i.e. implicitly assume that such a GUT model can indeed be constructed.

\subsection{Step 1: From GUT scale to $SU(2)_R$ breaking scale}

Just below the GUT scale the gauge group of the model is $SU(3)_c
\times SU(2)_L \times SU(2)_R \times U(1)_{B-L}$. In addition it
is assumed that parity is conserved, see below. The matter content 
of the model is given in table \ref{tab:particles-step1}. Here $Q$, 
$Q^c$, $L$ and $L^c$ are the quark and lepton superfields of the MSSM 
with the addition of (three) right-handed neutrino(s) $\nu^c$.

\begin{table}
\centering
\begin{tabular}{c c c c c c}
\hline
Superfield & generations & $SU(3)_c$ & $SU(2)_L$ & $SU(2)_R$ & $U(1)_{B-L}$ \\
\hline
$Q$ & 3 & 3 & 2 & 1 & $\frac{1}{3}$ \\
$Q^c$ & 3 & $\bar{3}$ & 1 & 2 & $-\frac{1}{3}$ \\
$L$ & 3 & 1 & 2 & 1 & -1 \\
$L^c$ & 3 & 1 & 1 & 2 & 1 \\
$\Phi$ & 2 & 1 & 2 & 2 & 0 \\
$\Delta$ & 1 & 1 & 3 & 1 & 2 \\
$\bar{\Delta}$ & 1 & 1 & 3 & 1 & -2 \\
$\Delta^c$ & 1 & 1 & 1 & 3 & -2 \\
$\bar{\Delta}^c$ & 1 & 1 & 1 & 3 & 2 \\
$\Omega$ & 1 & 1 & 3 & 1 & 0 \\
$\Omega^c$ & 1 & 1 & 1 & 3 & 0 \\
\hline
\end{tabular}
\caption{Matter content between the GUT scale and the $SU(2)_R$ breaking scale.}
\label{tab:particles-step1}
\end{table}

Two $\Phi$ superfields, bidoublets under $SU(2)_L \times SU(2)_R$, are
introduced. They contain the standard $H_d$ and $H_u$ MSSM Higgs
doublets. In this model, two copies are needed for a non-trivial 
CKM matrix. Although there are known attempts to build a realistic LR 
model with only one bidoublet generating the quark mixing angles at the 
loop level \cite{Babu:1998tm}, we will not rely on such a mechanism. 
Finally, the rest of the superfields in table \ref{tab:particles-step1} 
are introduced to break the LR symmetry, as explained above.

Table \ref{tab:particles-step1} shows also the gauge charges for the matter
content in the model. In particular, the last column shows the $B-L$
value for the different superfields. However, the following definition
for the electric charge operator will be used throughout this paper
\begin{equation}
Q = I_{3L} + I_{3R} + \frac{B-L}{2}
\end{equation}
and thus the $U(1)_{B-L}$ charge is actually $\frac{B-L}{2}$.

With the representations in table \ref{tab:particles-step1}, the most
general superpotential compatible with the gauge symmetry and parity
is
\begin{eqnarray} \label{eq:Wsuppot1}
{\cal W} &=& Y_Q Q \Phi Q^c 
          +  Y_L L \Phi L^c 
          - \frac{\mu}{2} \Phi \Phi
          +  f L \Delta L
          +  f^* L^c \Delta^c L^c \nonumber \\
         &+& a \Delta \Omega \bar{\Delta}
          +  a^* \Delta^c \Omega^c \bar{\Delta}^c
          + \alpha \Omega \Phi \Phi
          +  \alpha^* \Omega^c \Phi \Phi \nonumber \\
         &+& M_\Delta \Delta \bar{\Delta}
          +  M_\Delta^* \Delta^c \bar{\Delta}^c
          +  M_\Omega \Omega \Omega
          +  M_\Omega^* \Omega^c \Omega^c \thickspace.
\end{eqnarray}
Note that this superpotential is invariant under the parity transformations
$Q  \leftrightarrow  (Q^c)^*$, $L \leftrightarrow (L^c)^*$, 
$\Phi \leftrightarrow  \Phi^\dagger$, $\Delta \leftrightarrow 
(\Delta^c)^*$, $\bar{\Delta} \leftrightarrow (\bar{\Delta}^c)^*$, 
$\Omega \leftrightarrow (\Omega^c)^*$. This discrete symmetry fixes, 
for example, the $L^c \Delta^c L^c$ coupling to be $f^*$, the complex 
conjugate of the $L \Delta L$ coupling, thus reducing the number of free 
parameters of the model. 

Family and gauge indices have been omitted in eq. \eqref{eq:Wsuppot1}, 
more detailed expressions can be found in \cite{Aulakh:1997ba}. 
$Y_Q$ and $Y_L$ are quark and lepton Yukawa couplings. However, with 
two bidoublets there are two copies of them, and thus there are four 
$3 \times 3$ Yukawa matrices. Conservation of parity implies that they 
must be hermitian. $\mu$ is a $2 \times 2$ symmetric matrix, whose entries
have dimensions of mass, $f$ is a $3 \times 3$ (dimensionless) complex symmetric
matrix, and $\alpha$ is a $2 \times 2$ antisymmetric matrix, and thus
it only contains one (dimensionless) complex parameter, $\alpha_{12}$. 
The mass parameters $M_\Omega$ and $M_\Delta$ can be exchanged for
$v_R$ and $v_{BL}$, the vacuum expectation values of the scalar fields
that break the LR symmetry, see below. 

The soft terms of the model are
\begin{eqnarray} \label{eq:soft1}
- {\cal L}_{soft} &=& m_Q^2 \tilde{Q}^\dagger \tilde{Q} + m_{Q^c}^2
\tilde{Q^c}^\dagger \tilde{Q^c} + m_L^2 \tilde{L}^\dagger \tilde{L} +
m_{L^c}^2 \tilde{L^c}^\dagger \tilde{L^c} \nonumber \\ &+& m_{\Phi}^2
\Phi^\dagger \Phi + m_{\Delta}^2 \Delta^\dagger \Delta +
m_{\bar{\Delta}}^2 \bar{\Delta}^\dagger \bar{\Delta} + m_{\Delta^c}^2
{\Delta^c}^\dagger \Delta^c + m_{\bar{\Delta}^c}^2 \bar{\Delta}^{c \:
  \dagger} \bar{\Delta}^c \nonumber \\ 
&+& m_{\Omega}^2 \Omega^\dagger
\Omega + m_{\Omega^c}^2 {\Omega^c}^\dagger \Omega^c + \frac{1}{2}
\big[ M_1 \tilde{B}^0 \tilde{B}^0 + M_2 (\tilde{W_L} \tilde{W_L} +
  \tilde{W_R} \tilde{W_R}) + M_3 \tilde{g} \tilde{g} + h.c. \big]
\nonumber \\ 
&+& \big[ T_Q \tilde{Q} \Phi \tilde{Q^c} + T_L \tilde{L}
  \Phi \tilde{L^c} + T_f \tilde{L} \Delta \tilde{L} + T_f^*
  \tilde{L^c} \Delta^c \tilde{L^c} \nonumber \\ &+& T_a \Delta \Omega
  \bar{\Delta} + T_a^* \Delta^c \Omega^c \bar{\Delta^c} + T_\alpha
  \Omega \Phi \Phi + T_\alpha^* \Omega^c \Phi \Phi + h.c. \big]
\nonumber \\ 
&+& \big[ B_\mu \Phi \Phi + B_{M_\Delta} \Delta
  \bar{\Delta} + {B_{M_\Delta}}^* \Delta^c \bar{\Delta}^c +
  B_{M_\Omega} \Omega \Omega + {B_{M_\Omega}}^* \Omega^c \Omega^c +
  h.c. \big] \thickspace.
\end{eqnarray}
Again, family and gauge indices have been omitted for the sake of simplicity. 
The LR model itself does not, of course, fix the values of the soft SUSY breaking terms. 
In the numerical evaluation of the RGEs we will resort to mSugra-like 
boundary conditions, i.e. 
$m_0^2  \mathcal{I}_{3 \times 3} = m_Q^2 = m_{Q^c}^2 = m_L^2 = m_{L^c}^2$, 
$m_0^2  \mathcal{I}_{2 \times 2} = m_\Phi^2$, 
$m_0^2 = m_\Delta^2 = m_{\bar{\Delta}}^2 = m_{\Delta^c}^2 = 
m_{\bar{\Delta}^c}^2 = m_\Omega^2 = m_{\Omega^c}^2$, 
$M_{1/2} = M_1 = M_2 = M_3$, $T_Q = A_0 Y_Q, T_L = A_0 Y_L, T_f = A_0 f, 
T_a = A_0 a, T_\alpha = A_0 \alpha$, 
$B_\mu = B_0, B_{M_\Delta} = B_0 M_\Delta, B_{M_\Omega} = B_0 M_\Omega$. 
The superpotential couplings $f$, $Y_Q$ and $Y_L$ are fixed by the
low-scale fermion masses and mixing angles. Their values at the GUT
scale are obtained by RGE running. This will be discussed in more
detail in section \ref{sec:yukawas}.

The breaking of the LR gauge group to the MSSM gauge group takes place 
in two steps: $SU(2)_R \times U(1)_{B-L} \rightarrow U(1)_R \times 
U(1)_{B-L} \rightarrow U(1)_Y$. In the first step the neutral 
component of the triplet $\Omega$ takes a vev:
\begin{equation}
\langle \Omega^{c \: 0} \rangle = \frac{v_R}{\sqrt{2}}
\end{equation}
which breaks $SU(2)_R$. However, since $I_{3R} (\Omega^{c \: 0}) = 0$ 
there is a $U(1)_R$ symmetry left over. Next, the group 
$U(1)_R \times U(1)_{B-L}$ is broken by
\begin{equation}
\langle \Delta^{c \: 0} \rangle = \frac{v_{BL}}{\sqrt{2}} \thickspace, \qquad 
\langle \bar{\Delta}^{c \: 0} \rangle = \frac{\bar{v}_{BL}}{\sqrt{2}} \thickspace.
\end{equation}
The remaining symmetry is now $U(1)_Y$ with hypercharge defined as
$Y = I_{3R} + \frac{B-L}{2}$. 

The tadpole equations do not link $\Omega^c$, $\Delta^c$ and $\bar{\Delta}^c$ 
with their left-handed counterparts, due to supersymmetry. Thus, the left-handed
triplets can have vanishing vevs \cite{Aulakh:1997ba} and the model produces 
only a type-I seesaw.

Although a ``hierarchy'' between the two breaking scales may exist,
$v_{BL} \ll v_R$, one cannot neglect the effects of the second
breaking stage on the first one, since mass terms of $\Omega$ and 
$\Delta$ enter in both tadpole equations. If we assume
$\bar{v}_{BL} = v_{BL}$ the tadpole equations of the model can 
be written
\begin{eqnarray}
\frac{\partial V}{\partial v_R} &=& 4 |M_\Omega|^2 v_R + 
\frac{1}{2} |a|^2 v_{BL}^2 v_R - \frac{1}{2} v_{BL}^2 
\left[ a^* (M_\Delta + M_\Omega) + c.c \right] = 0 \label{tadpoleeqs-1} \thickspace, \\
\frac{\partial V}{\partial v_{BL}} &=& |M_{\Delta}|^2 v_{BL} + 
\frac{1}{4} |a|^2 (v_{BL}^2 + v_R^2) v_{BL} - \frac{1}{2} v_{BL} v_R 
\left[ a^* (M_\Delta + M_\Omega) + c.c \right]= 0 \label{tadpoleeqs-2} \thickspace.
\end{eqnarray}
In these equations (small) soft SUSY breaking terms have been
neglected.  Similarly, at this stage there are no electroweak symmetry
breaking vevs $v_d$ and $v_u$. From equations \eqref{tadpoleeqs-1} and
\eqref{tadpoleeqs-2} one sees that, in fact, there is an inverse
hierarchy between the vevs and the superpotential masses $M_\Delta$,
$M_\Omega$, given by
\begin{equation} \label{tadpolesol}
v_R = \frac{2 M_\Delta}{a} \thickspace, \qquad v_{BL} = \frac{2}{a} (2 M_\Delta M_\Omega)^{1/2} \thickspace.
\end{equation}
And so, $v_{BL} \ll v_R$ requires $M_\Delta \gg M_\Omega$, as has already 
been discussed in \cite{Aulakh:1997ba}.

\subsection{Step 2: From $SU(2)_R$ breaking scale to $U(1)_{B-L}$ breaking scale}

At this step the gauge group is $SU(3)_c \times SU(2)_L \times U(1)_R
\times U(1)_{B-L}$. The particle content of the model from the
$SU(2)_R$ breaking scale to the $U(1)_{B-L}$ breaking scale is given
in table \ref{tab:particles-step2}.

\begin{table}
\centering
\begin{tabular}{c c c c c c}
\hline
Superfield & generations & $SU(3)_c$ & $SU(2)_L$ & $U(1)_R$ & $U(1)_{B-L}$ \\
\hline
$Q$ & 3 & 3 & 2 & 0 & $\frac{1}{3}$ \\
$d^c$ & 3 & $\bar{3}$ & 1 & $\frac{1}{2}$ & $-\frac{1}{3}$ \\
$u^c$ & 3 & $\bar{3}$ & 1 & $-\frac{1}{2}$ & $-\frac{1}{3}$ \\
$L$ & 3 & 1 & 2 & 0 & $-1$ \\
$e^c$ & 3 & 1 & 1 & $\frac{1}{2}$ & $1$ \\
$\nu^c$ & 3 & 1 & 1 & $-\frac{1}{2}$ & $1$ \\
$H_d$ & 1 & 1 & 2 & $-\frac{1}{2}$ & 0 \\
$H_u$ & 1 & 1 & 2 & $\frac{1}{2}$ & 0 \\
$\Delta$ & 1 & 1 & 3 & 1 & 2 \\
$\bar{\Delta}$ & 1 & 1 & 3 & 1 & -2 \\
$\Delta^{c \: 0}$ & 1 & 1 & 1 & 1 & -2 \\
$\bar{\Delta}^{c \: 0}$ & 1 & 1 & 1 & -1 & 2 \\
$\Omega$ & 1 & 1 & 3 & 0 & 0 \\
$\Omega^{c \: 0}$ & 1 & 1 & 1 & 0 & 0 \\
\hline
\end{tabular}
\caption{Matter content from the $SU(2)_R$ breaking scale to the 
$U(1)_{B-L}$ breaking scale.}
\label{tab:particles-step2}
\end{table}

Some comments might be in order. Despite $M_{\Delta}$ being of the
order of $v_R$ (or larger), see eq.\eqref{tadpolesol}, not all
components of the $\Delta$ superfields receive large masses. The neutral
components of $\Delta^c$ and $\bar{\Delta}^c$ lie at the $v_{BL}$
scale.  One can easily check that the F-term contributions to their
masses vanish in the minimum of the scalar potential eq.~\eqref{tadpolesol}. Moreover, $\Omega^c$ does not generate D-terms
contributions to their masses. Therefore, contrary to the other
components of the $\Delta$ triplets, they only get masses at the
$v_{BL}$ scale. On the other hand, one might guess that all components
in the $\Omega$,$\Omega^c$ superfields should be retained at this
stage, since their superpotential mass $M_\Omega$ is required to be
below $v_{BL}$. However, some of their components get contributions
from $SU(2)_R$ breaking, and thus they become heavy. The charged 
components of $\Omega^c$ do develop large masses, in the case of the 
scalars through D-terms, while in the case of the fermions due to 
their mixing with the charged gauginos $\tilde{W}_R^\pm$, which have 
masses proportional to $v_R$. However, the neutral components of 
$\Omega^c$ do not get $SU(2)_R$ breaking contributions, since they 
have $I_{3R} (\Omega^{c  \: 0}) = 0$, and then they must be
included in this energy regime. See reference \cite{Aulakh:1997fq} for 
a more quantitative discussion.

After $SU(2)_R$ breaking the two bidoublets $\Phi_1$ and $\Phi_2$ get
split into four $SU(2)_L$ doublets. Two of them must remain light,
identified with the two Higgs doublets of the MSSM, responsible for EW
symmetry breaking, while, at the same time, the other two get masses
of the order of $v_R$. This strong hierarchy can be only obtained by
imposing a fine-tuning condition on the parameters involved in the
bidoublet sector.

The superpotential terms mixing the four $SU(2)_L$ doublets can be
rewritten as
\begin{equation}
{\cal W}_M = (H_d^f)^T M_H H_u^f
\end{equation}
where $H_d^f = ( H_d^1, H_d^2)$ and $H_u^f = ( H_u^1, H_u^2)$ are the
\emph{flavour eigenstates}. In this basis reads the matrix
\begin{equation}
M_H = 
\left(
\begin{array}{cc}
\mu_{11} & \mu_{12} + \alpha_{12} M_R \\
\mu_{12} - \alpha_{12} M_R & \mu_{22} 
\end{array}
\right) \thickspace,
\end{equation}
where the relations $\mu_{ij} = \mu_{ji}$ and $\alpha_{ij} = -
\alpha_{ji}$ have been used and $M_R = \frac{v_R}{2}$ has been
defined. In order to get two light doublets we impose the fine-tuning
condition \cite{Aulakh:1997fq}
\begin{equation} \label{fine-tuning}
{\rm Det}(M_H) = \mu_{11} \mu_{22} - (\mu_{12}^2 - \alpha_{12}^2 M_R^2) = 0 \thickspace.
\end{equation}
The result of eq.~\eqref{fine-tuning} is to split the two Higgs bidoublets
into two pairs of doublets $(H_d,H_u)_L$ and $(H_d,H_u)_R$, where
$(H_d,H_u)_L$ is the light pair that appears in table
\ref{tab:particles-step2}, and $(H_d,H_u)_R$ a heavy pair with mass of
order of $v_R$. In practice, equation \eqref{fine-tuning} implies that
one of the superpotential parameters must be chosen in terms of the
others. Since this fine-tuning condition is not protected by any
symmetry, the RGEs do not preserve it, and one must impose it at
the $SU(2)_R$ breaking scale. In our computation we chose to compute
$\mu_{11}$ in terms of the free parameters $\mu_{12}$, $\mu_{22}$,
$\alpha_{12}$ and $v_R$.

In order to compute the resulting couplings for the light Higgs
doublets one must rotate the original fields into their mass
basis. Since $M_H$ is not a symmetric matrix (unless $\alpha_{12} =0$)
one has to rotate independently $H_d^f$ and $H_u^f$, i.e.  $H_d^f =
D H_d^m$, $H_u^f = U H_u^m$, where $D$ and $U$ are orthogonal
matrices and $H_d^m = ( H_d^L, H_d^R)$ and $H_u^m = ( H_u^L, H_u^R)$
are the \emph{mass eigenstates}. This way one finds
\begin{equation}
{\cal W}_M = (H_d^f)^T M_H H_u^f = (H_d^m)^T D^T M_H U H_u^m =
(H_d^m)^T {\hat M}_H H_u^m
\end{equation}
where ${\hat M}_H$ is a diagonal matrix, with eigenvalues
\begin{eqnarray}\nonumber
\hat{M}_{H,1}^2 &=& 0 \thickspace, \\
\hat{M}_{H,2}^2 &=& \frac{1}{\mu_{22}^2}
\left(\alpha_{12}^4 M_R^4 + 2 \alpha_{12}^2 M_R^2 
(\mu_{22}^2-\mu_{12}^2) + (\mu_{22}^2+\mu_{12}^2)^2 \right) \thickspace.
\end{eqnarray}
The $D$ and $U$ rotations are, in general, different and  we parametrize them
 as
\begin{equation}
D = \left( \begin{array}{c c}
\cos \theta_1 & \sin \theta_1 \\
- \sin \theta_1 & \cos \theta_1
\end{array} \right) \thickspace, \qquad U = \left( \begin{array}{c c}
\cos \theta_2 & \sin \theta_2 \\
- \sin \theta_2 & \cos \theta_2
\end{array} \right)
\end{equation}
and get
\begin{eqnarray}\nonumber
H_d^1 &=& \cos \theta_1 H_d^L + \sin \theta_1 H_d^R \label{higgs-rot-1} \thickspace, \\
H_d^2 &=& - \sin \theta_1 H_d^L + \cos \theta_1 H_d^R \label{higgs-rot-2} \thickspace,
\end{eqnarray}
and similar for $H_u$.  In general the angles $\theta_1$ and
$\theta_2$ are different. However, they are connected to the same
matrix $M_H$ and can be calculated by diagonalizing $M_H (M_H)^T$ or
$(M_H)^T M_H$ and  one finds
\begin{eqnarray}
\tan \theta_{1,2} &=& 
\frac{\mu_{12} \pm \alpha_{12} M_R}{\mu_{22}} \label{ang1} \thickspace.
\end{eqnarray}
In these expressions ${\rm Det}(M_H)=0$ has been used to simplify the
result.  Exact ${\rm Det}(M_H)=0$ implies that the $\mu$-term of the MSSM is 
zero, so this condition can only be true up to small corrections, 
see the discussion below.
Note that there are two interesting limits. First, $\mu_{12} \gg
\alpha_{12} M_R$ : this implies $\tan \theta_1 = \tan \theta_2$ and
therefore $D = U$. This is as expected, since that limit makes $M_H$
symmetric. And, second, $\mu_{12} \ll \alpha_{12} M_R$ : this implies
$\tan \theta_1 = - \tan \theta_2$ and therefore $D = U^T$.

The superpotential at this stage is
\begin{eqnarray} \label{eq:Wsuppot2}
{\cal W} &=& Y_u Q H_u u^c + Y_d Q H_d d^c 
          +  Y_e L H_d e^c + Y_\nu L H_u \nu^c 
          +  \mu H_u H_d \nonumber \\
         &+& f_c^1 \nu^c \nu^c  \Delta^{c \: 0}
          +  M_{\Delta^c}^1 \Delta^{c \: 0} \bar{\Delta}^{c \: 0}
          +  a \Delta \Omega \bar{\Delta}
          +  a_c^1 \Delta^{c \: 0} \bar{\Delta}^{c \: 0} \Omega^{c \: 0} 
\nonumber \\
         &+& b \Omega H_d H_u
          +  b_c \Omega^{c \: 0} H_d H_u
          +  M_\Omega \Omega \Omega
          +  M_{\Omega^c} \Omega^{c \: 0} \Omega^{c \: 0} .
\end{eqnarray}
Particles belonging to the same $SU(2)_R$ gauge multiplets split due 
to their different $U(1)_R$ charges. At this stage both the LR group, 
that symmetrizes the $SU(2)_L$ and $SU(2)_R$ gauge interactions, and 
the discrete parity symmetry that we imposed on the couplings are broken. 

The soft terms are
\begin{eqnarray} \label{eq:soft2}
- {\cal L}_{soft} &=& m_Q^2 \tilde{Q}^\dagger \tilde{Q} + m_{u^c}^2
\tilde{u^c}^\dagger \tilde{u^c} + m_{d^c}^2 \tilde{d^c}^\dagger
\tilde{d^c} + m_L^2 \tilde{L}^\dagger \tilde{L} + m_{e^c}^2
\tilde{e^c}^\dagger \tilde{e^c} + m_{\nu^c}^2 \tilde{\nu^c}^\dagger
\tilde{\nu^c} \nonumber \\ &+& m_{H_u}^2 H_u^\dagger H_u + m_{H_d}^2
H_d^\dagger H_d + m_{\Delta^{c \: 0}}^2 {\Delta^{c \: 0}}^\dagger
\Delta^{c \: 0} + m_{\bar{\Delta}^{c \: 0}}^2 \bar{\Delta}^{c \: 0 \:
  \dagger} \bar{\Delta}^{c \: 0}\nonumber \\ &+& m_{\Omega}^2
\Omega^\dagger \Omega + m_{\Omega^{c \: 0}}^2 \Omega^{c \: 0 \:
  \dagger} \Omega^{c \: 0} + \frac{1}{2} \big[ M_1 \tilde{B}^0
  \tilde{B}^0 + M_L \tilde{W_L} \tilde{W_L} + M_R \tilde{W_R^0}
  \tilde{W_R^0} + M_3 \tilde{g} \tilde{g} + h.c. \big] \nonumber \\
&+& \big[ T_u \tilde{Q} H_u \tilde{u^c} + T_d \tilde{Q} H_d
  \tilde{d^c} + T_e \tilde{L} H_d \tilde{e^c} + T_\nu \tilde{L} H_u
  \tilde{\nu^c} \\ &+& T_{f_c}^1 \tilde{\nu^c} \tilde{\nu^c} \Delta^{c
    \: 0} + T_{a_c}^1 \Delta^{c \: 0} \Omega^{c \: 0} \bar{\Delta}^{c
    \: 0} + T_b \Omega H_d H_u + T_{b^c} \Omega^{c \: 0} H_d H_u +
  h.c. \big] \nonumber \\ &+& \big[ B_\mu H_u H_d + B_{M_{\Delta^c}^1}
  \Delta^{c \: 0} \bar{\Delta}^{c \: 0} + B_{M_\Omega} \Omega \Omega +
  B_{M_\Omega^c} \Omega^{c \: 0} \Omega^{c \: 0} + h.c. \big] \thickspace.
\nonumber
\end{eqnarray}
Again we  suppress gauge and family indices.

We must impose matching conditions at the $SU(2)_R$ breaking scale. 
These are for superpotential parameters given by
\begin{eqnarray}\nonumber
Y_d = Y_Q^1 \cos \theta_1 - Y_Q^2 \sin \theta_1 \thickspace,
&\qquad& Y_u = - Y_Q^1 \cos \theta_2 + Y_Q^2 \sin \theta_2  \thickspace, \\ \nonumber
Y_e = Y_L^1 \cos \theta_1 - Y_L^2 \sin \theta_1 \thickspace,
&\qquad& Y_\nu = - Y_L^1 \cos \theta_2 + Y_L^2 \sin \theta_2 \thickspace, \\ \nonumber
f_c^1 = - f^* \thickspace, &\qquad& a_c^1 = - \frac{a^*}{\sqrt{2}}  \thickspace, \\ \nonumber
M_{\Delta^c}^1 = M_\Delta^* \thickspace, &\qquad& M_{\Omega^c} = M_\Omega^* \thickspace, \\
b = 2 \alpha R \thickspace, &\qquad& b_c = \sqrt{2} \alpha^* R \thickspace,
\end{eqnarray}
where $R = \sin (\theta_1 - \theta_2)$. For the soft masses we have 
\begin{eqnarray}
m_{u^c}^2 = m_{d^c}^2 &=& m_{Q^c}^2 \thickspace,\\ \nonumber
m_{e^c}^2 = m_{\nu^c}^2 &=& m_{L^c}^2 \thickspace, \\ \nonumber
m_{\Delta^{c \: 0}}^2 &=& m_{\Delta^c}^2 \thickspace, \\ \nonumber
m_{\bar{\Delta}^{c \: 0}}^2 &=& m_{\bar{\Delta}^c}^2 \thickspace ,\\ \nonumber
m_{\Omega^{c \: 0}}^2 &=& m_{\Omega^c}^2 \thickspace, \\ \nonumber
M_L = M_R &=& M_2 \thickspace.
\end{eqnarray}
Soft trilinears matching follow corresponding conditions. 
In addition, one has
\begin{eqnarray}
m_{H_d}^2 &=& \cos^2 \theta_1 (m_\Phi^2)_{11} + \sin^2 \theta_1
(m_\Phi^2)_{22} - \sin \theta_1 \cos \theta_1 \left[ (m_\Phi^2)_{12} +
(m_\Phi^2)_{21} \right] \thickspace, \nonumber \\ 
m_{H_u}^2 &=& \cos^2 \theta_2
(m_\Phi^2)_{11} + \sin^2 \theta_2 (m_\Phi^2)_{22} - \sin \theta_2 \cos
\theta_2 \left[ (m_\Phi^2)_{12} + (m_\Phi^2)_{21} \right] \thickspace, \nonumber
\end{eqnarray}
as obtained when the operator $m_\Phi^2 \Phi^\dagger \Phi$ is
projected into the light Higgs doublets operators $(H_d^L)^\dagger
H_d^L$ and $(H_u^L)^\dagger H_u^L$. Gauge couplings are matched as 
$g_L = g_R = g_2$.

\subsection{Step 3: From $U(1)_{B-L}$ breaking scale to EW/SUSY scale}

We mention this stage only for completeness, since the last regime is 
just the usual MSSM. We need matching conditions in the
gauge sector. Since $U(1)_R \times U(1)_{B-L}$ breaks to $U(1)_Y$, the
MSSM gauge coupling $g_1$ will be a combination of $g_R$ and $g_{BL}$.
The resulting relationship is
\begin{equation}
g_1 = \frac{\sqrt{5} g_R g_{BL}}{\sqrt{2 g_R^2 + 3 g_{BL}^2}} \thickspace .
\end{equation}
Analogously, the following condition holds for gaugino masses
\begin{equation}
M_1({\rm MSSM}) = \frac{2 g_R^2 M_1 + 3 g_{BL}^2 M_R}{2 g_R^2 + 3 g_{BL}^2} \thickspace .
\end{equation}
Note that in the last two equations the gauge couplings are GUT-normalized. 
Electroweak symmetry breaking occurs as in the MSSM. We take the Higgs
doublet vevs
\begin{equation} \label{higgs-vevs}
\langle H_d^0 \rangle = \frac{v_d}{\sqrt{2}}  \thickspace,
\qquad \langle H_u^0 \rangle = \frac{v_u}{\sqrt{2}} \thickspace, 
\end{equation}
as free parameters and then solve the tadpole equations to find 
$\mu_{\rm MSSM}$ and $B^\mu$. $\mu_{\rm MSSM}$ must be different from zero, 
that is ${\rm Det}(M_H)$ can not be exactly zero. Instead the 
tuning must be exact up to ${\rm Det}(M_H)={\cal O}(\mu_{\rm MSSM}^2)$.
As usual $\tan \beta = \frac{v_u}{v_d}$ is used as a free parameter. 
Also the sign of $\mu_{\rm MSSM}$ is not constrained as usual.

\subsection{Neutrino masses, LFV and Yukawa couplings}
\label{sec:yukawas}

Neutrino masses are generated after $U(1)_{B-L}$ breaking through a
type-I seesaw mechanism. The matrix $f_c^1$ leads to Majorana
masses for the right-handed neutrinos once $\Delta^{c \: 0}$ gets a
vev. We define the seesaw scale as the lightest eigenvalue of 
\begin{equation}
M_S \equiv f_c^1 v_{BL} \thickspace .
\end{equation}

As usual, we can always rotate the fields to a basis where $M_S$ is diagonal. However, this will introduce lepton flavour violating entries in the $Y_{L_i}$ Yukawas, see discussion below. As mentioned above, contrary to non-supersymmetric LR models \cite{MohSen}, 
there is no type-II contribution to neutrino masses.
 
\begin{table}
\centering
\begin{tabular}{|c|c|c|}
\hline
parameter & best fit & $2$-$\sigma$ \\
\hline \hline
$\Delta m_{21}^2[10^{-5}\text{eV}^2]$ & $7.59^{+0.23}_{-0.18}$ & $7.22-8.03$\\
$|\Delta m_{31}^2|[10^{-3}\text{eV}^2]$ & $2.40^{+0.12}_{-0.11}$ & $2.18-2.64$\\
 $\sin^2\theta_{12}$ & $0.318^{+0.019}_{-0.016}$ & $0.29-0.36$ \\
 $\sin^2\theta_{23}$ & $0.50^{+0.07}_{-0.06}$ & $0.39-0.63$\\
 $\sin^2\theta_{13}$ & $0.013^{+0.013}_{-0.009}$ & $\leq 0.039$\\
\hline
\end{tabular}
\vspace{-2mm}
\caption{Best-fit values with $1$-$\sigma$ errors and $2$-$\sigma$ intervals 
($1$ d.o.f.) taken from the reference \cite{Schwetz:2008er}, which is updated 
continuously on the web.}
\label{tab:neutrinos}
\end{table}

Global fits to all available experimental data provide values for the
parameters involved in neutrino oscillations, see table \ref{tab:neutrinos} 
for updated results and ref.~\cite{Collaboration:2007zza,KamLAND2007} for experimental
results. As first observed in \cite{Harrison:2002er}, these 
data imply that the neutrino mass matrix can be diagonalized to a good 
approximation by the so-called tri-bimaximal mixing pattern:
\begin{equation}
\label{eq:UTBM}
U_{TBM} =
\left(\begin{array}{cccc}
\sqrt{\frac{2}{3}} & \sqrt{\frac{1}{3}} & 0 \cr
- \frac{1}{\sqrt{6}} &  \frac{1}{\sqrt{3}} & - \frac{1}{\sqrt{2}} \cr
- \frac{1}{\sqrt{6}} &  \frac{1}{\sqrt{3}} & \frac{1}{\sqrt{2}}
\end{array}\right).
\end{equation}
The matrix product $Y_\nu \cdot (f_c^1)^{-1} \cdot Y_\nu^T$ is constrained 
by this particular structure. LFV entries can be present in both $Y_\nu$ 
and $f_c^1$, see also the discussion about parameter counting in the 
next subsection. However, in the numerical section we will consider only 
two specific kinds of fits:
\begin{itemize}

\item $Y_\nu$-fit: flavour structure in $Y_\nu$ and diagonal $f_c^1$.

\item $f$-fit: flavour structure in $f_c^1$ and diagonal $Y_\nu$.

\end{itemize}
While at first it may seem either way of doing the fit is equivalent, 
$f_c^1$ and $Y_{\nu}$ in our setup can leave different traces in the 
soft slepton mass parameters if $v_{BL} \ll v_R$. This last condition 
is essential to distinguish between both possibilities, because otherwise 
one obtains the straightforward prediction that LFV entries in left and 
right slepton are equal, due to the assumed LR symmetry above $v_R$.

These two types of fit were already discussed in reference \cite{Babu:2002tb},
which investigates low energy LFV signatures in a supersymmetric seesaw model
where the right-handed neutrino mass is generated from a term of the form
$f \Delta^c \nu^c \nu^c$. When the scalar component of $\Delta^c$ acquires a vev
a type-I seesaw is obtained, generating masses for the light neutrinos.
Therefore, this model has the ingredients to accommodate a $Y_\nu$-fit,
named as \emph{Dirac LFV} in \cite{Babu:2002tb}, or a $f$-fit, named as
\emph{Majorana LFV}. Note, however, that the left-right symmetry, central in
our work, is missing in this reference, thus implying different signatures at
the electroweak scale.

The difference in phenomenology of the two fits can be easily understood 
considering approximated expressions for the RGEs for $m_L^2$ and 
$m_{e^c}^2$. In the first step, from the GUT scale to the $v_R$ 
scale RGEs at 1-loop order can be written in leading-log approximation 
as \cite{Chao:2007ye}
\begin{eqnarray}\nonumber
\Delta m_L^2 &=& - \frac{1}{4 \pi^2} \left( 3 f f^\dagger + Y_L^{(k)} Y_L^{(k) 
\: \dagger} \right) (3 m_0^2 + A_0^2) \ln \left( \frac{m_{GUT}}{v_R} \right) \thickspace,
\label{apprge1} \\
\Delta m_{L^c}^2 &=& - \frac{1}{4 \pi^2} \left( 3 f^\dagger f + Y_L^{(k) 
\: \dagger} Y_L^{(k)} \right) (3 m_0^2 + A_0^2)
\ln \left( \frac{m_{GUT}}{v_R} \right) \thickspace. \label{apprge2}
\end{eqnarray}
Of course, also the $A$ parameters develop LFV off-diagonals in the 
running. We do not give the corresponding approximated equations for 
brevity. After parity breaking at the $v_R$ scale the Yukawa coupling 
$Y_L$ splits into $Y_e$, the charged lepton Yukawa, and $Y_\nu$, the
neutrino Yukawa. The later contributes to LFV entries in the running
down to the $v_{BL}$ scale. Thus,
\begin{eqnarray}
\Delta m_L^2 &\sim & - \frac{1}{8 \pi^2} Y_\nu Y_\nu^\dagger
  \left( m_L^2|_{v_R} + A_e^2|_{v_R}\right)
\ln \left( \frac{v_R}{v_{BL}} \right) \thickspace, \nonumber  \\
\Delta m_{e^c}^2 & \sim & 0  \thickspace, \label{apprge4}
\end{eqnarray}
where $m_L^2|_{v_R}$ is the matrix $m_L^2$ at the scale $v_R$ and 
$A_e^2|_{v_R}$ is defined as $T_e = Y_e A_e$ and also has to be taken 
at $v_R$. 
In order to understand the main difference between the two fits, let
us first consider the $f$-fit.  This assumes that $Y_{\nu}$ is
diagonal at the seesaw scale and thus the observed low energy mismatch
between the neutrino and charged lepton sectors is due to a
non-trivial flavour structure in $f_c^1$. Of course, non-diagonal 
entries in $f$ generate in the running also non-diagonal entries in 
$Y_{\nu}$ and $Y_e$, but these can be neglected in first approximation. 
In this case, equations \eqref{apprge2} and \eqref{apprge4} show that 
the LR symmetry makes $m_L^2$ and $m_{e^c}^2$ run with the same flavour 
structure and the magnitudes of their off-diagonal entries at the SUSY 
scale are similar. If, on the other hand, $Y_{\nu}$ is non-trivial 
($Y_{\nu}$-fit), while $f$ is diagonal, the running from the GUT scale 
to the $v_R$ scale induces again the same off-diagonal entries in $m_L^2$ 
and $m_{L^c}^2$. However, from $v_R$ to $v_{BL}$ the off-diagonals entries 
in $m_L^2$ continue to run, while those in $m_{e^c}^2$ do not. This effect, 
generated by the right-handed neutrinos via the $Y_\nu$ Yukawas, induces 
additional flavour violating effects in the L sector compared to the R sector. 
Seeing LFV in both left and right slepton sectors thus allows us to 
indirectly learn about the high energy theory. We will study this in 
some detail in the numerical section below.

\subsection{Parameter counting}

Let us briefly summarize the free parameters of the model. With the 
assumption of mSugra (or better: mSugra-like) boundary conditions, 
in the SUSY breaking sector we only have the standard parameters 
$m_0$, $M_{1/2}$, $A_0$, $\tan \beta$, $sign(\mu_{\rm MSSM})$. Thus, 
we count 4+1 parameters in the soft terms. We note in passing that the 
soft terms of the heavy sector, of course, do not have to follow strictly 
the conditions outlined in equation \eqref{eq:soft1}, as long as these 
parameters are small compared to $v_{BL}$ there are no changes compared 
to the above discussion.

In the superpotential we have $a$, $\alpha$, $\mu$, $M_{\Delta}$ and 
$M_{\Omega}$. This leaves, at first sight, 7 parameters free. However, 
we can reduce them to 4+2 parameters as follows. Since $\alpha_{ij} 
= - \alpha_{ji}$, $\alpha$ only contains one free parameter: $\alpha_{12}$. 
The matrix $\mu$ has 3 entries, but one of them, $\mu_{11}$, is fixed 
by the fine-tuning condition ${\rm Det}(M_H)={\cal O}(\mu_{\rm MSSM}^2)$. This 
leaves two free parameters, $\mu_{12}$, $\mu_{22}$. We have traded 
$M_{\Delta}$ and $M_{\Omega}$ for the vevs $v_R$, $v_{BL}$, since 
$\ln ( \frac{v_R}{v_{BL}})$ and $\ln ( \frac{v_{GUT}}{v_{R}})$ enter 
into the RGEs and thus can, at least in principle, be determined from 
low-energy spectra. There are then in summary 6 parameters, four  
independent of low-energy constraints and two which could be fixed 
from LFV data, see below.

In addition, in the superpotential we have the Yukawa matrices $Y_{Q_i}$, 
$Y_{L_i}$ and $f$. Let's consider the quark sector first. Since we 
can always go to a basis in which one of the $Y_{Q_i}$ is diagonal 
with only real entries, there are 12 parameters. Ten of them are 
fixed by six quark masses, three CKM angles and the CKM phase, leaving 
two phases undetermined. 

In the lepton sector we have the symmetric matrices, $Y_{L_1}$ and 
$Y_{L_2}$. As with the quark sector, a basis change shows that there 
are only 12 free parameters. $f$ is symmetric and thus counts as another 
9 parameters. Going to a basis in which $f$ is diagonal does not reduce 
the number of free parameters, since in this basis we can no longer 
assume one of the $Y_{L_i}$ to be diagonal. In summary there are thus 
free 21 parameters in these three matrices. 

In the simple, pure seesaw type-I with three generations of right-handed 
neutrinos the number of free parameters is 21. Only 12 of them can be 
fixed from low-energy data: three neutrino and three charged lepton masses, 
three leptonic mixing angles and three phases (two Majorana and one 
Dirac phase). However, as pointed out in \cite{Ellis:2002fe}, in principle, 
$m_L^2$ contains 9 observable entries and thus, if the normalization (i.e. 
$m_0$, $A_0$, $\tan\beta$ etc.) is known from other sfermion measurements, 
one could re-construct the type-I seesaw parameters \footnote{Of course, 
this discussion is slightly academic, since at least one of the Majorana 
phases will never be measured in praxis.}.

How does the SUSY LR model compare to this? We have, as discussed above, 
also 21 parameters in the three coupling matrices, but neutrino masses 
depend also on $v_{BL}$. However, in principle, we have 9 more observables 
in $m_{e^c}^2$, assuming again that the soft SUSY breaking terms 
can be extracted 
from other measurements. Since in the RGEs also $v_R$ appears we have 
in total 23 parameters which need to be determined. The number of observables, 
on the other hand is fixed to 30 in total, as we have 12 (low-energy lepton 
sector) plus 9 (left sleptons) plus 9 (right sleptons) possible 
measurements.

\section{Numerical results}

\subsection{Procedure for numerics}

All necessary, analytical expressions were calculated with SARAH. For
this purpose, two different model files for the model above the two
threshold scales were created and used to calculate the full set of
2-loop RGEs. SARAH calculates the RGEs using the generic expressions
of \cite{Martin:1993zk} in the most general form respecting the
complete flavour structure. These RGEs were afterwards exported to
Fortran code and implemented in SPheno. As starting point for the RGE
running, the gauge and Yukawa couplings at the electroweak scale are
used. In the calculation of the gauge and Yukawa couplings we follow
closely the procedure described in ref.\ \cite{Porod:2003um}: the
values for the Yukawa couplings giving mass to the SM fermions and the
gauge couplings are determined at the scale \(M_Z\) based on the
measured values for the quark, lepton and vector boson masses as well
as for the gauge couplings.  Here, we have included the 1-loop
corrections to the mass of W- and Z-boson as well as the SUSY
contributions to \(\delta_{VB}\) for calculating the gauge
couplings. Similarly, we have included the complete 1-loop
corrections to the self-energies of SM fermions
\cite{Pierce:1996zz}. Moreover, we have resummed the $\tan\beta$
enhanced terms for the calculation of the Yukawa couplings of the
$b$-quark and the $\tau$-lepton as in \cite{Porod:2003um}. 
 The vacuum expectation values \(v_d\) and
\(v_u\) are calculated with respect to the given value of
\(\tan\beta\) at \(M_Z\). Since we are working with two distinct
threshold scales, not all heavy fields are integrated out at their
mass and the corresponding 1-loop boundary conditions at the
threshold scales are needed. It is known that these particles cause a
finite shift in the gauge couplings and gaugino masses.  The general
expressions are \cite{Hall:1980kf}
\begin{eqnarray}
\label{eq:shift1}
 g_i & \rightarrow & g_i \left( 1\pm \frac{1}{16 \pi^2} g_i^2 I^i_2(r) 
\ln\left(\frac{M^2}{M_T^2}\right)\right)  \thickspace ,\\
\label{eq:shift2}
 M_i & \rightarrow & M_i \left( 1\pm \frac{1}{16 \pi^2} g_i^2 I^i_2(r) 
\ln\left(\frac{M^2}{M_T^2}\right)\right) \thickspace .
\end{eqnarray}
\(I^i_2(r)\) is the Dynkin index of a field transforming as
representation \(r\) with respect to the gauge group belonging to the
gauge coupling \(g_i\), \(M\) is the mass of this particle and \(M_T\)
is the threshold scale. When evaluating the RGEs from the low to the
high scale, the contribution is positive, when running down, it is
negative. The different masses used for calculating the finite shifts
are the eigenvalues of the full tree-level mass matrix of the charged,
heavy particles removed from the spectrum.  The correct mass spectrum
is calculated in an iterative way. The GUT scale is defined as the
scale at which \(g_{BL} = g_2 = g_{GUT}\) holds. Generally, there is
difference with \(g_3\) to \(g_{GUT}\) in the percent
range, the actual numerical mismatch depending on the scales $v_{BL}$ 
and $v_R$ and being larger for lower values of $v_{BL}$ and $v_R$. It 
has been stressed in particular in \cite{Kopp:2009xt} that within 
supersymmetric LR models, the LR symmetry breaking scale has to be 
close to the GUT scale, otherwise this mismatch will grow too large. 
However, in \cite{Majee:2007uv} it was pointed out that, among other 
possibilities, GUT thresholds - unknown unless the GUT model, including 
the complete Higgs sector used to break the GUT symmetry, is specified - 
can lead to important corrections, accounting for this apparent 
non-unification. (For a discussion of these effects in
the context of \(SU(5)\) see \cite{Martens:2010nm}.) We simply use 
$g_{BL} = g_2 = g_{GUT}$ and attribute departures from complete 
unification to (unknown) thresholds. After applying the
GUT scale boundary conditions, the RGEs are evaluated down to the low scale
and the mass spectrum of the MSSM is calculated. The MSSM masses are,
in general, calculated at the 1-loop level in the
\(\overline{\mbox{DR}}\) scheme using on-shell external momenta. For
the Higgs fields also the most important 2-loop contributions are taken into
account. We note that the corresponding Fortran routines are also
written by SARAH but they are equivalent to the routines
included in the public version of SPheno based on
\cite{Pierce:1996zz}.  The iteration stops when the largest change
in the calculation of the SUSY and Higgs boson masses at
\(M_{SUSY}\) is below one per-mille  between two iterations.

\subsection{Mass spectrum}

\begin{figure}
\begin{center}
\vspace{5mm}
\includegraphics[width=0.47\textwidth]{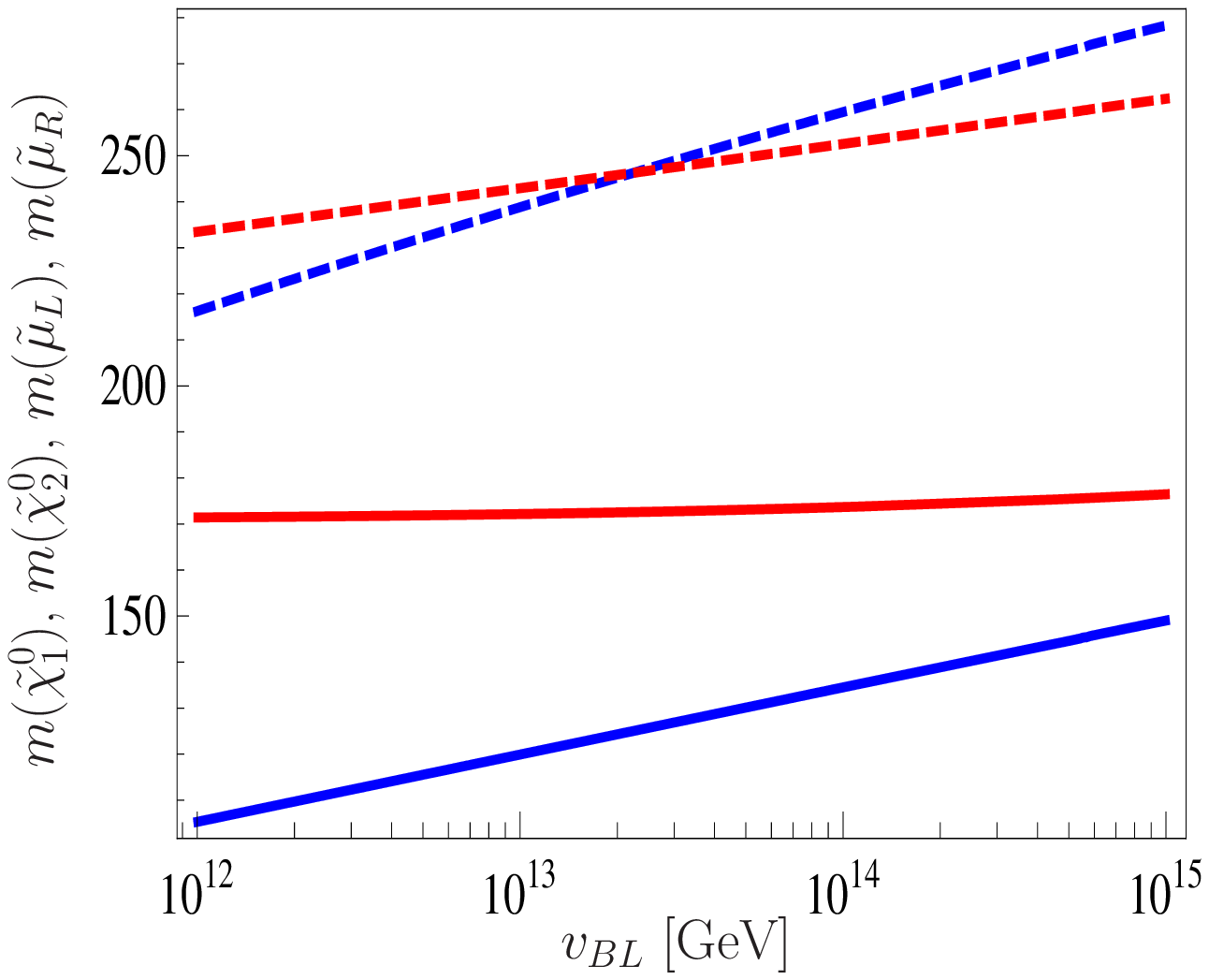}
\hspace{5mm}
\includegraphics[width=0.47\textwidth]{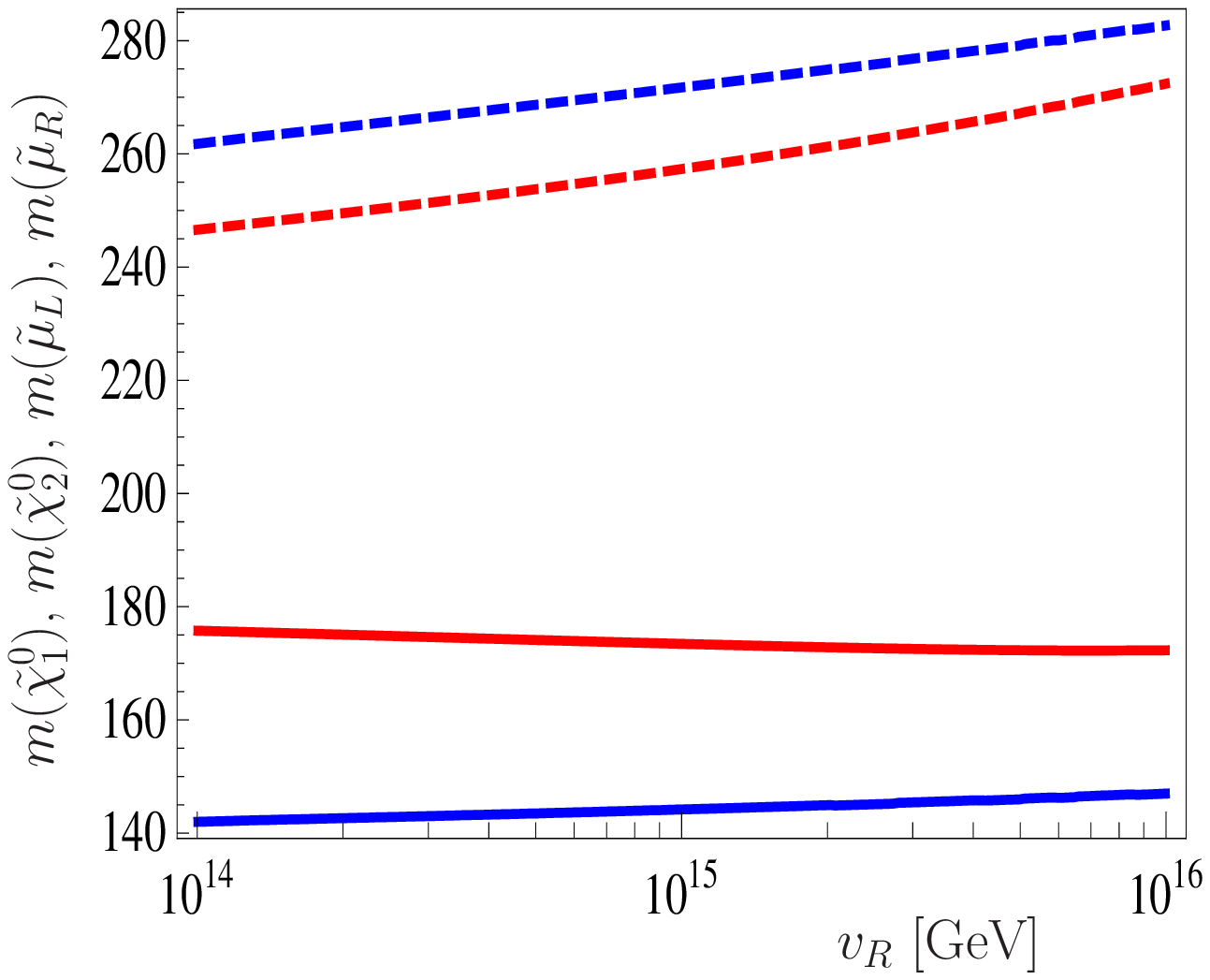}
\end{center}
\vspace{-5mm}
\caption{Example of spectra at the SUSY scale and its dependence on
$v_{BL}$ (left side) and $v_R$ (right side). The masses of four states
are shown: $\tilde{\chi}_1^0$ (blue line), $\tilde{\chi}_2^0$ (blue
dashed line), $\tilde{\mu}_R$ (red line) and $\tilde{\mu}_L$ (red
dashed line). In both panels the mSugra parameters have been taken as
in the SPS3 benchmark point.}
\label{fig:masses-scales}
\end{figure}

The appearance of charged particles at scales between the electroweak
scale and the GUT scale leads to changes in the beta functions
of the gauge couplings \cite{Rossi:2002zb,Buckley:2006nv}.  
This does not only change the evolution of the gauge couplings but 
also the evolution of the gaugino and scalar mass parameters 
\cite{Buckley:2006nv,Hirsch:2008gh}. The LR model contains additional 
triplets, and similar to what is observed in the seesaw models 
\cite{OurTypeIII} the mass spectrum at low energies is shifted with 
respect to mSugra expectations. Two examples of this behaviour are 
shown in figure \ref{fig:masses-scales}. In this figure we show the 
two lightest neutralino masses and the masses of the left and right 
smuons versus $v_{BL}$ (left side) and $v_R$ (right side). We note 
that also all other sfermion and gaugino masses show the same dependence and
in general smaller values are obtained for lower values of $v_{BL}$ and $v_R$. 
One finds that gaugino masses depend stronger on $v_{BL}$ and $v_R$ 
than sfermion masses and that right sleptons are the sfermions for
which the sensitivity to these vevs is smallest. 

\begin{figure}
\begin{center}
\vspace{5mm}
\includegraphics[width=0.47\textwidth]{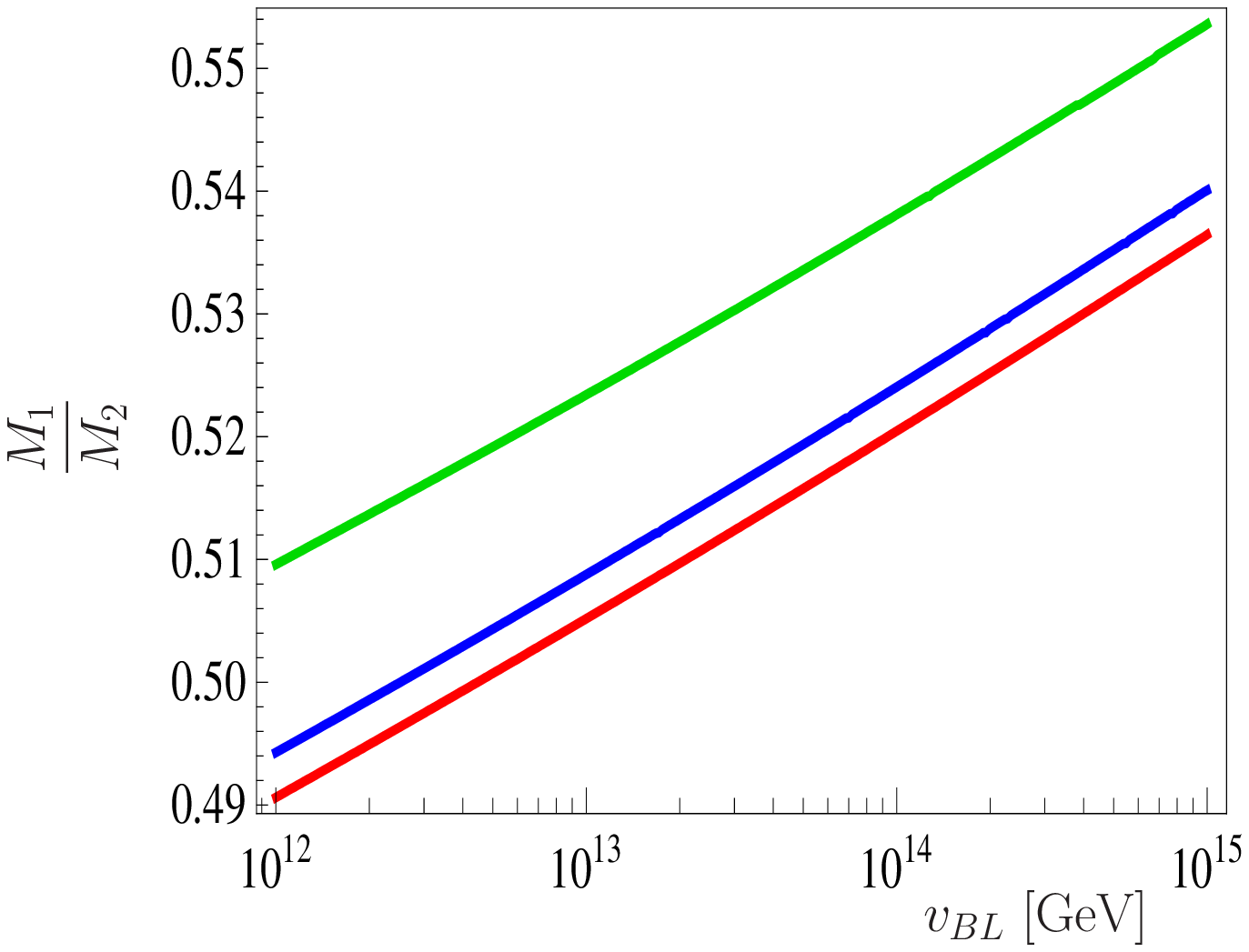}
\hspace{5mm}
\includegraphics[width=0.47\textwidth]{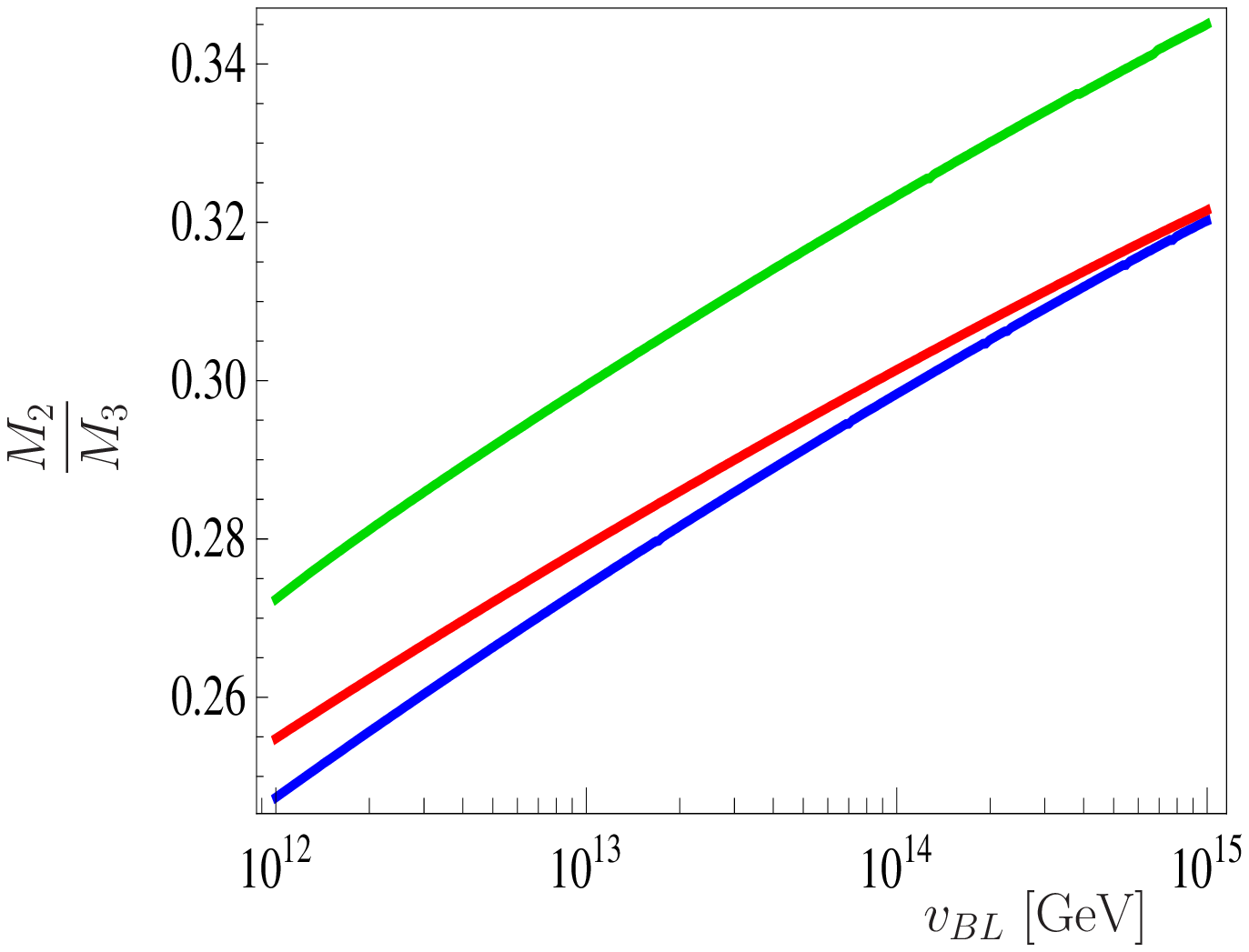}
\end{center}
\vspace{-5mm}
\caption{Gaugino mass ratios as a function of $v_{BL}$ for the fixed
value $v_R = 10^{15}$ GeV. To the left, $M_1 / M_2$, whereas to the
right $M_2 / M_3$. In both figures the three coloured lines correspond
to three mSugra benchmark points: SPS1a' (blue), SPS3 (green) and SPS5
(red). Note the small variation in the numbers on the Y axis.}
\label{fig:ratiogauginosvBL}
\end{figure}

\begin{figure}
\begin{center}
\vspace{5mm}
\includegraphics[width=0.47\textwidth]{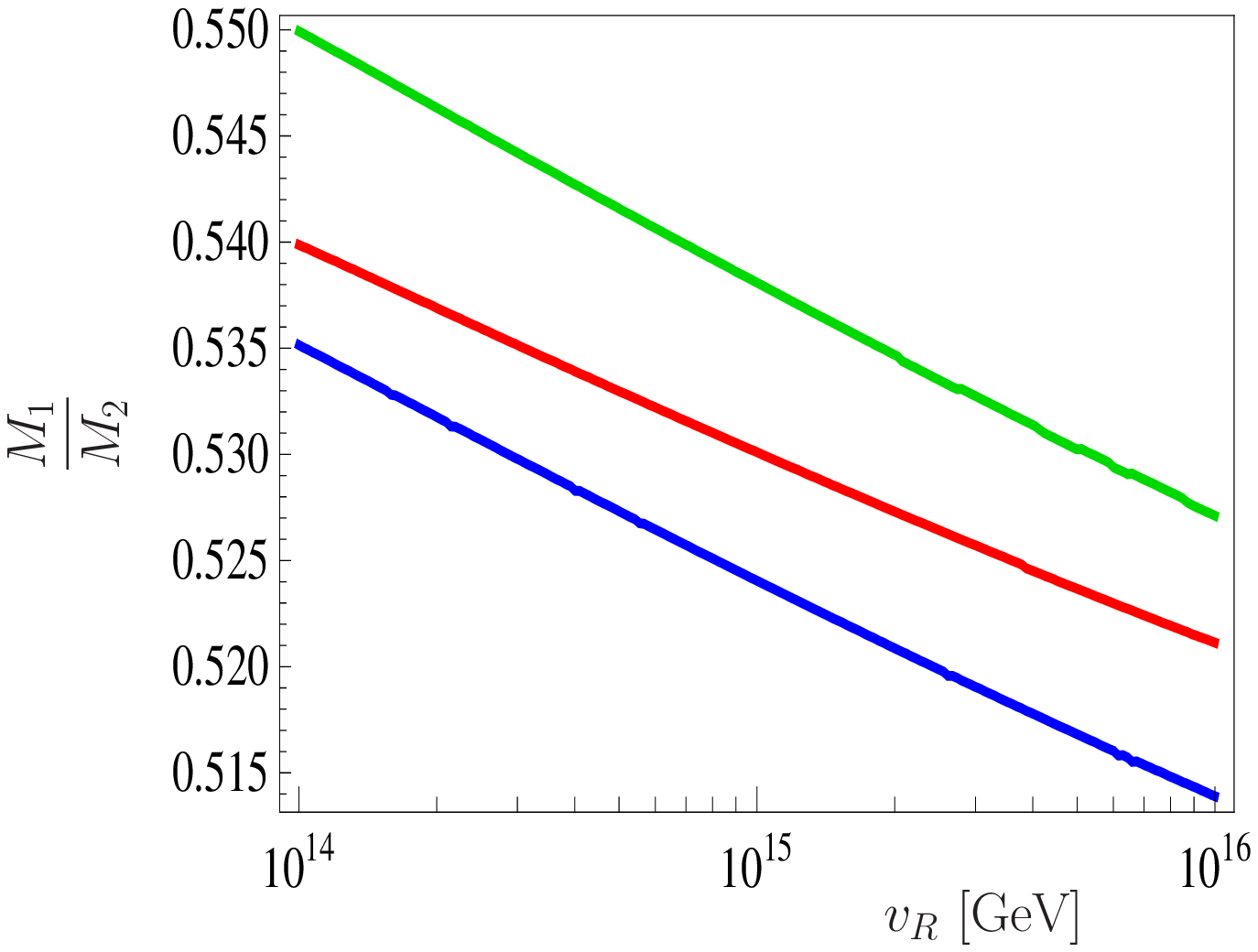}
\hspace{5mm}
\includegraphics[width=0.47\textwidth]{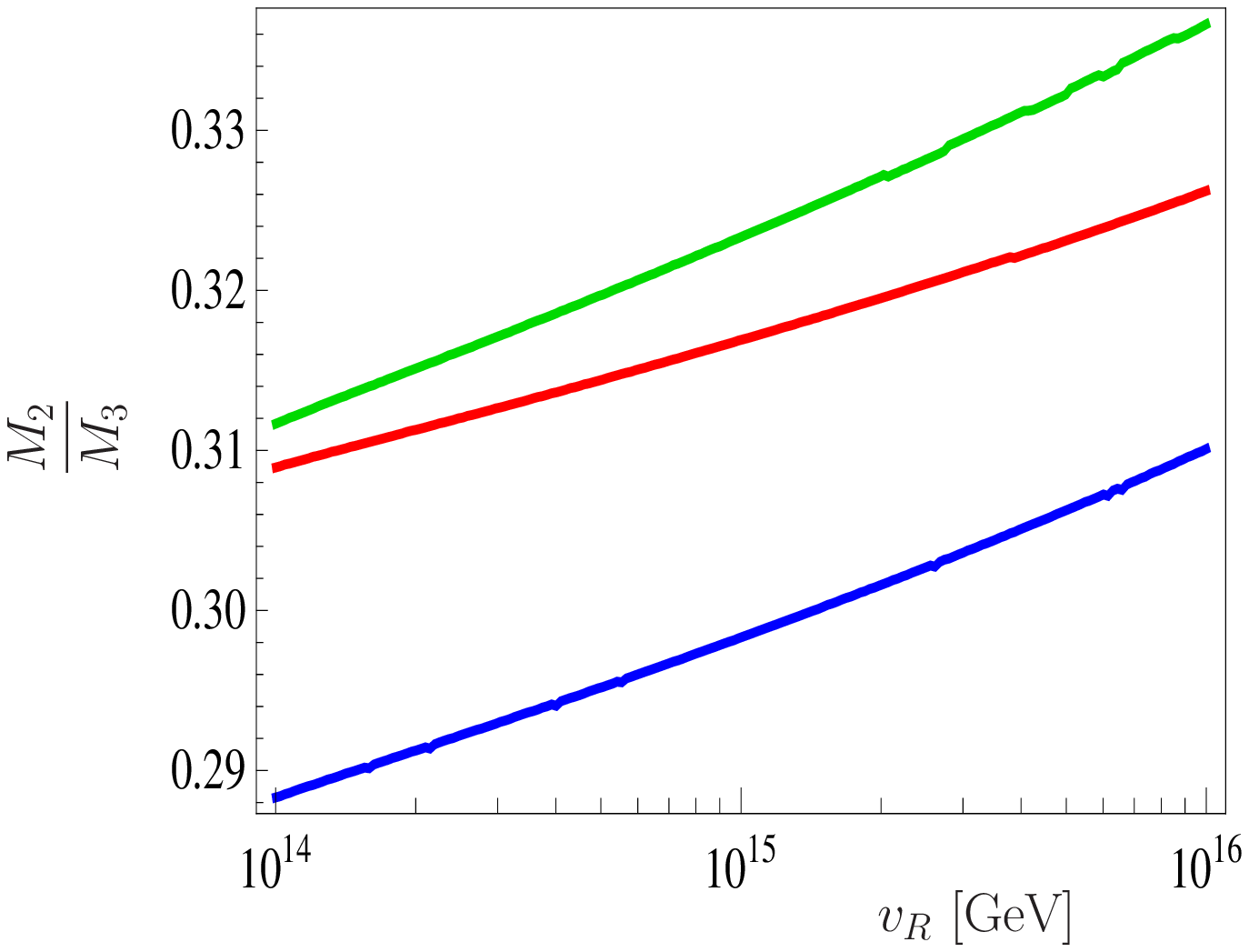}
\end{center}
\vspace{-5mm}
\caption{Gaugino mass ratios as a function of $v_R$ for the fixed
value $v_{BL} = 10^{14}$ GeV. To the left, $M_1 / M_2$, whereas to the
right $M_2 / M_3$. In both figures the three coloured lines correspond
to three mSugra benchmark points: SPS1a' (blue), SPS3 (green) and SPS5
(red). Note the small variation in the numbers on the Y axis.}
\label{fig:ratiogauginosvR}
\end{figure}

The change in the low energy spectrum, however, maintains to a good 
degree the standard mSugra expectation for the ratios of gaugino 
asses, as shown in figures \ref{fig:ratiogauginosvBL} and 
\ref{fig:ratiogauginosvR}. Here, figure \ref{fig:ratiogauginosvBL} 
shows the ratios $M_1/M_2$ and $M_2/M_3$ versus $v_{BL}$, while 
figure \ref{fig:ratiogauginosvR} shows the same ratios versus $v_R$. 
Shown are the results for three different SUSY points, which in 
the limit of $v_R,v_{BL} \to m_{GUT}$ approach the standard SPS 
points SPS1a' \cite{AguilarSaavedra:2005pw}, SPS3 and SPS5 
\cite{Allanach:2002nj}. For example, the ratio $M_1/M_2$ is 
expected to be $(5/3)\tan^2\theta_W \simeq 0.5$ at 1-loop order 
in mSugra. The exact ratio, however, depends on higher order corrections,
and thus on the SUSY spectrum. The LR model will thus appear rather mSugra 
like, if these ratios are measured. Only with very high precision 
on mass measurements, possible only at a linear collider, can one 
hope to find any (indirect) dependence on $v_{BL}$ and $v_R$.

\subsection{LFV of leptons}

Lepton flavour violation in charged lepton decays has attracted a lot
of attention for decades. Processes like $\mu \to e \gamma$ are highly
suppressed in the standard model (plus non-zero neutrino masses) due 
to the GIM mechanism \cite{Glashow:1970gm}, and thus the observation 
of these rare decays would imply new physics. The MEG experiment \cite{meg} 
is currently the most advanced experimental setup in the search for 
$\mu^+ \to e^+ \gamma$. This rare decay will be observed if its branching 
ratio is above the MEG expected sensitivity, around $Br(\mu \to e \gamma) \sim
10^{-13}$. 

LFV decays like $l_i \to l_j \gamma$ are induced by 1-loop diagrams
with the exchange of neutralinos and sleptons. They can be described
by the effective Lagrangian, see for example the review \cite{Kuno:1999jp},
\begin{equation}
\mathcal{L}_{eff} = e \frac{m_i}{2} \bar{l}_i \sigma_{\mu \nu} F^{\mu \nu} 
(A_L^{ij} P_L + A_R^{ij} P_R) l_j + h.c. \thickspace.
\end{equation}

Here $P_{L,R} = \frac{1}{2}(1 \mp \gamma_5)$ are the usual chirality
projectors and therefore the couplings $A_L$ and $A_R$ are generated
by loops with left and right sleptons, respectively. In our 
numerical calculation we use exact expressions for $A_L$ and $A_R$. 
However, for an easier understanding of the numerical results, 
we note that the relation between these couplings and the slepton soft 
masses is very approximately given by 
\begin{equation} \label{A-dependence}
A_L^{ij} \sim \frac{(m_L^2)_{ij}}{m_{SUSY}^4} \quad , \quad A_R^{ij} 
\sim \frac{(m_{e^c}^2)_{ij}}{m_{SUSY}^4} \thickspace, 
\end{equation}
where $m_{SUSY}$ is a typical supersymmetric mass. Here it has been 
assumed that (a) chargino/neutralino masses are similar to slepton 
masses and (b) A-terms mixing left-right transitions are negligible. 
Therefore, due to the negligible off-diagonal entries in 
$m_{e^c}^2$, a pure seesaw model predicts $A_R \simeq 0$.

The branching ratio for $l_i \to l_j \gamma$ can be calculated from
the previous formulas. The result is
\begin{equation} \label{brLLG}
Br(l_i \to l_j \gamma) = \frac{48 \pi^3 \alpha}{G_F^2} 
\left( |A_L^{ij}|^2 + |A_R^{ij}|^2 \right) Br(l_i \to l_j \nu_i \bar{\nu}_j) \thickspace .
\end{equation}

\begin{figure}
\begin{center}
\vspace{5mm}
\includegraphics[width=0.49\textwidth]{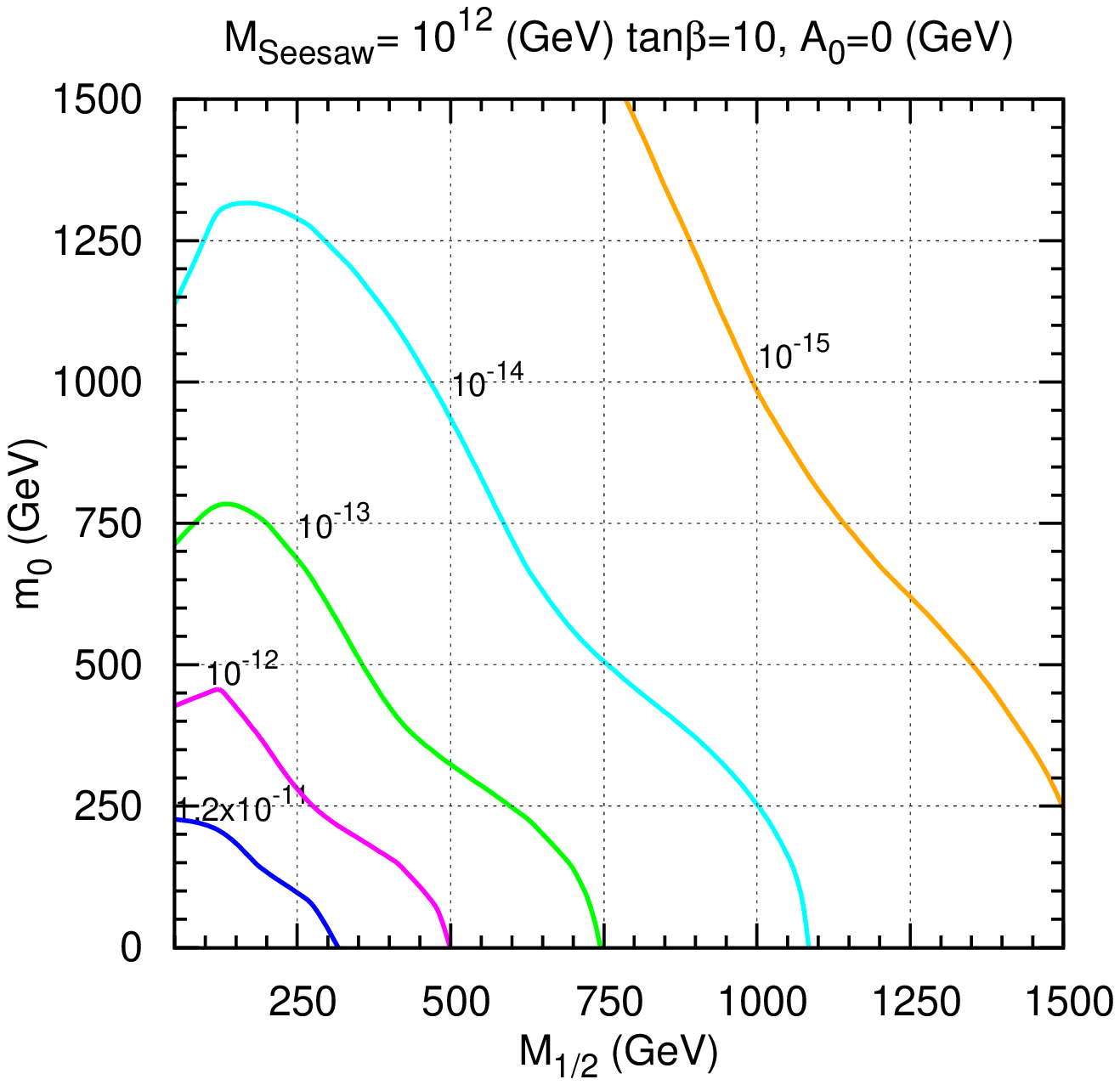}
\includegraphics[width=0.49\textwidth]{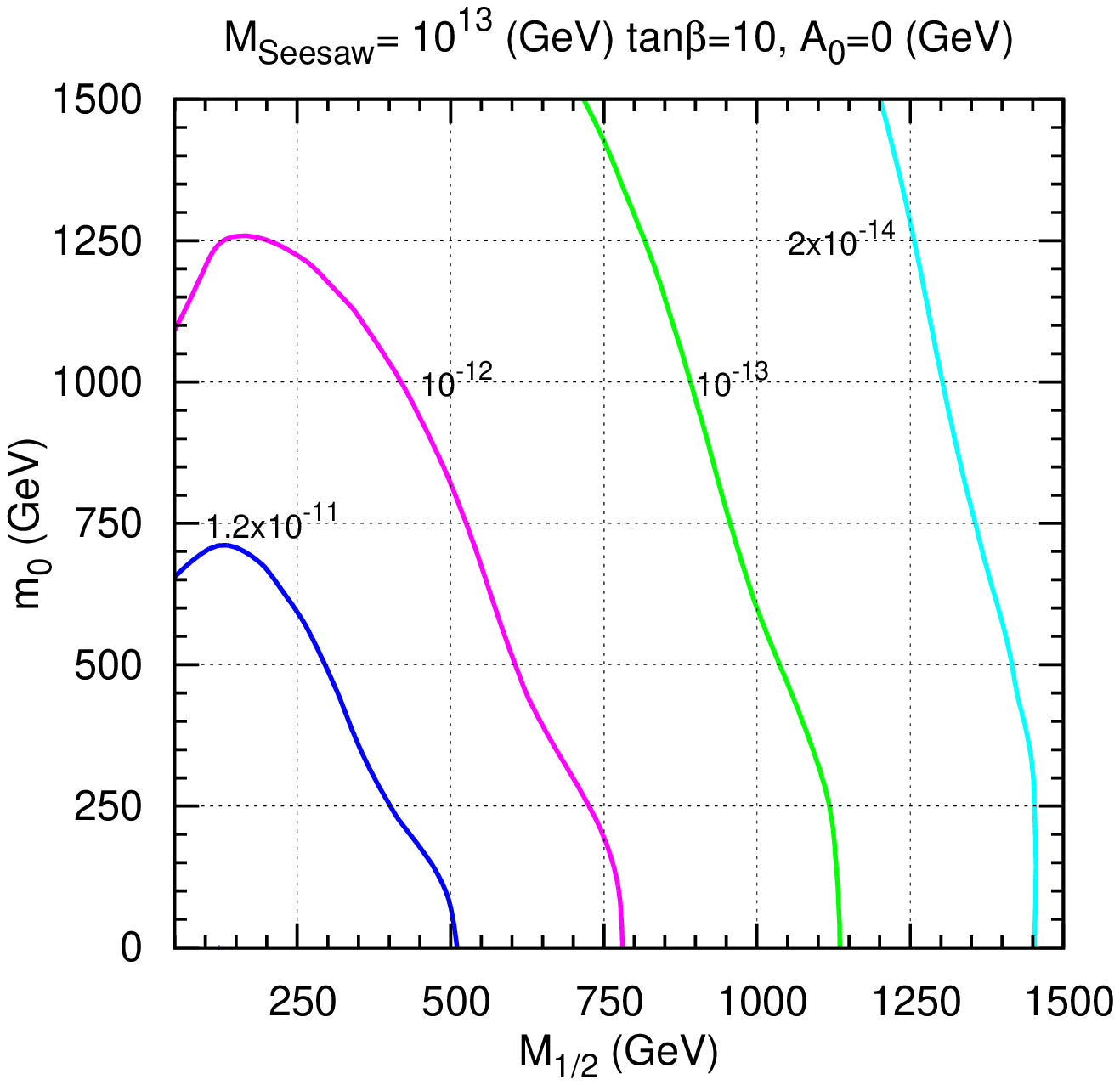}
\end{center}
\vspace{-5mm}
\caption{Contours of $Br(\mu \to e \gamma)$ in the $m_0,M_{1/2}$ plane
for $v_{BL} = 10^{14}$ GeV and $v_R = 10^{15}$ GeV. To the left $M_S =
10^{12}$ GeV, whereas to the right $M_S = 10^{13}$ GeV. Neutrino
oscillation data have been fitted with the $Y_\nu$ fit.}
\label{fig:contourMuEG}
\end{figure}

Figure \ref{fig:contourMuEG} shows two examples for $Br(\mu \to e
\gamma)$ in the $m_0,M_{1/2}$ plane. Here, we have fixed $v_{BL} = 10^{14}$ 
GeV and $v_R = 10^{15}$ GeV and show to the left $M_S =10^{12}$ GeV, 
whereas to the right $M_S = 10^{13}$ GeV. Here we have assumed a
degenerate spectrum right-handed neutrinos which we denote by $M_S = M_{R i}$.
 Once Yukawas are fitted 
to explain the observed neutrino masses, the branching ratio shows 
an approximately quadratic dependence on the seesaw scale, with lower 
$M_S$ giving smaller $Br(\mu \to e\gamma)$. As expected, the branching ratio
also strongly decreases as $m_0$ and/or $M_{1/2}$ increase. This is because
the superparticles in the loops leading to $\mu \to e \gamma$ become
heavier in these directions, suppressing the decay rate. In fact, from
equations \eqref{A-dependence} and \eqref{brLLG} one easily finds the
dependence
\begin{equation}
Br(\mu \to e \gamma) \sim \frac{48 \pi^3 \alpha}{G_F^2} 
\frac{(m_{L,\tilde{e}^c}^2)_{ij}^2}{m_{SUSY}^8} \thickspace,
\end{equation}
which shows that $Br(\mu \to e \gamma)$ decreases as $m_{SUSY}^{-8}$.

It is also remarkable that for a given seesaw scale, $Br(\mu \to e \gamma)$ 
is sizeably larger in the LR model than in a pure seesaw type-I model, 
see for example \cite{Esteves:2009vg}. 
The explanation of this is that right sleptons contribute 
significantly in the LR model to $Br(\mu \to e \gamma)$ and these 
contributions are absent in seesaw models. 

As already discussed, a pure seesaw model predicts simply $A_R \simeq
0$. However, in the LR model we expect a more complicated picture. 
Left-right symmetry implies that, above the parity breaking scale, 
non-negligible flavour violating entries are generated in 
$m_{e^c}^2$. Therefore, $A_R \ne 0$ is obtained at 
low energy. The angular distribution of the outgoing positron at, 
for example, the MEG experiment could be used to discriminate between 
left- and right-handed polarized states \cite{Okada:1999zk,Hisano:2009ae}. 
If MEG is able to measure the positron polarization asymmetry, defined as

\begin{equation}
\mathcal{A}(\mu^+ \to e^+ \gamma) = \frac{|A_L|^2-|A_R|^2}{|A_L|^2+|A_R|^2},
\end{equation}

\noindent there will be an additional observable to distinguish from minimal 
seesaw models. In a pure seesaw model one expects $\mathcal{A} \simeq +1$
to a very
good accuracy. However, the LR model typically leads to significant 
departures from this expectation, giving an interesting signature 
of the high energy restoration of parity.

\begin{figure}
\begin{center}
\vspace{5mm}
\includegraphics[width=0.49\textwidth]{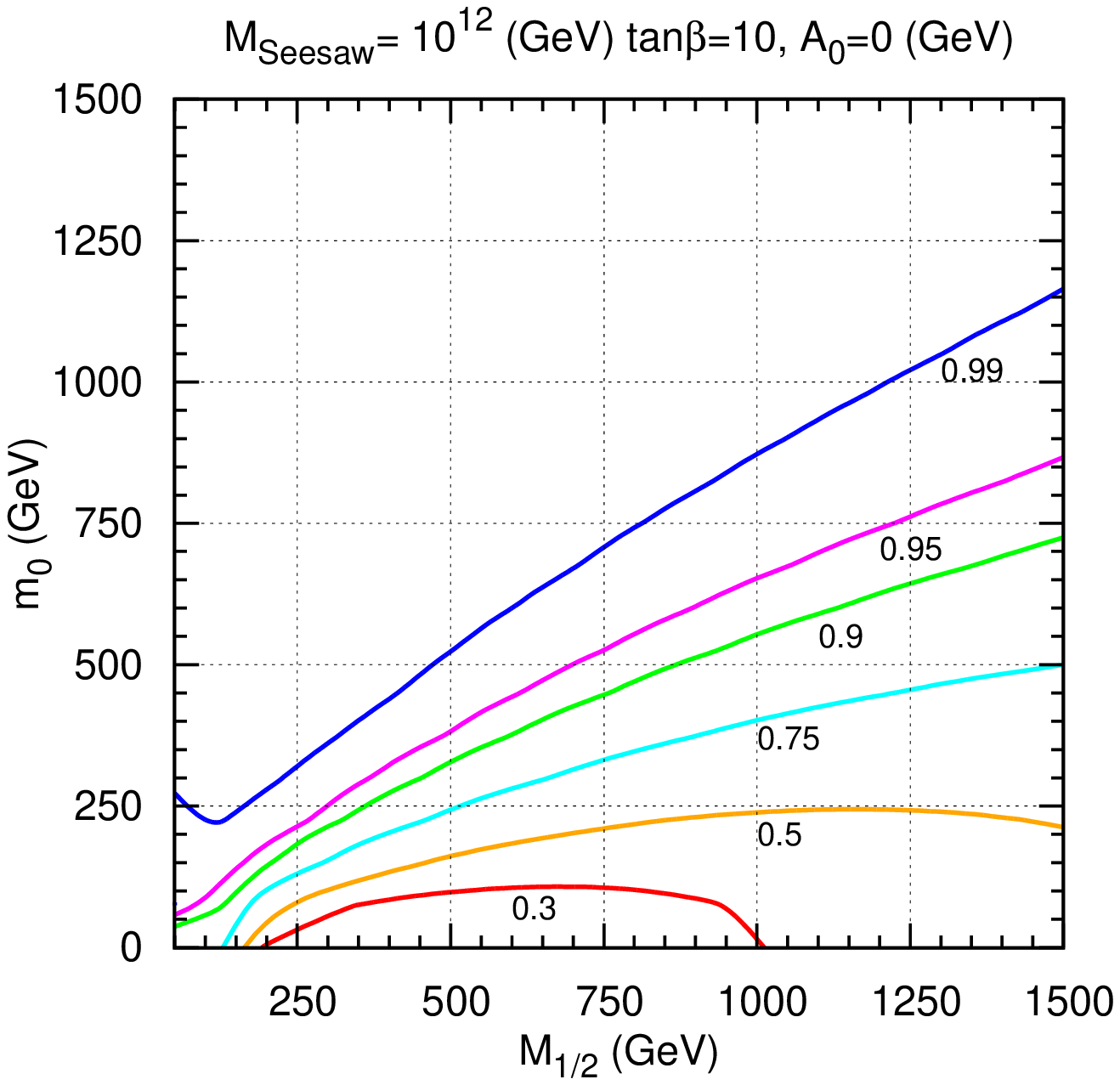}
\includegraphics[width=0.49\textwidth]{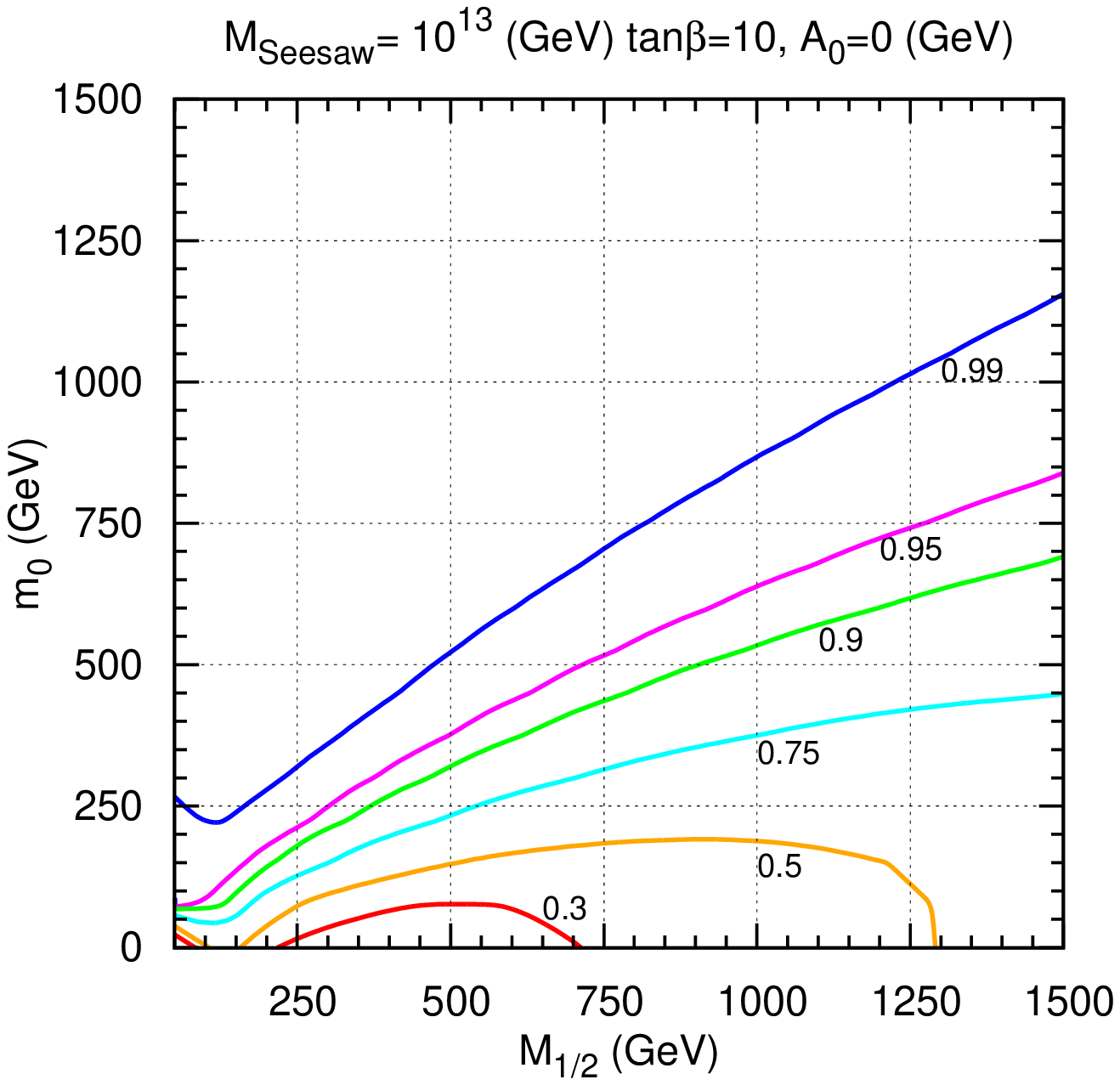}
\end{center}
\vspace{-5mm}
\caption{Contours of $\mathcal{A}(\mu^+ \to e^+ \gamma)$ in the
$m_0,M_{1/2}$ plane. To the left $M_S = 10^{12}$ GeV, whereas to the
right $M_S = 10^{13}$ GeV. The parameters have been chosen as in
figure \ref{fig:contourMuEG}.}
\label{fig:contourALR}
\end{figure}

Figure \ref{fig:contourALR} shows contours for $\mathcal{A}(\mu^+ \to
e^+ \gamma)$ in the $m_0,M_{1/2}$ plane. For the corresponding branching 
ratios see figure \ref{fig:contourMuEG}. Note the rather strong dependence 
on $m_0$. The latter can be understood as follows. Since $v_{BL}$ in 
these examples is one order of magnitude smaller than $v_R$,
and the $Y_\nu$ fit has been used, the LFV 
mixing angles in the left slepton sector are larger than the corresponding 
LFV entries in the right sleptons. At very large values of $m_0$, were the 
masses of right and left sleptons are of comparable magnitude, therefore 
``left'' LFV is more important and the model approaches the pure seesaw 
expectation. At smaller values of $m_0$, right sleptons are lighter 
than left sleptons, and due to the strong dependence of $\mu\to e\gamma$ 
on the sfermion masses entering the loop calculation, see eq. 
(\ref{A-dependence}), $A_R$ and $A_L$ can become comparable, despite 
the smaller LFV entries in right slepton mass matrices. In the limit 
of very small right slepton masses the model then approaches 
$\mathcal{A} \sim 0$. We have not explicitly searched for regions of 
parameter space with $\mathcal{A} < 0$, but one expects that negative 
values for $\mathcal{A}$ are possible if $v_{BL}$ is not much below 
$v_R$ and sleptons are light at the same time, i.e. small values of 
$m_0$ and $M_{1/2}$. Note that, again due to the LR symmetry above to $v_R$,
the model can never approach the limit $\mathcal{A} =-1$ exactly.

\begin{figure}
\begin{center}
\vspace{5mm}
\includegraphics[width=0.49\textwidth]{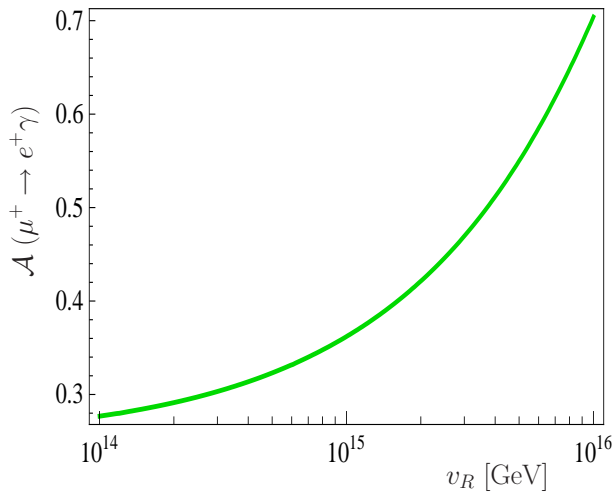}
\end{center}
\vspace{-5mm}
\caption{Positron polarization asymmetry $\mathcal{A}(\mu^+ \to e^+
\gamma)$ as a function of $v_R$ for the parameter choice $M_S =
10^{13}$ GeV and $v_{BL} = 10^{14}$ GeV. The mSugra parameters have
been taken as in the SPS3 benchmark point and neutrino oscillation
data have been fitted with the $Y_\nu$ fit, assuming degenerate
right-handed neutrinos.}
\label{fig:SPS3-megpolvR-Yv}
\end{figure}

The positron polarization asymmetry is very sensitive to the high
energy scales. Figure \ref{fig:SPS3-megpolvR-Yv} shows $\mathcal{A}$
as a function of $v_R$ for $M_S = 10^{13}$ GeV, $v_{BL} = 10^{14}$ GeV
and the mSugra parameters as in the SPS3 benchmark point. The plot has
been obtained using the $Y_\nu$ fit. This example shows that as $v_R$
approaches $m_{GUT}$ the positron polarization $\mathcal{A}$
approaches $+1$, which means $A_L$ dominates the calculation. This is 
because, in the $Y_\nu$ fit, the right-handed LFV soft slepton masses, 
and thus the corresponding $A_R$ coupling, only run from $m_{GUT}$ to $v_R$. 

\begin{figure}
\begin{center}
\vspace{5mm}
\includegraphics[width=0.5\textwidth]{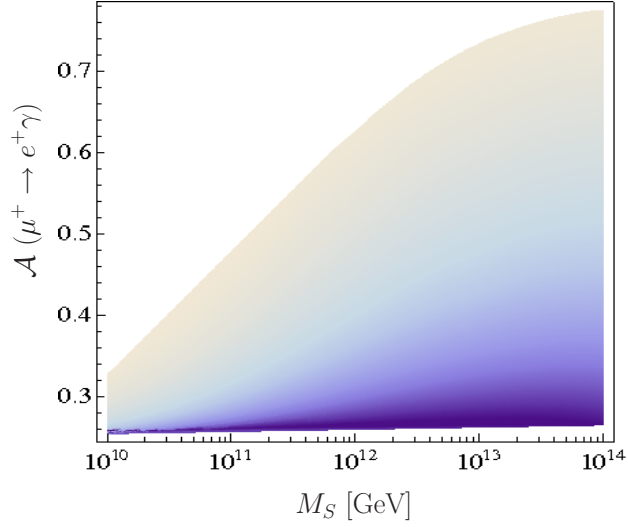}
\end{center}
\vspace{-5mm}
\caption{Positron polarization asymmetry $\mathcal{A}(\mu^+ \to e^+
\gamma)$ as a function of the seesaw scale, defined as the mass of the
lightest right-handed neutrino, for the parameter choice $v_{BL} =
10^{15}$ GeV and $v_R \in [10^{15},10^{16}]$ GeV. Lighter colours mean
higher values of $v_R$. The mSugra parameters have been taken as in
the SPS3 benchmark point and neutrino oscillation data have been
fitted with the $Y_\nu$ fit, assuming degenerate right-handed
neutrinos.}
\label{fig:megpol-MR}
\end{figure}

$\mathcal{A}(\mu^+ \to e^+ \gamma)$ also has an important dependence
on the seesaw scale. This is shown in figure \ref{fig:megpol-MR},
where $\mathcal{A}$ is plotted as a function of the lightest
right-handed neutrino mass. This dependence can be easily understood from
the seesaw formula for neutrino masses. It implies that larger $M_S$ 
requires larger Yukawa parameters in order to fit neutrino masses which, 
in turn, leads to larger flavour violating soft terms due to RGE
running. However, note that, for very small seesaw scales all lepton flavour 
violating effects are negligible and no asymmetry is produced, since $A_L \sim
A_R \sim 0$.

In addition, figure \ref{fig:megpol-MR} shows again the relevance of
$v_R$, which determines the parity breaking scale at which the LFV
entries in the right-handed slepton sector essentially stop running. 
Lighter colours indicate larger $v_R$. As shown already in figure
\ref{fig:SPS3-megpolvR-Yv} for a particular point, the positron
polarization approaches $+1$ as $v_R$ is increased. 

\begin{figure}
\begin{center}
\vspace{5mm}
\includegraphics[width=0.5\textwidth]{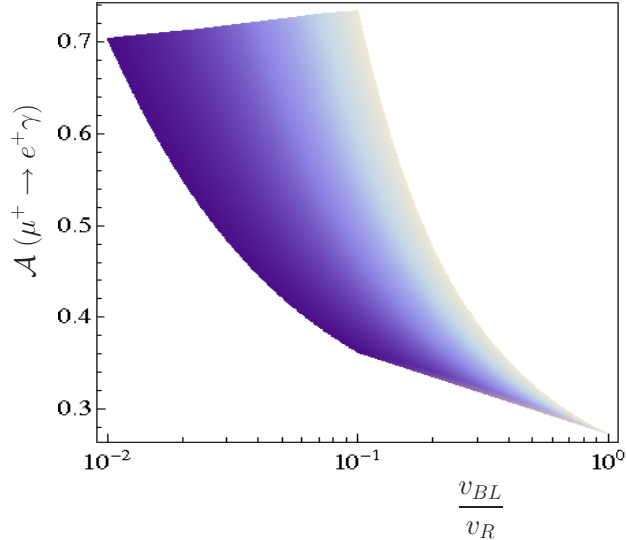}
\end{center}
\vspace{-5mm}
\caption{Positron polarization asymmetry $\mathcal{A}(\mu^+ \to e^+
\gamma)$ as a function of the ratio $v_{BL} / v_R$. The seesaw scale
$M_S$ has been fixed to $10^{13}$ GeV, whereas $v_{BL}$ and $v_R$ take
values in the ranges $v_{BL} \in [10^{14},10^{15}]$ GeV and $v_R \in
[10^{15},10^{16}]$ GeV. Lighter colours indicate larger $v_{BL}$. The
mSugra parameters have been taken as in the SPS3 benchmark point and
neutrino oscillation data have been fitted with the $Y_\nu$ fit,
assuming degenerate right-handed neutrinos.}
\label{fig:megpol-ratio}
\end{figure}

Below the $SU(2)_R$ breaking scale parity is broken and left and
right slepton soft masses evolve differently. The approximate
solutions to the RGEs in equations \eqref{apprge2} and \eqref{apprge4}
show that, if neutrino data is fitted according to the $Y_\nu$ fit,
the left-handed ones keep running from the $SU(2)_R$ breaking scale to
the $U(1)_{B-L}$ scale. In this case one expects larger flavour
violating effects in the left-handed slepton sector and a correlation
with the ratio $v_{BL}/v_R$, which measures the difference between the
breaking scales. This correlation, only present in the $Y_\nu$ fit, 
is shown in figure \ref{fig:megpol-ratio}. On the one hand, one finds 
that as $v_{BL}$ and $v_R$ become very different, $v_{BL}/v_R \ll 1$, 
the positron asymmetry approaches $\mathcal{A}=+1$. On the other hand, 
when the two breaking scales are close, $v_{BL}/v_R \sim 1$, this effect 
disappears and the positron polarization asymmetry approaches
$\mathcal{A}=0$. Note that the $Y_\nu$ fit does not usually produce a 
negative value for $\mathcal{A}$ since the LFV terms in the right
slepton sector never run more than the corresponding terms in the
left-handed sector. The only possible exception to this general rule 
is, as discussed above, in the limit of very small $m_0$ and 
$v_{BL}/v_R \sim 1$. 

\begin{figure}
\begin{center}
\vspace{5mm}
\includegraphics[width=0.5\textwidth]{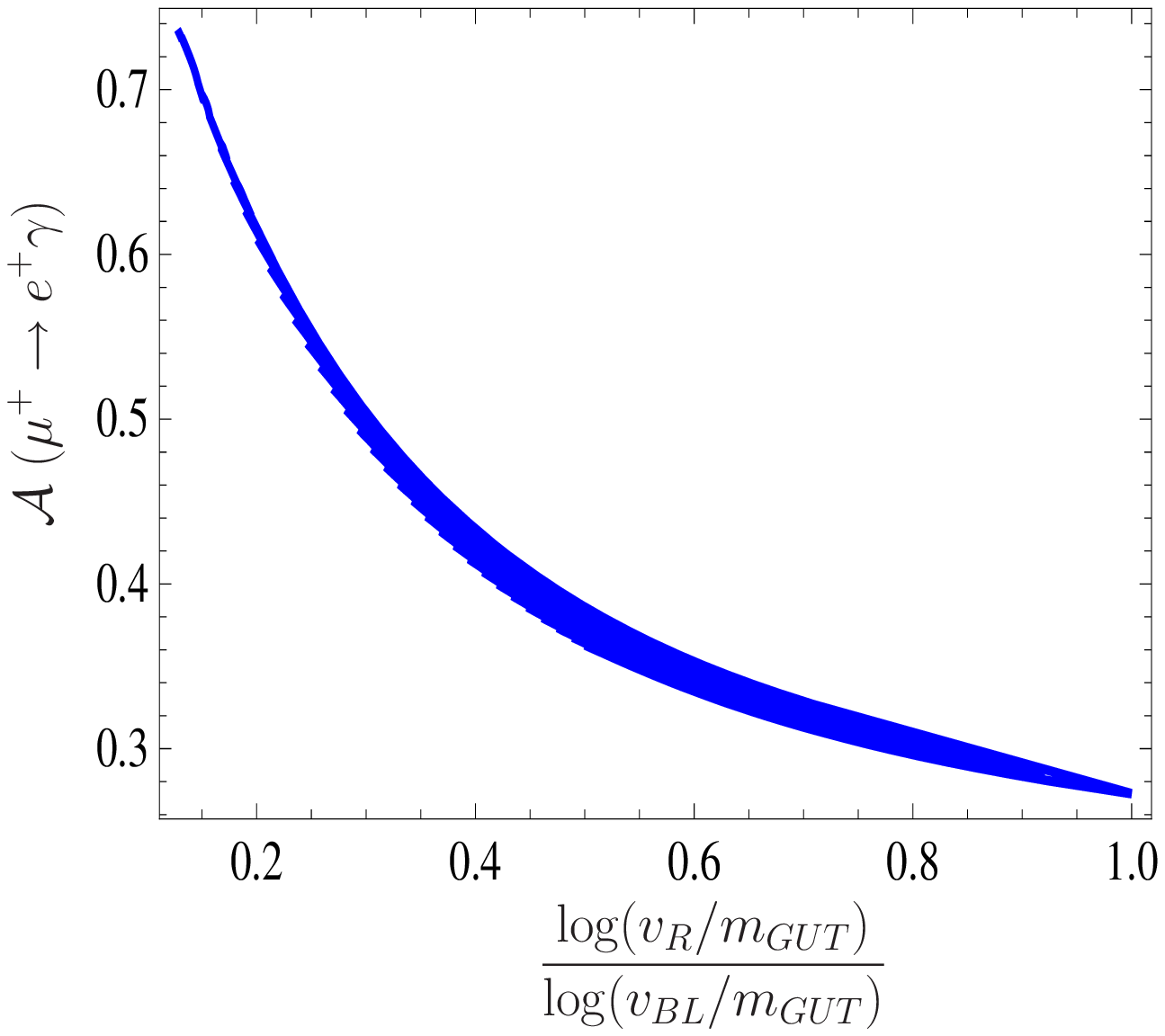}
\end{center}
\vspace{-5mm}
\caption{Positron polarization asymmetry $\mathcal{A}(\mu^+ \to e^+
\gamma)$ as a function of $\log (v_R/m_{GUT}) / \log
(v_{BL}/m_{GUT})$. The parameters have been chosen as in figure
\ref{fig:megpol-ratio}.}
\label{fig:megpol-ratio2}
\end{figure}

The determination of the ratio $v_{BL}/v_R$ from figure
\ref{fig:megpol-ratio} is shown to be very inaccurate. This is due to
the fact that other parameters, most importantly $m_{GUT}$ (which itself 
has an important dependence on the values of $v_{BL}$ and $v_R$), have a
strong impact on the results. Therefore, although it would be possible
to constrain the high energy structure of the theory, a precise
determination of the ratio $v_{BL}/v_R$ will require additional
input. Figure \ref{fig:megpol-ratio2}, on the other hand, shows that
the polarization asymmetry $\mathcal{A}(\mu^+ \to e^+ \gamma)$ is much
better correlated with the quantity $\log (v_R/m_{GUT}) / \log
(v_{BL}/m_{GUT})$. This is as expected from equations \eqref{apprge2} and
\eqref{apprge4} and confirms the validity of this approximation.

\begin{figure}
\begin{center}
\vspace{5mm}
\includegraphics[width=0.49\textwidth]{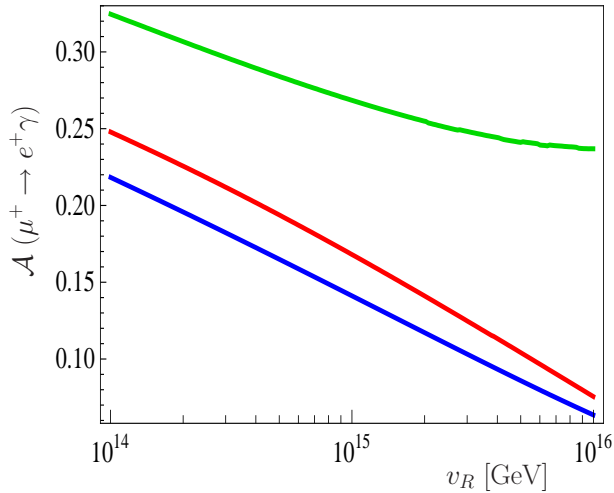}
\end{center}
\vspace{-5mm}
\caption{Positron polarization asymmetry $\mathcal{A}(\mu^+ \to e^+
\gamma)$ as a function of $v_R$ for three different mSugra benchmark
points: SPS1a' (blue line), SPS3 (green line) and SPS5 (red
line). In this figure a fixed value $v_{BL} = 10^{14}$ is taken.
Neutrino oscillation data have been fitted with the $f$ fit.}
\label{fig:megpolvR-f}
\end{figure}

We close our discussion on the positron polarization asymmetry with
some comments on the $f$ fit. Since this type of fit leads to $\Delta
m_L^2 \sim \Delta m_{e^c}^2 \sim 0$ in the $v_{BL} - v_R$
energy region, there is little dependence on these symmetry breaking
scales. This is illustrated in figure \ref{fig:megpolvR-f}, where the
asymmetry $\mathcal{A}$ is plotted as a function of $v_R$ for three
different mSugra benchmark points: SPS1a' (blue line), SPS3 (green
line) and SPS5 (red line). One clearly sees that the dependence on $v_R$
is quite weak compared to the $Y_\nu$ fit. In fact, the variations in 
this figure are mostly due to the changes in the low energy supersymmetric 
spectrum due to different $v_R$ values. In the case of the $f$-fit one 
then typically finds $\mathcal{A} \in [0.0-0.3]$.

\subsection{LFV at LHC/ILC}

Lepton flavour violation might show up at collider experiments as
well. Although the following discussion is focused on the LHC
discovery potential for LFV signatures, let us emphasize that a future
linear collider will be able to determine the relevant
observables with much higher precision.

\begin{figure}
\begin{center}
\vspace{5mm}
\includegraphics[width=0.47\textwidth]{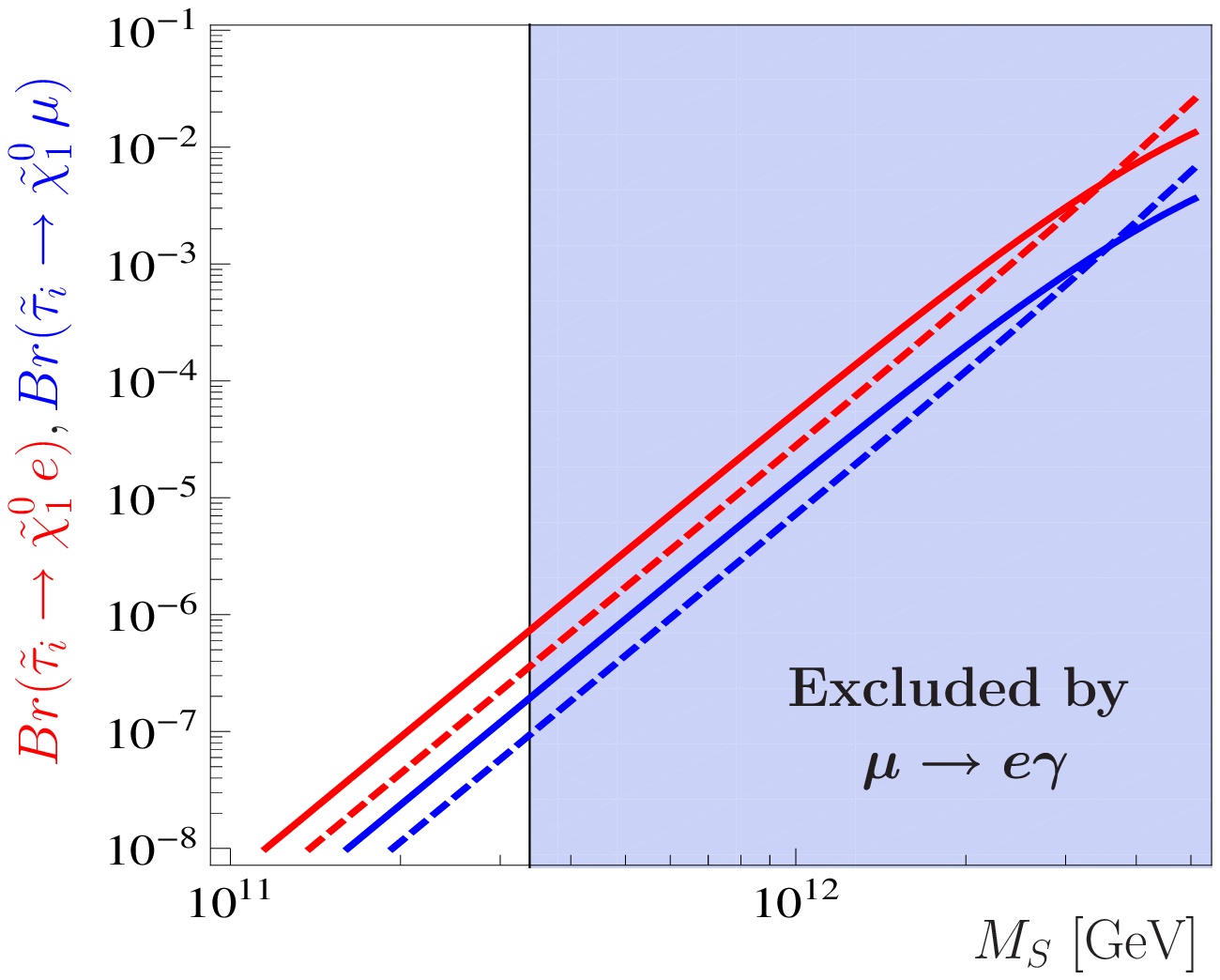}
\hspace{5mm}
\includegraphics[width=0.47\textwidth]{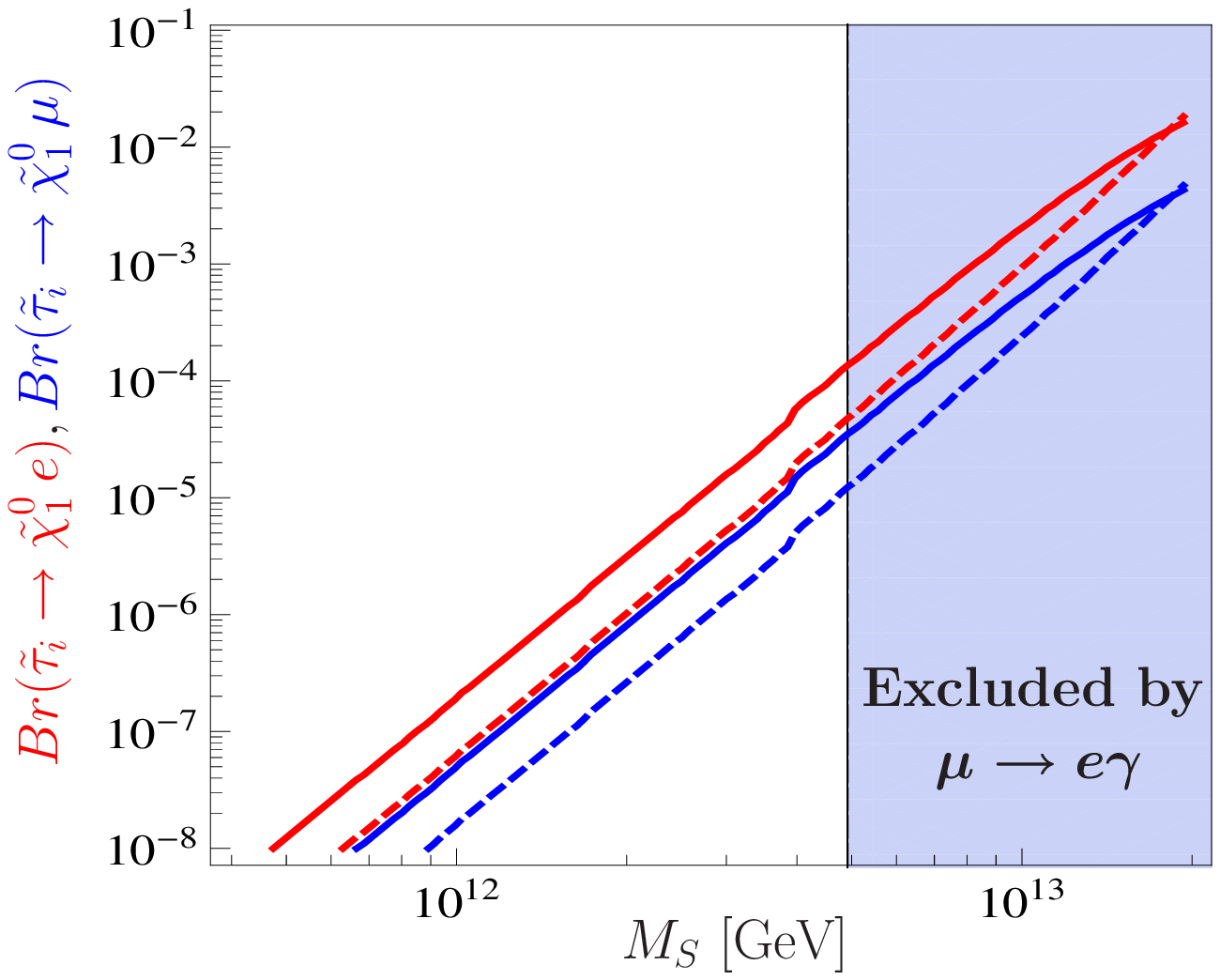}
\end{center}
\vspace{-5mm}
\caption{$Br(\tilde{\tau}_i \to \tilde{\chi}_1^0 \: e)$ and
$Br(\tilde{\tau}_i \to \tilde{\chi}_1^0 \: \mu)$ as a function of the
seesaw scale, defined as the mass of the lightest right-handed
neutrino, for the parameter choice $v_{BL} = 10^{15}$ GeV and $v_R = 5
\cdot 10^{15}$ GeV. The dashed lines correspond to $\tau_1 \simeq
\tau_R$ and the solid ones to $\tau_2 \simeq \tau_L$. To the left, the
mSugra parameters have been taken as in the SPS1a' benchmark point,
whereas to the right as in the SPS3 benchmark point. In both figures
neutrino oscillation data have been fitted according to the $f$ fit,
with non-degenerate right-handed neutrinos. The blue shaded regions
are excluded by $\mu \to e \gamma$.}
\label{fig:stauFV}
\end{figure}

Figure \ref{fig:stauFV} shows $Br(\tilde{\tau}_i \to \tilde{\chi}_1^0
\: e)$ and $Br(\tilde{\tau}_i \to \tilde{\chi}_1^0 \: \mu)$ as a
function of the seesaw scale. The dashed lines correspond to $\tau_1 \simeq
\tau_R$ and the solid ones to $\tau_2 \simeq \tau_L$.
As in the case of $\mu\to e \gamma$, 
see figure \ref{fig:contourMuEG}, lower seesaw scales imply less 
flavour violating effects due to smaller Yukawa couplings.
Moreover, figure \ref{fig:stauFV} presents the same results for two
different benchmark points, SPS1a' and SPS3. As already shown in
figure \ref{fig:contourMuEG}, $\mu \to e \gamma$ is strongly dependent
on the SUSY spectrum. For lighter supersymmetric particles, as in the
benchmark point SPS1a', $\mu\to e\gamma$ is large, setting strong limits 
on the seesaw scale and thus on the possibility to observe LFV at 
colliders. In the case of heavier spectrums, as in SPS3, $\mu
\to e \gamma$ is still the most stringent constraint, but larger 
values of the seesaw scale and thus LFV violating branching ratios 
become allowed. Whether decays such as $Br(\tilde{\tau}_i \to \tilde{\chi}_1^0
\: e)$ and $Br(\tilde{\tau}_i \to \tilde{\chi}_1^0 \: \mu)$ are 
observable at the LHC or not, thus depends very sensitively on the 
unknown $m_0$, $M_{1/2}$ and $M_S$. 

Furthermore, the right panel of figure \ref{fig:stauFV} also shows
that right staus can also have LFV decays with sizable rates.
Of course, as 
emphasized already above, this is the main novelty of the LR model 
compared to pure seesaw models. This is
direct consequence of parity restoration at high energies.

\begin{figure}
\begin{center}
\vspace{5mm}
\includegraphics[width=0.5\textwidth]{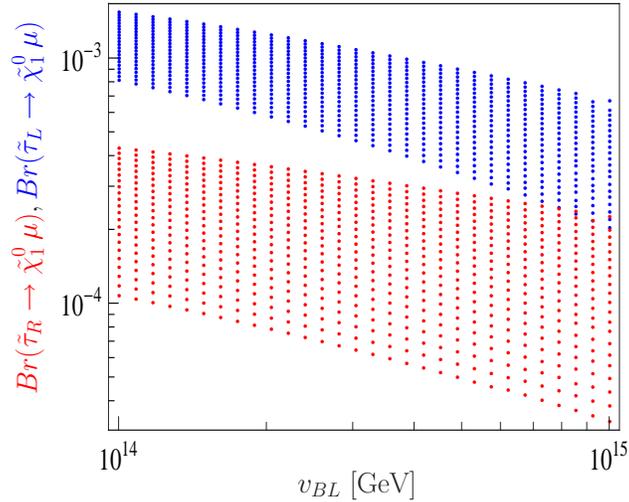}
\end{center}
\vspace{-5mm}
\caption{$Br(\tilde{\tau}_L \to \tilde{\chi}_1^0 \: \mu)$ and
$Br(\tilde{\tau}_R \to \tilde{\chi}_1^0 \: \mu)$ as a function of
$v_{BL}$ for $M_S = 10^{13}$ GeV and $v_R \in [10^{15},5 \cdot
10^{15}]$ GeV. Red dots correspond to $\tau_1 \simeq \tau_R$, whereas
the blue ones correspond to $\tau_2 \simeq \tau_L$. The mSugra
parameters have been taken as in the SPS3 benchmark point and neutrino
oscillation data have been fitted with the $Y_\nu$ fit, assuming
degenerate right-handed neutrinos.}
\label{fig:difLR}
\end{figure}

Moreover, as in our analysis of the positron polarization asymmetry,
one expects to find that if the difference between $v_R$ and $v_{BL}$
is increased, the difference between the LFV entries in the L and R
sectors gets increased as well. This property of the $Y_\nu$ fit is
shown in figure \ref{fig:difLR}, which shows branching ratios for the
LFV decays of the staus as a function of $v_{BL}$ for $v_R \in
[10^{15},5 \cdot 10^{15}]$ GeV. As the figure shows, the theoretical 
expectation is confirmed numerically: the difference between 
$Br(\tilde{\tau}_L)$ and $Br(\tilde{\tau}_R)$ strongly depends on 
the difference between $v_R$ and $v_{BL}$.

\begin{figure}
\begin{center}
\vspace{5mm}
\includegraphics[width=0.5\textwidth]{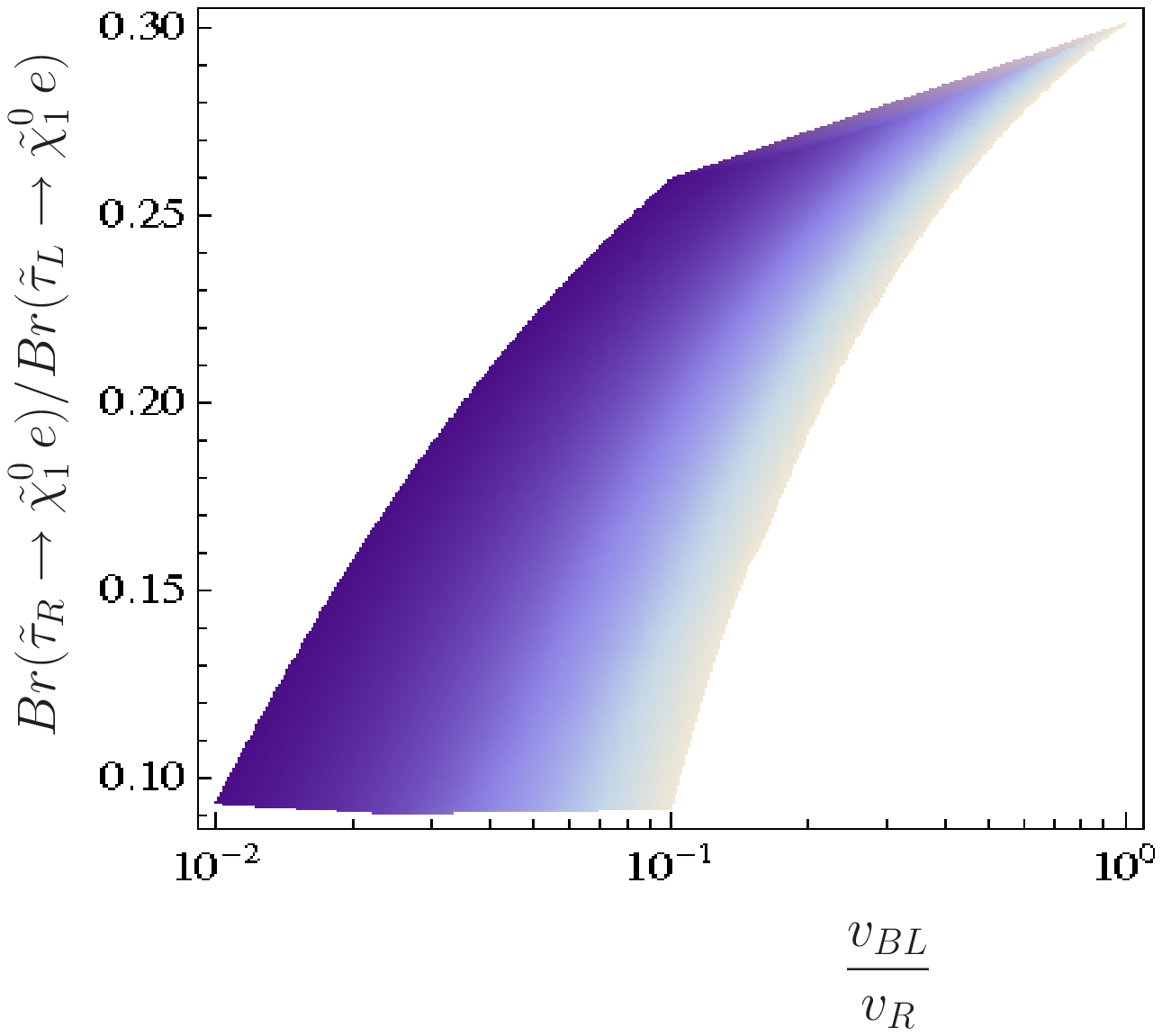}
\end{center}
\vspace{-5mm}
\caption{$Br(\tilde{\tau}_R \to \tilde{\chi}_1^0 \: \mu) /
Br(\tilde{\tau}_L \to \tilde{\chi}_1^0 \: \mu)$ as a function of
$v_{BL} / v_R$. The seesaw scale $M_S$ has been fixed to $10^{13}$
GeV, whereas $v_{BL}$ and $v_R$ take values in the ranges $v_{BL} \in
[10^{14},10^{15}]$ GeV and $v_R \in [10^{15},10^{16}]$ GeV. Lighter
colours indicate larger $v_{BL}$. The rest of the parameters have been
chosen as in figure \ref{fig:difLR}.}
\label{fig:compLR}
\end{figure}

The question arises whether one can determine the ratio $v_{BL}/v_R$
by measuring both $Br(\tilde{\tau}_L)$ and $Br(\tilde{\tau}_R)$ at
colliders. Figure \ref{fig:compLR} attempts to answer this. Here the ratio
$Br(\tilde{\tau}_R \to \tilde{\chi}_1^0 \: e) / Br(\tilde{\tau}_L \to
\tilde{\chi}_1^0 \: e)$ is plotted as a function of $v_{BL} / v_R$. A
measurement of both branching ratios would allow to constrain the
ratio $v_{BL} / v_R$ and increase our knowledge on the high energy
regimes. For the sake of brevity we do not present here the
analogous plots for other LFV slepton decays and/or other lepton final 
states, since they show very similar correlations with
$v_{BL}/v_R$. For example, in principle, one could also use the ratio
$Br(\tilde{\mu}_R \to \tilde{\chi}_1^0 \: \tau) / Br(\tilde{\mu}_L \to
\tilde{\chi}_1^0 \: \tau)$ to determine the ratio between the two high
scales.

\begin{figure}
\begin{center}
\vspace{5mm}
\includegraphics[width=0.5\textwidth]{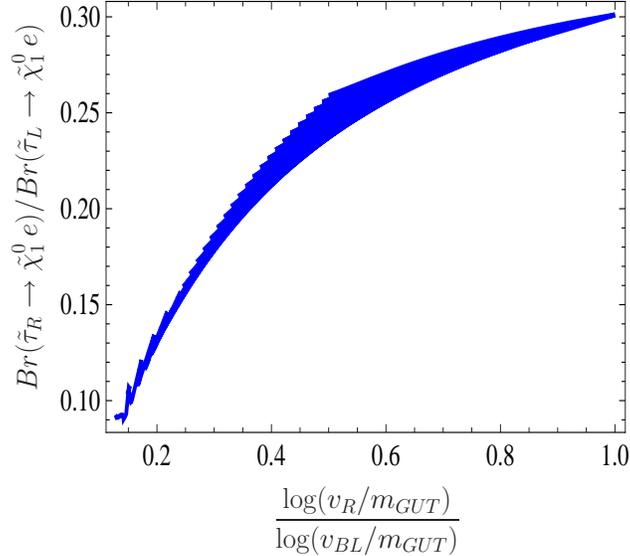}
\end{center}
\vspace{-5mm}
\caption{$Br(\tilde{\tau}_R \to \tilde{\chi}_1^0 \: e) /
Br(\tilde{\tau}_L \to \tilde{\chi}_1^0 \: e)$ as a function of $\log
(v_R/m_{GUT}) / \log (v_{BL}/m_{GUT})$. The parameters have been
chosen as in figure \ref{fig:compLR}.}
\label{fig:compLR2}
\end{figure}

However, as observed also in the polarization asymmetry for $\mu\to e \gamma$ 
there is an important dependence on other parameters of the model, 
especially the exact value of $m_{GUT}$. This implies a theoretical 
uncertainty in the determination of $v_{BL} / v_R$. Again, as for 
$\mathcal{A}$,  a much better correlation with 
$\log (v_R/m_{GUT}) / \log (v_{BL}/m_{GUT})$ is found, 
see figure \ref{fig:compLR2}.

In conclusion, to the determine $v_{BL}$ and $v_R$ individually more 
{\em theoretical} input is needed, such as the GUT scale thresholds, 
which are needed to fix the value of $m_{GUT}$. Recall, that we did 
not specify the exact values of these thresholds in our numerical 
calculation. This leads to a ``floating'' value of $m_{GUT}$ when 
$v_R$ and $v_{BL}$ are varied. Also more experimental data is needed 
to make more definite predictions. Especially SUSY mass spectrum 
measurements, which may or may not be very precise at the LHC, 
depending on the SUSY point realized in nature, will be of great 
importance. Recall that, if in reach of a linear collider, 
slepton mass and branching ratio measurements can be highly precise. 

So far only slepton decays have been discussed. This served to illustrate 
the most interesting signatures of the model, namely, lepton
flavour violation in the right slepton sector. However, LHC 
searches for lepton flavour violation usually concentrate on the decay chain
\cite{Hinchliffe:2000np,Carvalho:2002jg,Carquin:2008gv}
\begin{displaymath}
\tilde{\chi}_2^0 \to \tilde{l}^\pm l^\mp \to \tilde{\chi}_1^0 l^\pm l^\mp \thickspace.
\end{displaymath}
This well known signature has been widely studied due to the accurate
information it can provide about the particle spectrum
\cite{Paige:1996nx,Hinchliffe:1996iu,Bachacou:1999zb,Ball:2007zza,Aad:2009wy}. 
Note that in this decay one assumes usually that the $\tilde{\chi}_2^0$ 
themselves stem from the decay chain ${\tilde q}_L \to q\tilde{\chi}_2^0$. 
If the mass ordering $m_{\tilde{\chi}_2^0} > m_{\tilde{l}} >
m_{\tilde{\chi}_1^0}$ is realized, the dilepton invariant mass
\cite{Bachacou:1999zb,Allanach:2000kt}, defined as $m^2 (l^+ l^-) =
(p_{l^+} + p_{l^-})^2$, has an edge structure with a prominent
kinematical endpoint at
\begin{equation} \label{def-edge}
\left[ m^2 (l^+ l^-) \right]_{max} \equiv m_{ll}^2 = 
\frac{(m_{\tilde{\chi}_2^0}^2-m_{\tilde{l}}^2)
(m_{\tilde{l}}^2-m_{\tilde{\chi}_1^0}^2)}{m_{\tilde{l}}^2}  \thickspace,
\end{equation}
where the masses of the charged leptons have been neglected. The
position of this edge can be measured with impressively high precision 
at the LHC \cite{Paige:1996nx,Hinchliffe:1996iu,Bachacou:1999zb}, 
implying also an accurate determination of the intermediate slepton masses.

In fact, if two different sleptons $\tilde{l}_{1,2}$ have sufficiently
high event rates for $\tilde{\chi}_2^0 \to \tilde{l}_{1,2}^\pm l^\mp_j \to
\tilde{\chi}_1^0 l^\pm_i l^\mp_j$ and their masses allow these chains to 
be on-shell, two different dilepton edge distributions are expected 
\cite{Bartl:2005yy,Allanach:2008ib}. This presents a powerful tool 
to measure slepton mass splittings, which in turn allows to discriminate 
between the standard mSugra expectation, with usually negligible mass 
splittings for the first two generations, and extended models with 
additional sources of flavour violation.

The relation between the slepton mass splitting and the variation in
the position of the kinematical is edge is found to be
\cite{Allanach:2008ib}
\begin{equation} \label{edge-split}
\frac{\Delta m_{ll}}{\bar{m}_{ll}} = \frac{\Delta
m_{\tilde{l}}}{\bar{m}_{\tilde{l}}} \frac{m_{\tilde{\chi}_1^0}^2
m_{\tilde{\chi}_2^0}^2 -
\bar{m}_{\tilde{l}}^4}{(\bar{m}_{\tilde{l}}^2-m_{\tilde{\chi}_1^0}^2)
(\bar{m}_{\tilde{l}}^2-m_{\tilde{\chi}_2^0}^2)} \thickspace .
\end{equation}
Here $\Delta m_{ll} (i,j) = m_{l_i l_i} - m_{l_j l_j}$ is the
difference between two edge positions, $\Delta m_{\tilde{l}} =
m_{\tilde{l}_i} - m_{\tilde{l}_j}$ the difference between slepton
masses and $\bar{m}_{ll}$ and $\bar{m}_{\tilde{l}}$ average values of
the corresponding quantities. Note that higher order contributions of
$\frac{\Delta m_{\tilde{l}}}{\bar{m}_{\tilde{l}}}$ have been neglected
in equation \eqref{edge-split}.

A number of studies about the dilepton mass distribution have been performed
\cite{Paige:1996nx,Hinchliffe:1996iu,Bachacou:1999zb}, concluding that
the position of the edges can be measured at the LHC with an accuracy
up to $10^{-3}$. Moreover, as shown in reference
\cite{Allanach:2008ib}, this can be generally translated into a
similar precision for the relative $\tilde{e}-\tilde{\mu}$ mass
splitting, with some regions of parameter space where values as small 
as $10^{-4}$ might be measurable. Since this mass splitting is usually 
negligible in a pure mSugra scenario, it is regarded as an interesting 
signature of either lepton flavour violation or non-universality in 
the soft terms. Furthermore, in the context of this paper, it is important 
to emphasize that pure seesaw models can have this signature only in 
the left slepton sector \cite{Abada:2010kj}.

\begin{figure}
\begin{center}
\vspace{5mm}
\includegraphics[width=0.47\textwidth]{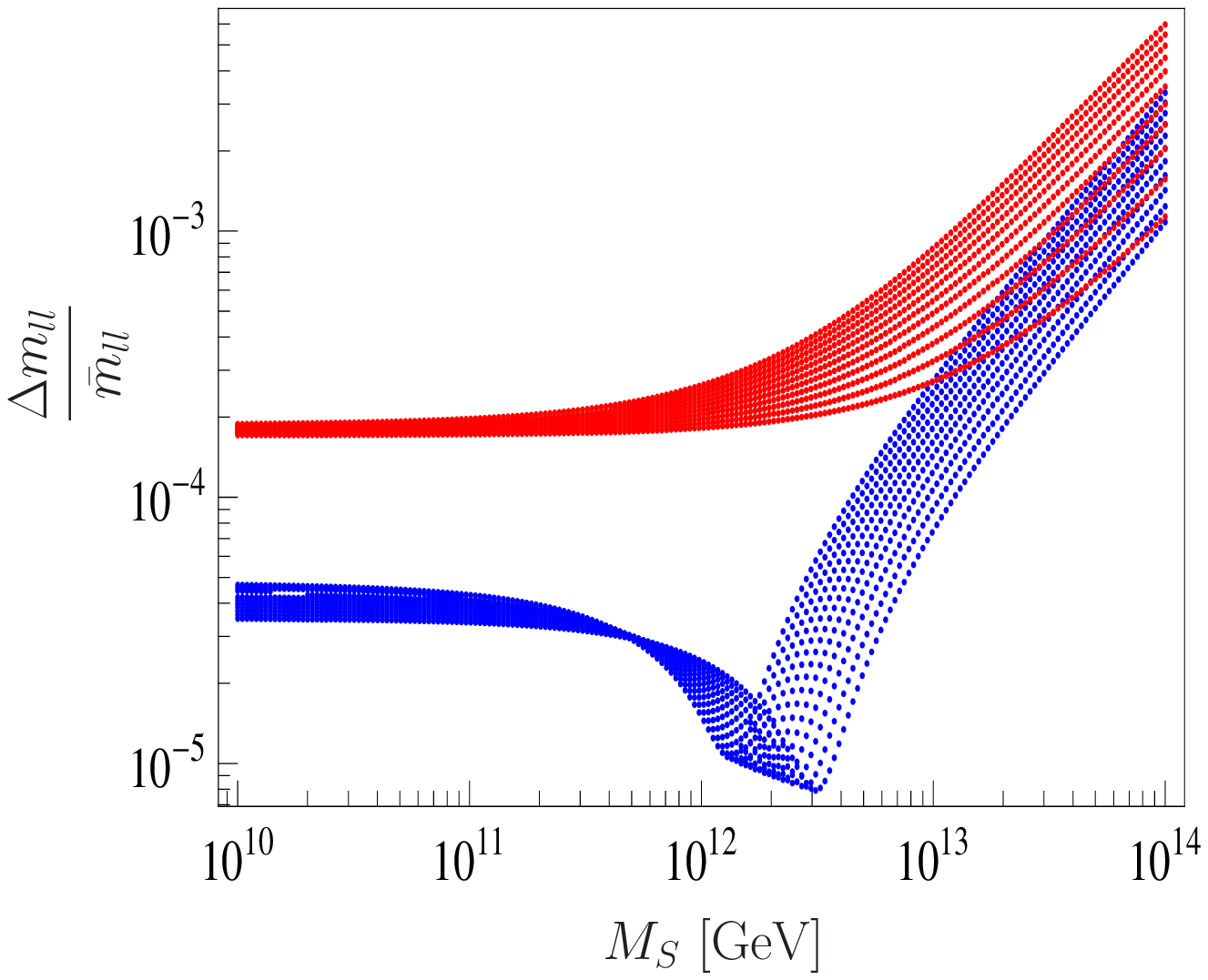}
\hspace{5mm}
\includegraphics[width=0.47\textwidth]{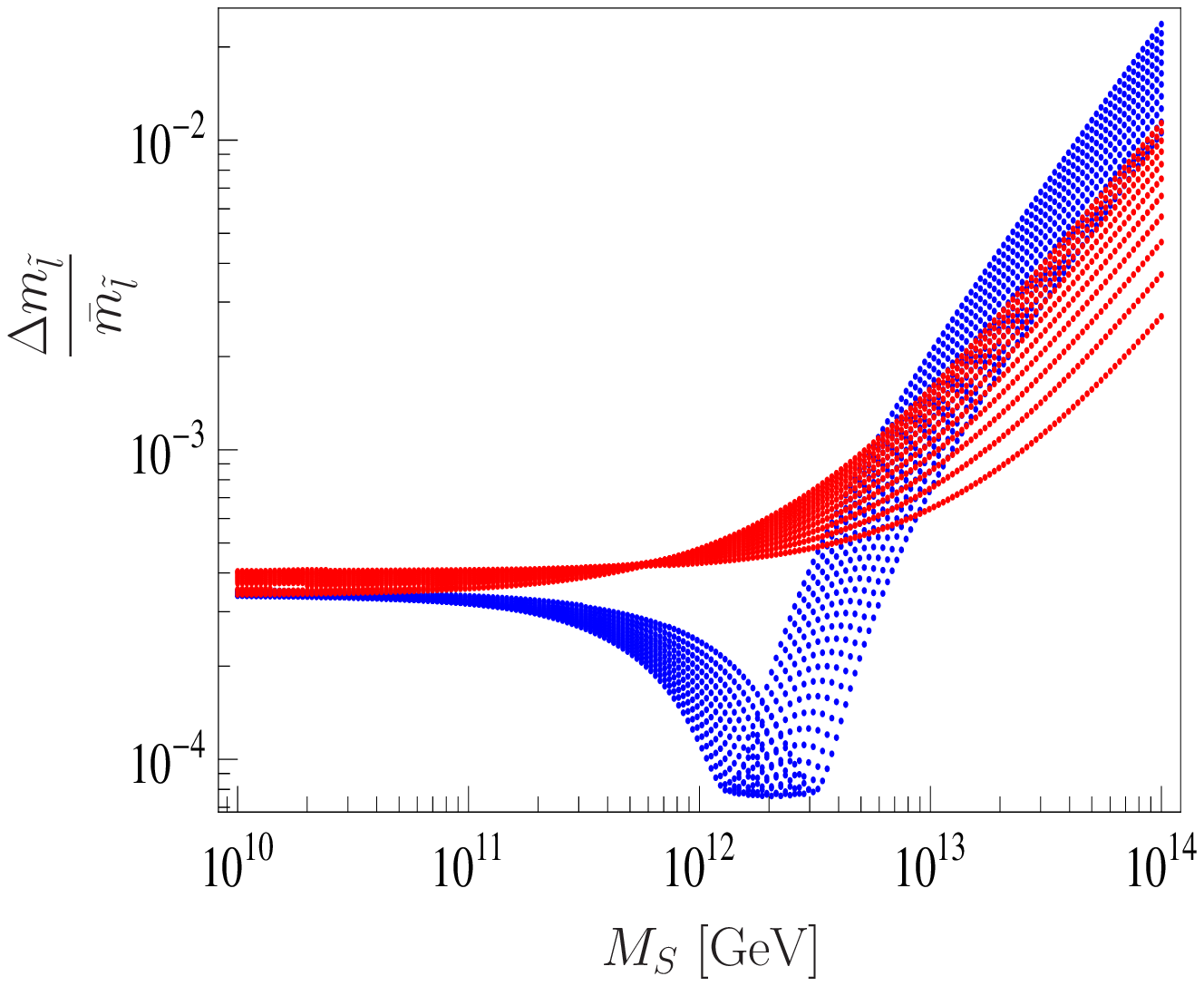}
\end{center}
\vspace{-5mm}
\caption{$\frac{\Delta m_{ll}}{\bar{m}_{ll}}$ (left-hand side) and
$\frac{\Delta m_{\tilde{l}}}{\bar{m}_{\tilde{l}}}$ (right-hand side)
as a function of the seesaw scale, defined as the mass of the lightest
right-handed neutrino, for the parameter choice $v_{BL} = 10^{15}$ GeV
and $v_R \in [10^{15},10^{16}]$ GeV. Blue dots correspond to the mass
distribution generated by intermediate left sleptons whereas
red dots correspond to the mass distribution generated by the
right ones. The mSugra parameters have been taken as in the
SPS3 benchmark point and neutrino oscillation data have been fitted
according to the $Y_\nu$ fit, with degenerate right-handed neutrinos.}
\label{fig:edge-split}
\end{figure}

Figure \ref{fig:edge-split} shows our results for the observables
$\frac{\Delta m_{ll}}{\bar{m}_{ll}}$ and $\frac{\Delta
m_{\tilde{l}}}{\bar{m}_{\tilde{l}}}$ as a function of the seesaw
scale. Large values for $M_S$ lead to sizable deviations from the mSugra
expectation, with a distinctive multi-edge structure in the dilepton 
mass distribution. Moreover, this effect is 
found in both left- and right- mediated decays. Observing this 
affect would clearly point towards a non-minimal seesaw model, 
such as the LR model we discuss.

\begin{figure}
\begin{center}
\vspace{5mm}
\includegraphics[width=0.47\textwidth]{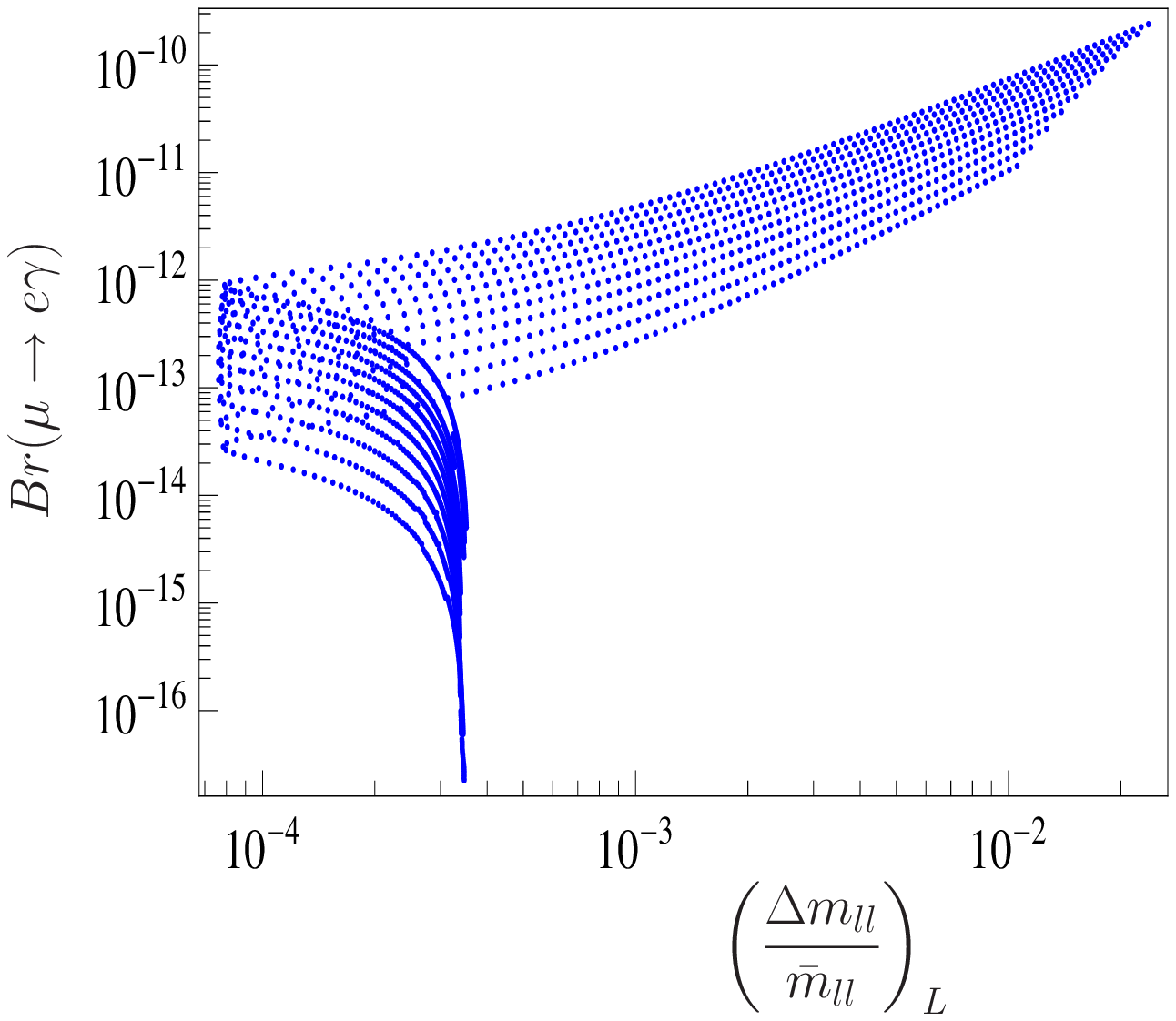}
\hspace{5mm}
\includegraphics[width=0.47\textwidth]{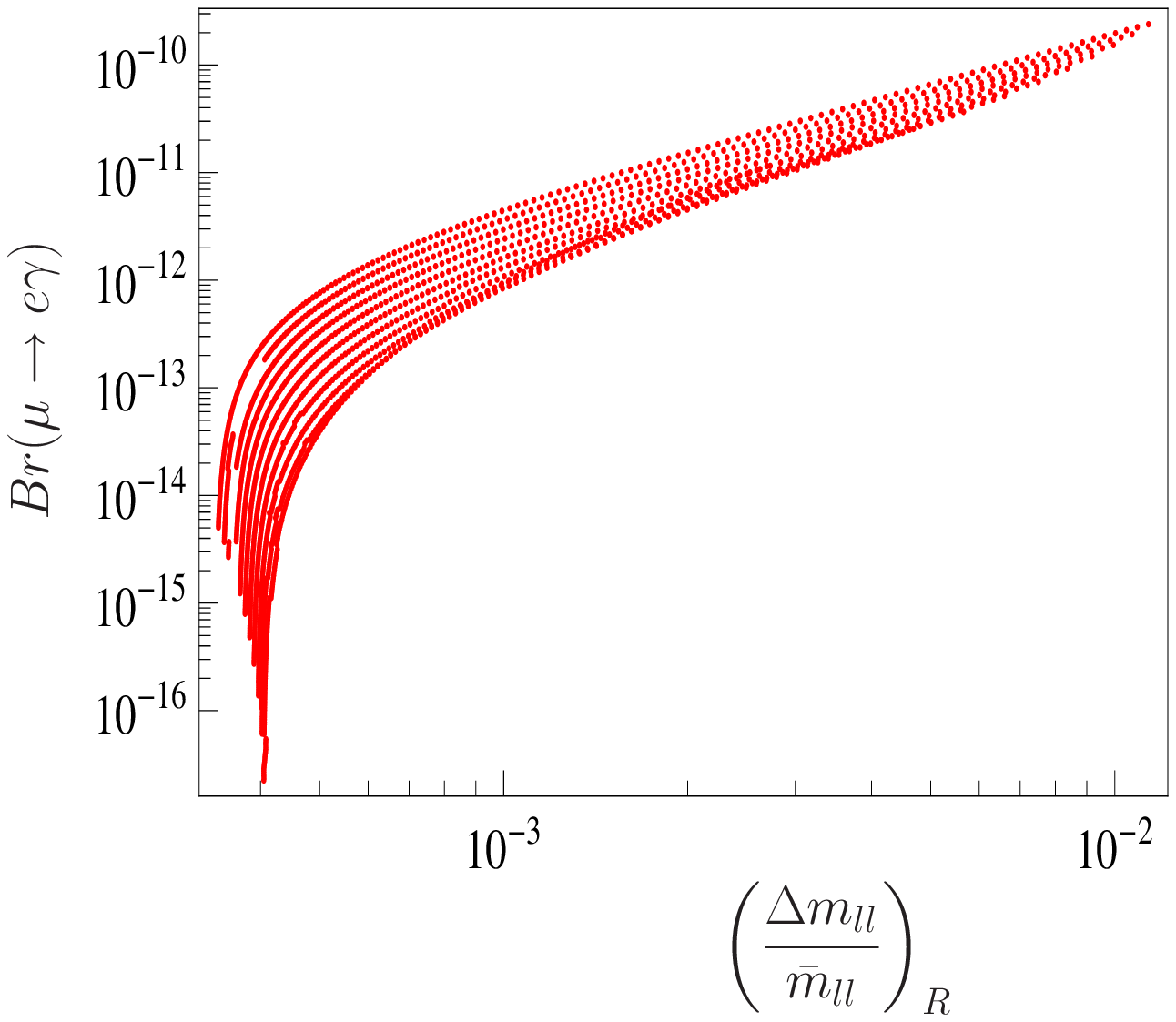}
\end{center}
\vspace{-5mm}
\caption{$Br(\mu \to e \gamma)$ as a function of $\left(\frac{\Delta
m_{ll}}{\bar{m}_{ll}}\right)_L$ (left-hand side) and
$\left(\frac{\Delta m_{ll}}{\bar{m}_{ll}}\right)_R$ (right-hand
side). The parameters are chosen as in figure \ref{fig:edge-split}.}
\label{fig:corredge}
\end{figure}

As expected, these observables are correlated with other LFV signals
\cite{Buras:2009sg,Abada:2010kj}. Figure \ref{fig:corredge} shows
$Br(\mu \to e \gamma)$ as a function of $\left(\frac{\Delta
m_{ll}}{\bar{m}_{ll}}\right)_L$ (mass distribution with intermediate L
sleptons) and $\left(\frac{\Delta m_{ll}}{\bar{m}_{ll}}\right)_R$
(mass distribution with intermediate R sleptons). Again, the main
novelty with respect to the usual seesaw implementations is the
correlation in the right sector, not present in the minimal
case \cite{Abada:2010kj}.

Furthermore, the process $\tilde{\chi}_2^0 \to \tilde{\chi}_1^0 l_i^+
l_j^-$ might provide additional LFV signatures if the rate for decays
with $l_i \neq l_j$ is sufficiently high. Reference
\cite{Andreev:2006sd} has investigated this possibility in great
detail, performing a complete simulation of the CMS detector in the
LHC for the decay $\tilde{\chi}_2^0 \to \tilde{\chi}_1^0 e \mu$. The
result is given in terms of the quantity
\begin{equation}
K_{e \mu} = \frac{Br(\tilde{\chi}_2^0 \to \tilde{\chi}_1^0 e \mu)}{Br(\tilde{\chi}_2^0 \to \tilde{\chi}_1^0 e e) + Br(\tilde{\chi}_2^0 \to \tilde{\chi}_1^0 \mu \mu)} \thickspace,
\end{equation}
which parametrizes the amount of flavour violation in
$\tilde{\chi}_2^0$ decays. The study, focused on the CMS test point
LM1 ($m_0 = 60$ GeV, $M_{1/2} = 250$ GeV, $A_0 = 0$ GeV, $\tan \beta =
10$, $sign(\mu) = +$) \cite{Ball:2007zza}, concludes that LFV can be
discovered at the LHC at $5 \sigma$ level with an integrated
luminosity of $10 fb^{-1}$ if $K_{e \mu} \ge K_{e \mu}^{min} = 0.04$.

\begin{figure}
\begin{center}
\vspace{5mm}
\includegraphics[width=0.6\textwidth]{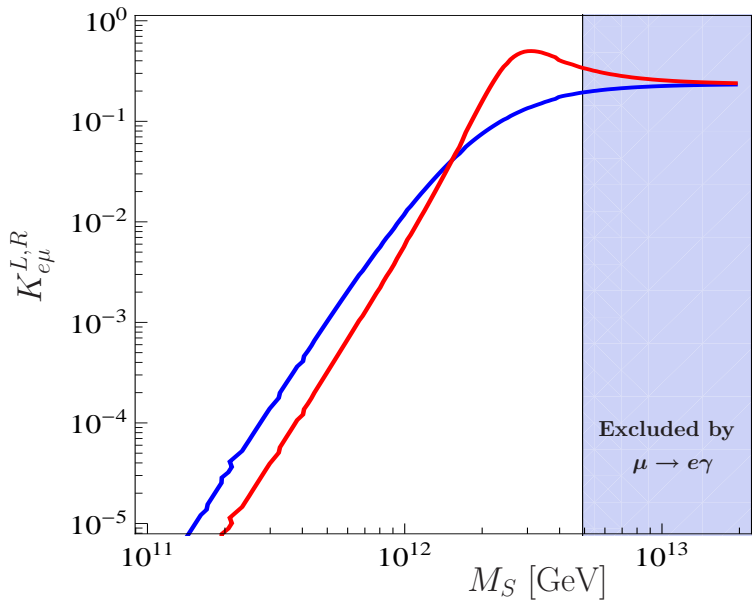}
\end{center}
\vspace{-5mm}
\caption{$K_{e \mu}$ as a function of the lightest right-handed
neutrino mass, for the parameter choice $v_{BL} = 10^{15}$ GeV and
$v_R = 5 \cdot 10^{15}$ GeV. The blue curve corresponds to
contributions from intermediate L sleptons, whereas the red one
corresponds to intermediate R sleptons. The mSugra parameters have
been taken as in the SPS3 benchmark point, which satisfies
$m(\tilde{\chi}_2^0) > m(\tilde{l_i})> m(\tilde{\chi}_1^0)$, and thus
the intermediate L and R sleptons can be produced on-shell. Neutrino
oscillation data have been fitted according to the $f$ fit, with
non-degenerate right-handed neutrinos. The blue shaded region is
excluded by $\mu \to e \gamma$.}
\label{fig:Kemu-SPS3}
\end{figure}

Figure \ref{fig:Kemu-SPS3} shows our computation of $K_{e \mu}$ as a
function of the lightest right-handed neutrino mass, for the parameter
choice $v_{BL} = 10^{15}$ GeV and $v_R = 5 \cdot 10^{15}$ GeV. The
results are shown splitting the contributions from intermediate left
(blue) and right (red) sleptons. Although the selected mSugra
parameters belong to the SPS3 point, and not to LM1 as in reference
\cite{Andreev:2006sd}, a similar sensitivity for $K_{e \mu}^{min}$ is
expected\footnote{Moreover, the LM1 point, being very similar to
  SPS1a', is strongly constrained by $\mu \to e
  \gamma$.}. This is because the reduction in the cross-section due to
the slightly heavier supersymmetric spectrum is possibly partially
compensated by the corresponding reduction in the SM background and
thus a limiting value $K_{e \mu}^{min}$ of a similar order is
expected. Moreover, \cite{Andreev:2006sd} uses 10 $fb^{-1}$ and with
larger integrated luminosities even smaller $K_{e \mu}^{min}$ should
become accessible at the LHC.

The main result in figure \ref{fig:Kemu-SPS3} is that for large $M_S$
values the rates for LFV $\tilde{\chi}_2^0$ decays are measurable for
both left and right intermediate sleptons. In fact, for $M_S
\gtrsim 10^{12}$ GeV the parameter $K_{e \mu}$ is above its minimum
value for the $5 \sigma$ discovery of $\tilde{\chi}_2^0 \to
\tilde{\chi}_1^0 e \mu$. See references
\cite{Andreev:2006sd,delAguila:2008iz} for more details on the LHC
discovery potential in the search for LFV in this channel.

\section{Conclusions}

We have studied a supersymmetric left-right symmetric model. Our 
motivation for studying this setup was twofold. First, LR models 
are theoretically attractive, since they contain all the necessary 
ingredients to {\em generate} a seesaw mechanism, instead of adding 
it by hand as is so often done. And, second, in a setup where the 
SUSY LR is supplemented by flavour blind supersymmetry breaking 
boundary conditions, different from all pure seesaw setups, lepton 
flavour violation occurs in both, the left and the right slepton 
sectors. 

We have calculated possible low-energy signals of this SUSY LR model, 
using full 2-loop RGEs for all parameters. We have found that low-energy 
lepton flavour violating decays, such as $\mu\to e \gamma$ are 
(a) expected to be larger than for the corresponding mSugra points 
in parameter space of seesaw type-I models and (b) the polarization 
asymmetry $\mathcal{A}$ of the outgoing positron is found to differ 
significantly from the pure seesaw prediction of $\mathcal{A}=+1$ 
in large regions of parameter space. We have also discussed possible 
collider signatures of the SUSY LR model for LHC and a possible ILC. 
Mass splittings between smuons and selectrons and LFV violating 
slepton decays should occur in both the left and the right slepton sector, 
again different from the pure seesaw expectations. 

We think therefore that the SUSY LR model is a good example of a
``beyond'' minimal, pure seesaw and offers many interesting novelties.
For example, the impact of the intermediate scales on dark matter
relic density and on certain mass combinations and the influence of
the right-handed neutrino spectrum on low energy observables, are
topics that certainly deserve further studies.

\section*{Acknowledgements}

W.P.\ thanks IFIC/C.S.I.C. for hospitality.
This work was supported by the Spanish MICINN under grants
FPA2008-00319/FPA, by the MULTIDARK Consolider CAD2009-00064, by
Prometeo/2009/091, by the EU grant UNILHC PITN-GA-2009-237920 and
FPA2008-04002-E/PORTU. A.V. thanks the Generalitat Valenciana for
financial support and the people at CFTP in Lisbon for hospitality.
The work of J.~N.~E.  has been supported by {\it Funda\c c\~ao para a
Ci\^encia e a Tecnologia} through the fellowship SFRH/BD/29642/2006.
J.~N.~E. and J.~C.~R. also acknowledge the financial support from {\it
Funda\c{c}\~ao para a Ci\^encia e a Tecnologia} grants CFTP-FCT
UNIT~777 and CERN/FP/109305/2009.
W.P.\ is partially supported
by the German Ministry of Education and Research (BMBF) under contract
05HT6WWA and by the Alexander von Humboldt Foundation. F.S.\ has been
supported by the DFG research training group GRK1147.

\appendix

\section{RGEs}
We present in the following appendices our results for the RGEs of the model above the $U(1)_{B-L}$ breaking scale. We will only show the \(\beta\)-functions for the gauge couplings and the anomalous dimensions of all chiral superfields. We briefly discuss in this section how these results were calculated. Furthermore, we show how they can be used to calculate the other \(\beta\)-functions of the models and give as example the 1-loop results for the soft SUSY breaking masses of the sleptons. The complete results are given online on this site
\begin{verbatim}
http://theorie.physik.uni-wuerzburg.de/~fnstaub/supplementary.html
\end{verbatim}
In addition, the corresponding model files for SARAH are also given on this web page. \\

\subsection{Calculation of supersymmetric RGEs}
\label{sec:genericRGEs}
For a  general $N=1$ supersymmetric gauge theory with superpotential  
\begin{equation}
 W (\phi) = \frac{1}{2}{\mu}^{ij}\phi_i\phi_j + \frac{1}{6}Y^{ijk}
\phi_i\phi_j\phi_k
\end{equation}
the  soft SUSY-breaking scalar terms are given by
\begin{equation}
V_{\hbox{soft}} = \left(\frac{1}{2}b^{ij}\phi_i\phi_j
+ \frac{1}{6}h^{ijk}\phi_i\phi_j\phi_k +\hbox{c.c.}\right)
+(m^2)^i{}_j\phi_i\phi_j^* \thickspace.
\end{equation}
The anomalous dimensions are given by \cite{Martin:1993zk}
\begin{align}
 \gamma_i^{(1)j} = & \frac{1}{2} Y_{ipq} Y^{jpq} - 2 \delta_i^j g^2 C_2(i) \thickspace, \\
 \gamma_i^{(2)j}  = &  -\frac{1}{2} Y_{imn} Y^{npq} Y_{pqr} Y^{mrj} + g^2 Y_{ipq} Y^{jpq} [2C_2(p)- C_2(i)] \nonumber \\
 & \; \;  + 2 \delta_i^j g^4 [ C_2(i) S(R)+ 2 C_2(i)^2 - 3 C_2(G) C_2(i)] \thickspace,
\end{align}
and the \(\beta\)-functions for the gauge couplings are given by
\begin{align}
 \beta_g^{(1)}  =  & g^3 \left[S(R) - 3 C_2(G) \right] \thickspace,\\
 \beta_g^{(2)}  =  & g^5 \left\{ - 6[C_2(G)]^2 + 2 C_2(G) S(R) + 4 S(R) C_2(R) \right\}
    - g^3 Y^{ijk} Y_{ijk}C_2(k)/d(G) \thickspace .
\end{align}
Here, \(C_2(i)\) is the quadratic Casimir for a specific superfield and $C_2(R),C_2(G)$ are the quadratic Casimirs for the matter and adjoint  representations, respectively. \(d(G)\) is the dimension of the adjoint representation.  \\
The $\beta$-functions for the superpotential parameters can be obtained by using superfield technique. The obtained expressions are \cite{West:1984dg,Jones:1984cx}. 
\begin{eqnarray}
 \beta_Y^{ijk} &= & Y^{p(ij} {\gamma_p}^{k)} \thickspace, \\
 \beta_{\mu}^{ij} &= & \mu^{p(i} {\gamma_p}^{j)} \thickspace .
\end{eqnarray}
The $(..)$ in the superscripts denote symmetrization. Most of the \(\beta\)-functions of the models can be derived from these results using the procedure given in \cite{Jack:1997eh} based on the spurion formalism \cite{Yamada:1994id}. In the following, we briefly summarize the basic ideas of this calculation for completeness. \\

The  exact results for the soft $\beta$-functions are given by \cite{Jack:1997eh}:
\begin{eqnarray}
\label{eq:betaM}
\beta_M &=& 2{\cal O} \left[\frac{\beta_g}{g}\right] \thickspace, \\
\beta_{h}^{ijk} &=& h{}^{l(jk}\gamma^{i)}{}_l -
2Y^{l(jk}\gamma_1{}^{i)}{}_l \thickspace, \cr
\beta_{b}^{ij} &=&   
b{}^{l(i}\gamma^{j)}{}_l-2\mu{}^{l(i}\gamma_1{}^{j)}{}_l \thickspace,\cr
\left(\beta_{m^2}\right){}^i{}_j &=& \Delta\gamma^i{}_j \thickspace.
\label{eq:betam2}
\end{eqnarray}
where we defined
\begin{eqnarray}
{\cal O}  &=& Mg^2\frac{\partial}{\partial g^2}-h^{lmn}
\frac{\partial}{\partial Y^{lmn}} \thickspace, \\
(\gamma_1)^i{}_j  &=& {\cal O}\gamma^i{}_j \thickspace, \\
\Delta &=& 2{\cal O} {\cal O}^* +2MM^* g^2 \frac{\partial}{\partial g^2}
+\left[{\tilde Y}^{lmn} \frac{\partial}{\partial Y^{lmn}} + \hbox{c.c.}\right]
+X \frac{\partial}{\partial g} \thickspace.
\end{eqnarray}
Here, $M$ is the gaugino mass and  
${\tilde Y}^{ijk} = (m^2)^i{}_lY^{jkl} +  (m^2)^j{}_lY^{ikl} + (m^2)^k{}_lY^{ijl}.$
Eqs.~(\ref{eq:betaM})-(\ref{eq:betam2}) hold  in a class of renormalization schemes that includes the DRED$'$-one \cite{Jack:1994rk}. We take the known contributions of $X$ from \cite{Jack:1998iy}:
\begin{eqnarray}
X^{\mathrm{DRED}'(1)}&=&-2g^3S \thickspace, \\
X^{\mathrm{DRED}'(2)}&=& (2r)^{-1}g^3 \mathrm{tr} [ W C_2(R)]
-4g^5C_2(G)S-2g^5C_2(G)QMM^* \thickspace,\end{eqnarray}
where
\begin{eqnarray}
S &=&  r^{-1} \mathrm{tr}[m^2C_2(R)] -MM^* C_2(G) \thickspace,  \\
W^j{}_i&=&{1\over2}Y_{ipq}Y^{pqn}(m^2)^j{}_n+ \frac{1}{2}Y^{jpq}Y_{pqn}(m^2)^n{}_i
+2Y_{ipq}Y^{jpr}(m^2)^q{}_r 
+h_{ipq}h^{jpq}-8g^2MM^*C_2(R)^j{}_i \thickspace .
\nonumber \\
\end{eqnarray}
With $Q = T(R) - 3C_2(G)$, and $T(R) = \mathrm{tr} \left[C_2(R)\right]$, $r$ being  the number of group generators. \\

\subsection{From GUT scale to $SU(2)_R$ breaking scale}
In the following sections we will use the definitions
\begin{equation}
 Y^{ij}_{Q_k} = Y^{ijk}_Q \thickspace, \hspace{1cm} Y^{ij}_{L_k} = Y^{ijk}_L
\end{equation}
and in the same way \(T^{ij}_{Q_k}\) and \(T^{ij}_{L_k}\). We will also assume summation of repeated indices.

\subsubsection{Anomalous Dimensions}
\label{sec:Ana1}
{\allowdisplaybreaks 
\begin{align} 
\gamma_{\hat{Q}}^{(1)} & = 
2 Y_{Q_k}^{*}  Y_{Q_k}^{T} -\frac{1}{12} \Big(18 g_{2}^{2}  + 32 g_{3}^{2}  + g_{BL}^{2}\Big){\bf 1} \\ 
\gamma_{\hat{Q}}^{(2)} & =  
+\frac{1}{144} \Big(-128 g_{3}^{4}  + 2052 g_{2}^{4}  + 289 g_{BL}^{4}  + 36 g_{2}^{2} \Big(32 g_{3}^{2}  + g_{BL}^{2}\Big) + 64 g_{3}^{2} g_{BL}^{2} \Big){\bf 1} \nonumber \\ 
 &+ Y_{Q_m}^{*} \Big( 6 g_2^2 \delta_{mn} + \frac{27}{4} \mbox{Tr}\Big({\alpha  \alpha^*}\Big) \delta_{mn} -2 \mbox{Tr}\Big({Y_{L_n}^{*}  Y_{L_m}^{T}}\Big) - 6 \mbox{Tr}\Big({Y_{Q_n}^{*}  Y_{Q_m}^{T}}\Big) \Big) Y_{Q_n}^{T} \nonumber \\ 
& -32 Y_{Q_m}^{*} Y_{Q_n}^\dagger Y_{Q_n} Y_{Q_m}^{T} \\
\gamma_{\hat{Q}^c}^{(1)} & = 
2 Y_{Q_k}^{\dagger}  Y_{Q_k} -\frac{1}{12} \Big(18 g_{2}^{2}  + 32 g_{3}^{2}  + g_{BL}^{2}\Big){\bf 1} \\ 
\gamma_{\hat{Q}^c}^{(2)} & =  
+\frac{1}{144} \Big(-128 g_{3}^{4}  + 2052 g_{2}^{4}  + 289 g_{BL}^{4}  + 36 g_{2}^{2} \Big(32 g_{3}^{2}  + g_{BL}^{2}\Big) + 64 g_{3}^{2} g_{BL}^{2} \Big){\bf 1} \nonumber \\ 
 &+ Y_{Q_m}^{\dagger} \Big( 6 g_2^2 \delta_{mn} + \frac{27}{4} \mbox{Tr}\Big({\alpha  \alpha^*}\Big) \delta_{mn} -2 \mbox{Tr}\Big({Y_{L_n}^{*}  Y_{L_m}^{T}}\Big) - 6 \mbox{Tr}\Big({Y_{Q_n}^{*}  Y_{Q_m}^{T}}\Big) \Big) Y_{Q_n} \nonumber \\ 
& -32 Y_{Q_m}^{\dagger} Y_{Q_n} Y_{Q_m}^T Y_{Q_n}^{*} \\
\gamma_{\hat{L}}^{(1)} & =  
2 \Big(3 {f^\dagger  f}  + {Y_{L_k}^{*}  Y_{L_k}^{T}}\Big) -\frac{3}{4} \Big(2 g_{2}^{2}  + g_{BL}^{2}\Big){\bf 1} \\ 
\gamma_{\hat{L}}^{(2)} & =  
\frac{3}{16} \Big(12 g_{2}^{2} g_{BL}^{2}  + 76 g_{2}^{4}  + 99 g_{BL}^{4} \Big){\bf 1} +3 {f^\dagger  f} \Big(-3 |a|^2  -4 \mbox{Tr}\Big({f  f^\dagger}\Big)  + 6 g_{BL}^{2}  + 8 g_{2}^{2} \Big)\nonumber \\ 
&+ Y_{L_m}^{*} \Big( 6 g_2^2 \delta_{mn} + \frac{27}{4} \mbox{Tr}\Big({\alpha  \alpha^*}\Big) \delta_{mn} -2 \mbox{Tr}\Big({Y_{L_n}^{*}  Y_{L_m}^{T}}\Big) - 6 \mbox{Tr}\Big({Y_{Q_n}^{*}  Y_{Q_m}^{T}}\Big) \nonumber \\
& - 11 f^\dagger f \delta_{mn} \Big) Y_{L_n}^{T} -4 Y_{L_m}^{*} Y_{L_n}^\dagger Y_{L_n} Y_{L_m}^{T} - 2 f^\dagger \Big( 17 f f^\dagger + 3 Y_{L_k} Y_{L_k}^\dagger \Big) f \nonumber \\
& - 6 f^\dagger Y_{L_k} f Y_{L_k}^* \\
\gamma_{\hat{L}^c}^{(1)} & =  
2 \Big(3 {f f^\dagger}  + {Y_{L_k}^{\dagger}  Y_{L_k}}\Big) -\frac{3}{4} \Big(2 g_{2}^{2}  + g_{BL}^{2}\Big){\bf 1} \\ 
\gamma_{\hat{L}^c}^{(2)} & =  
\frac{3}{16} \Big(12 g_{2}^{2} g_{BL}^{2}  + 76 g_{2}^{4}  + 99 g_{BL}^{4} \Big){\bf 1} +3 {f f^\dagger} \Big(-3 |a|^2  -4 \mbox{Tr}\Big({f  f^\dagger}\Big)  + 6 g_{BL}^{2}  + 8 g_{2}^{2} \Big)\nonumber \\ 
&+ Y_{L_m}^{\dagger} \Big( 6 g_2^2 \delta_{mn} + \frac{27}{4} \mbox{Tr}\Big({\alpha  \alpha^*}\Big) \delta_{mn} -2 \mbox{Tr}\Big({Y_{L_n}^{*}  Y_{L_m}^{T}}\Big) - 6 \mbox{Tr}\Big({Y_{Q_n}^{*}  Y_{Q_m}^{T}}\Big) \nonumber \\
& - 11 f f^\dagger \delta_{mn} \Big) Y_{L_n} -4 Y_{L_m}^{\dagger} Y_{L_n} Y_{L_m}^T Y_{L_n}^{*} - 2 f \Big( 17 f^\dagger f + 3 Y_{L_k}^T Y_{L_k}^* \Big) f^\dagger \nonumber \\
& - 6 Y_{L_k}^\dagger Y_{L_k} f f^\dagger \\
(\gamma_{\hat{\Phi}}^{(1)})_{ij} & =  
-3 g_{2}^{2} {\bf 1} -\frac{3}{2} \Big({\alpha  \alpha^*} + {\alpha^*  \alpha}\Big) +\delta_{im} \delta_{jn} \Big(3 \mbox{Tr}\Big({Y_{Q_m}^{*}  Y_{Q_n}^{T}}\Big)  + \mbox{Tr}\Big({Y_{L_m}^{*}  Y_{L_n}^{T}}\Big)\Big) \\ 
(\gamma_{\hat{\Phi}}^{(2)})_{ij} & = 
33 g_{2}^{4} {\bf 1} -9 \Big( 2 (\alpha \alpha \alpha^* \alpha^* + \alpha^* \alpha^* \alpha \alpha) + 3 (\alpha \alpha^* \alpha \alpha^* + \alpha^* \alpha \alpha^* \alpha) \Big) \nonumber \\
& -24 (\alpha \alpha^* + \alpha^* \alpha) \Big( 2 g_2^2 + 2 \mbox{Tr}\Big( \alpha \alpha^* \Big) -|a|^2 \Big) \nonumber \\
& - \frac{3}{2} \Big( \alpha_{jm} \alpha_{in}^* + \alpha_{jm}^* \alpha_{in} \Big) \Big( 3 \mbox{Tr}\Big({Y_{Q_m}^{*}  Y_{Q_n}^{T}}\Big) + \mbox{Tr}\Big({Y_{L_m}^{*}  Y_{L_n}^{T}}\Big) \Big) \nonumber \\
& - \frac{1}{2} \delta_{im} \delta_{in} \Big( -3 g_{BL}^2 \mbox{Tr}\Big({Y_{L_m}^{*}  Y_{L_n}^{T}}\Big) - (32 g_3^2 + g_{BL}^2) \mbox{Tr}\Big({Y_{Q_m}^{*}  Y_{Q_n}^{T}}\Big) \nonumber \\
& +2 \Big( 5 \mbox{Tr}\Big({ f f^\dagger Y_{L_n} Y_{L_m}^\dagger }\Big) + \mbox{Tr}\Big({ f Y_{L_n}^* Y_{L_m}^T f^\dagger }\Big)+ 6 \mbox{Tr}\Big({ f Y_{L_m}^* Y_{L_n}^T f^\dagger }\Big) \Big) \nonumber \\
& +4 \Big( 2 \mbox{Tr}\Big({ Y_{L_m}^\dagger Y_{L_n} Y_{L_n}^T Y_{L_n}^* }\Big) + \mbox{Tr}\Big({ Y_{L_m}^\dagger Y_{L_m} Y_{L_n}^T Y_{L_m}^* }\Big) + \mbox{Tr}\Big({ Y_{L_m}^\dagger Y_{L_n} Y_{L_m}^T Y_{L_m}^* }\Big) \Big) \nonumber \\
& +12 \Big( 2 \mbox{Tr}\Big({ Y_{Q_m}^\dagger Y_{Q_n} Y_{Q_n}^T Y_{Q_n}^* }\Big) + \mbox{Tr}\Big({ Y_{Q_m}^\dagger Y_{Q_m} Y_{Q_n}^T Y_{Q_m}^* }\Big) \nonumber \\
& + \mbox{Tr}\Big({ Y_{Q_m}^\dagger Y_{Q_n} Y_{Q_m}^T Y_{Q_m}^* }\Big) \Big) \Big) \\
\gamma_{\hat{\Delta}}^{(1)} & =  
2 \mbox{Tr}\Big({f  f^\dagger}\Big)  -3 g_{BL}^{2}  -4 g_{2}^{2}  + \frac{3}{2} |a|^2 \\ 
\gamma_{\hat{\Delta}}^{(2)} & =  
48 g_{2}^{4} +24 g_{2}^{2} g_{BL}^{2} +81 g_{BL}^{4} \nonumber \\
& +\frac{3}{2} |a|^2 \Big( 4 g_2^2 + \mbox{Tr}\Big( \alpha \alpha^* \Big) - \frac{7}{2} |a|^2 \Big) - \Big(2 g_{2}^{2}  + 3 g_{BL}^{2} \Big)\mbox{Tr}\Big({f  f^\dagger}\Big) \nonumber \\ 
 &-24 \mbox{Tr}\Big({f  f^\dagger  f  f^\dagger}\Big) -6 \mbox{Tr}\Big({f  f^\dagger  Y_{L_k}  Y_{L_k}^{\dagger}}\Big) -2 \mbox{Tr}\Big({f  Y_{L_k}^{*}  Y_{L_k}^{T}  f^\dagger}\Big) \\ 
\gamma_{\hat{\bar{\Delta}}}^{(1)} & =  
-3 g_{BL}^{2}  -4 g_{2}^{2}  + \frac{3}{2} |a|^2 \\ 
\gamma_{\hat{\bar{\Delta}}}^{(2)} & =  
\frac{3}{4} \Big(4 \Big(16 g_{2}^{4}  + 27 g_{BL}^{4}  + 8 g_{2}^{2} g_{BL}^{2} \Big) \nonumber \\
& + |a|^2 \Big(2 \mbox{Tr}\Big({\alpha  \alpha^*}\Big)  -3 \mbox{Tr}\Big({f  f^\dagger}\Big)  -7 |a|^2  + 8 g_{2}^{2} \Big)\Big)\\ 
\gamma_{\hat{\Delta}^c}^{(1)} & =  
2 \mbox{Tr}\Big({f  f^\dagger}\Big)  -3 g_{BL}^{2}  -4 g_{2}^{2}  + \frac{3}{2} |a|^2 \\ 
\gamma_{\hat{\Delta}^c}^{(2)} & =  
48 g_{2}^{4} +24 g_{2}^{2} g_{BL}^{2} +81 g_{BL}^{4} \nonumber \\
& +\frac{3}{2} |a|^2 \Big( 4 g_2^2 + \mbox{Tr}\Big( \alpha \alpha^* \Big) - \frac{7}{2} |a|^2 \Big) - \Big(2 g_{2}^{2}  + 3 g_{BL}^{2} \Big)\mbox{Tr}\Big({f  f^\dagger}\Big) \nonumber \\ 
 &-24 \mbox{Tr}\Big({f  f^\dagger  f  f^\dagger}\Big) -8 \mbox{Tr}\Big({f  Y_{L_k}^T Y_{L_k}^* f^\dagger}\Big) \\ 
\gamma_{\hat{\bar{\Delta}}^c}^{(1)} & =  
-3 g_{BL}^{2}  -4 g_{2}^{2}  + \frac{3}{2} |a|^2 \\ 
\gamma_{\hat{\bar{\Delta}}^c}^{(2)} & =  
\frac{3}{4} \Big(4 \Big(16 g_{2}^{4}  + 27 g_{BL}^{4}  + 8 g_{2}^{2} g_{BL}^{2} \Big) \nonumber \\
& + |a|^2 \Big(2 \mbox{Tr}\Big({\alpha  \alpha^*}\Big)  -3 \mbox{Tr}\Big({f  f^\dagger}\Big)  -7 |a|^2  + 8 g_{2}^{2} \Big)\Big)\\ 
\gamma_{\hat{\Omega}}^{(1)} & =  
2 |a|^2  -4 g_{2}^{2} \\ 
\gamma_{\hat{\Omega}}^{(2)} & =  
3 \mbox{Tr}\Big({\alpha  (\alpha  \alpha^* -\alpha^* \alpha) \alpha^*}\Big)  + 48 g_{2}^{4}  + |a|^2 \Big(12 g_{BL}^{2}  -3 \mbox{Tr}\Big({f  f^\dagger}\Big)  -6 |a|^2  + 8 g_{2}^{2} \Big) \\
\gamma_{\hat{\Omega}^c}^{(1)} & =  
2 |a|^2  -4 g_{2}^{2} \\ 
\gamma_{\hat{\Omega}^c}^{(2)} & =  
3 \mbox{Tr}\Big({\alpha  (\alpha  \alpha^* -\alpha^* \alpha) \alpha^*}\Big)  + 48 g_{2}^{4}  + |a|^2 \Big(12 g_{BL}^{2}  -3 \mbox{Tr}\Big({f  f^\dagger}\Big)  -6 |a|^2  + 8 g_{2}^{2} \Big)
\end{align} }

Note that the previous formulas are totally general and can be applied with any number of bidoublets. Nevertheless, if two bidoublets are considered $\alpha \alpha^* = \alpha^* \alpha$ and further simplifications are possible.

\subsubsection{Beta functions for soft breaking masses of sleptons}

Using the procedure explained in sec.~\ref{sec:genericRGEs}, we can calculate the soft breaking masses for the sleptons. The results are

\begin{eqnarray}
16 \pi^2 \frac{d}{dt} m_L^2 &=& 6 f f^\dagger m_L^2 + 12 f m_L^2 f^\dagger + 6 m_L^2 f f^\dagger + 12 m_\Delta^2 f f^\dagger \nonumber \\
&& + 2 Y_{L_k} Y_{L_k}^\dagger m_L^2 + 2 m_L^2 Y_{L_k} Y_{L_k}^\dagger + 4 Y_{L_k} m_{L^c}^2 Y_{L_k}^\dagger \nonumber \\
&& + 4 (m_\Phi^2)_{mn} Y_L^{(m)} Y_L^{(n) \: \dagger} + 12 T_f T_f^\dagger + 4 T_{L_k} T_{L_k}^\dagger \nonumber \\
&& - (3 g_{BL}^2 |M_1|^2 + 6 g_2^2 |M_2|^2 + \frac{3}{2} g_{BL}^2 S_1) {\bf 1} \\
16 \pi^2 \frac{d}{dt} m_{L^c}^2 &=& 6 f^\dagger f m_{L^c}^2 + 12 f^\dagger m_{L^c}^2 f + 6 m_{L^c}^2 f^\dagger f + 12 m_{\Delta^c}^2 f^\dagger f \nonumber \\
&& + 2 Y_{L_k}^\dagger Y_{L_k} m_{L^c}^2 + 2 m_{L^c}^2 Y_{L_k}^\dagger Y_{L_k} + 4 Y_{L_k}^\dagger m_L^2 Y_{L_k} \nonumber \\
&& + 4 (m_\Phi^2)_{mn} Y_L^{(m) \: \dagger} Y_L^{(n)} + 12 T_f^\dagger T_f + 4 T_{L_k}^\dagger T_{L_k} \nonumber \\
&& - (3 g_{BL}^2 |M_1|^2 + 6 g_2^2 |M_2|^2 - \frac{3}{2} g_{BL}^2 S_1) {\bf 1}
\end{eqnarray}

where
\begin{eqnarray}
S_1 &=& 3 (m_\Delta^2 - m_{\bar{\Delta}}^2 - m_{\Delta^c}^2 + m_{\bar{\Delta}^c}^2) \nonumber \\
&& + \sum_{m,n} \left[ (m_Q^2)_{mn} - (m_{Q^c}^2)_{mn} - (m_L^2)_{mn} + (m_{L^c}^2)_{mn} \right]
\end{eqnarray}

\subsubsection{Beta functions for gauge couplings}
\label{sec:Beta1}
{\allowdisplaybreaks  \begin{align} 
\beta_{g_{BL}}^{(1)} & =  
24 g_{BL}^{3} \\ 
\beta_{g_{BL}}^{(2)} & =  
\frac{1}{2} g_{BL}^{3} \Big(-192 |a|^2 -93 \mbox{Tr}\Big({f  f^\dagger}\Big) +2 \Big(115 g_{BL}^{2}  + 162 g_{2}^{2}  \nonumber \\ 
 &-2 \mbox{Tr}\Big({Y_{Q_k}^*  Y_{Q_k}^{T}}\Big) -6 \mbox{Tr}\Big({Y_{L_k}^*  Y_{L_k}^{T}}\Big) + 8 g_{3}^{2} \Big)\Big)\\ 
\beta_{g_2}^{(1)} & =  
8 g_{2}^{3} \\ 
\beta_{g_2}^{(2)} & =  
\frac{1}{6} g_{2}^{3} \Big(660 g_{2}^{2} +144 g_{3}^{2} +162 g_{BL}^{2} -192 |a|^2 +108 \mbox{Tr}\Big({\alpha  \alpha^*}\Big) -73 \mbox{Tr}\Big({f  f^\dagger}\Big) \nonumber \\ 
 &-24 \mbox{Tr}\Big({Y_{L_k}^*  Y_{L_k}^{T}}\Big) -72 \mbox{Tr}\Big({Y_{Q_k}^*  Y_{Q_k}^{T}}\Big) \Big)\\ 
\beta_{g_3}^{(1)} & =  
-3 g_{3}^{3} \\ 
\beta_{g_3}^{(2)} & =  
g_{3}^{3} \Big(14 g_{3}^{2}  + 18 g_{2}^{2}  -8 \mbox{Tr}\Big({Y_{Q_k}^*  Y_{Q_k}^{T}}\Big) + g_{BL}^{2}\Big)
\end{align}}

\subsection{From $SU(2)_R$ breaking scale to $U(1)_{B-L}$ breaking scale}
\subsubsection{Anomalous Dimensions}
\label{sec:Ana2}
{\allowdisplaybreaks \begin{align} 
\gamma_{\hat{Q}}^{(1)} & =  
-\frac{1}{12} \Big(18 g_{L}^{2}  + 32 g_{3}^{2}  + g_{BL}^{2}\Big){\bf 1}  + {Y_d^*  Y_{d}^{T}} + {Y_u^*  Y_{u}^{T}}\\ 
\gamma_{\hat{Q}}^{(2)} & =  
+\frac{1}{144} \Big(109 g_{BL}^{4}  -128 g_{3}^{4}  + 36 g_{BL}^{2} g_{L}^{2}  + 64 g_{3}^{2} \Big(18 g_{L}^{2}  + g_{BL}^{2}\Big) + 972 g_{L}^{4} \Big){\bf 1} \nonumber \\ 
 &-2 \Big({Y_d^*  Y_{d}^{T}  Y_d^*  Y_{d}^{T}} + {Y_u^*  Y_{u}^{T}  Y_u^*  Y_{u}^{T}}\Big)\nonumber \\ 
 &+{Y_d^*  Y_{d}^{T}} \Big(-3 \mbox{Tr}\Big({Y_d  Y_{d}^{\dagger}}\Big)  - |b_c|^2  -\frac{3}{2} |b|^2  - \mbox{Tr}\Big({Y_e  Y_{e}^{\dagger}}\Big)  + g_{R}^{2}\Big)\nonumber \\ 
 &+{Y_u^*  Y_{u}^{T}} \Big(-3 \mbox{Tr}\Big({Y_u  Y_{u}^{\dagger}}\Big)  - |b_c|^2  -\frac{3}{2} |b|^2  - \mbox{Tr}\Big({Y_\nu  Y_{\nu}^{\dagger}}\Big)  + g_{R}^{2}\Big)\\ 
\gamma_{\hat{d}^c}^{(1)} & =  
2 {Y_{d}^{\dagger}  Y_d}  -\frac{1}{12} \Big(32 g_{3}^{2}  + 6 g_{R}^{2}  + g_{BL}^{2}\Big){\bf 1} \\ 
\gamma_{\hat{d}^c}^{(2)} & =  
-\frac{1}{144} \Big(-109 g_{BL}^{4}  + 128 g_{3}^{4}  -12 g_{BL}^{2} g_{R}^{2}  -64 g_{3}^{2} \Big(6 g_{R}^{2}  + g_{BL}^{2}\Big) -684 g_{R}^{4} \Big){\bf 1} \nonumber \\ 
 &-2 \Big({Y_{d}^{\dagger}  Y_d  Y_{d}^{\dagger}  Y_d} + {Y_{d}^{\dagger}  Y_u  Y_{u}^{\dagger}  Y_d}\Big)\nonumber \\ 
 &- {Y_{d}^{\dagger}  Y_d} \Big(2 |b_c|^2  + 2 \mbox{Tr}\Big({Y_e  Y_{e}^{\dagger}}\Big)  + 3 |b|^2  -6 g_{L}^{2}  + 6 \mbox{Tr}\Big({Y_d  Y_{d}^{\dagger}}\Big) \Big)\\ 
\gamma_{\hat{u}^c}^{(1)} & =  
2 {Y_{u}^{\dagger}  Y_u}  -\frac{1}{12} \Big(32 g_{3}^{2}  + 6 g_{R}^{2}  + g_{BL}^{2}\Big){\bf 1} \\ 
\gamma_{\hat{u}^c}^{(2)} & =  
-\frac{1}{144} \Big(-109 g_{BL}^{4}  + 128 g_{3}^{4}  -12 g_{BL}^{2} g_{R}^{2}  -64 g_{3}^{2} \Big(6 g_{R}^{2}  + g_{BL}^{2}\Big) -684 g_{R}^{4} \Big){\bf 1} \nonumber \\ 
 &-2 \Big({Y_{u}^{\dagger}  Y_d  Y_{d}^{\dagger}  Y_u} + {Y_{u}^{\dagger}  Y_u  Y_{u}^{\dagger}  Y_u}\Big)\nonumber \\ 
 &- {Y_{u}^{\dagger}  Y_u} \Big(2 |b_c|^2  + 2 \mbox{Tr}\Big({Y_\nu  Y_{\nu}^{\dagger}}\Big)  + 3 |b|^2  -6 g_{L}^{2}  + 6 \mbox{Tr}\Big({Y_u  Y_{u}^{\dagger}}\Big) \Big)\\ 
\gamma_{\hat{L}}^{(1)} & =  
-\frac{3}{4} \Big(2 g_{L}^{2}  + g_{BL}^{2}\Big){\bf 1}  + {Y_e^*  Y_{e}^{T}} + {Y_\nu^*  Y_{\nu}^{T}}\\ 
\gamma_{\hat{L}}^{(2)} & =  
+\frac{9}{16} \Big(12 g_{L}^{4}  + 13 g_{BL}^{4}  + 4 g_{BL}^{2} g_{L}^{2} \Big){\bf 1} -2 \Big({Y_e^*  Y_{e}^{T}  Y_e^*  Y_{e}^{T}} + {Y_\nu^*  Y_{\nu}^{T}  Y_\nu^*  Y_{\nu}^{T}} \Big) \nonumber \\ 
 &- {Y_\nu^*  F_c  Y_{\nu}^{T}}  \nonumber \\ 
 &+{Y_e^*  Y_{e}^{T}} \Big(-3 \mbox{Tr}\Big({Y_d  Y_{d}^{\dagger}}\Big)  - |b_c|^2  -\frac{3}{2} |b|^2  - \mbox{Tr}\Big({Y_e  Y_{e}^{\dagger}}\Big)  + g_{R}^{2}\Big)\nonumber \\ 
 &+{Y_\nu^*  Y_{\nu}^{T}} \Big(-3 \mbox{Tr}\Big({Y_u  Y_{u}^{\dagger}}\Big)  - |b_c|^2  -\frac{3}{2} |b|^2  - \mbox{Tr}\Big({Y_\nu  Y_{\nu}^{\dagger}}\Big)  + g_{R}^{2}\Big)\\ 
\gamma_{\hat{e}^c}^{(1)} & =  
2 {Y_{e}^{\dagger}  Y_e}  -\frac{1}{4} \Big(2 g_{R}^{2}  + 3 g_{BL}^{2} \Big){\bf 1} \\ 
\gamma_{\hat{e}^c}^{(2)} & =  
+\frac{1}{16} \Big(117 g_{BL}^{4}  + 12 g_{BL}^{2} g_{R}^{2}  + 76 g_{R}^{4} \Big){\bf 1} -2 \Big({Y_{e}^{\dagger}  Y_e  Y_{e}^{\dagger}  Y_e} + {Y_{e}^{\dagger}  Y_\nu  Y_{\nu}^{\dagger}  Y_e}\Big)\nonumber \\ 
 &- {Y_{e}^{\dagger}  Y_e} \Big(2 |b_c|^2  + 2 \mbox{Tr}\Big({Y_e  Y_{e}^{\dagger}}\Big)  + 3 |b|^2  -6 g_{L}^{2}  + 6 \mbox{Tr}\Big({Y_d  Y_{d}^{\dagger}}\Big) \Big)\\ 
\gamma_{\hat{\nu}^c}^{(1)} & =  
2 {Y_{\nu}^{\dagger}  Y_\nu}  -\frac{1}{4} \Big(2 g_{R}^{2}  + 3 g_{BL}^{2} \Big){\bf 1}  + F_c^* \\ 
\gamma_{\hat{\nu}^c}^{(2)} & =  
\frac{1}{16} \Big(117 g_{BL}^{4}  + 12 g_{BL}^{2} g_{R}^{2}  + 76 g_{R}^{4} \Big){\bf 1} - (f_c^{1 \dagger} + f_c^{1 *}) F_c (f_c^{1} + f_c^{1 T}) \nonumber \\
& -2 Y_{\nu}^{\dagger}  \Big( Y_e  Y_{e}^{\dagger}  +  Y_\nu  Y_{\nu}^{\dagger}  \Big) Y_\nu -2 (f_c^{1 \dagger} + f_c^{1 *}) Y_{\nu}^{T}  Y_\nu^*  (f_c^{1} + f_c^{1 T}) \nonumber \\ 
&- {Y_{\nu}^{\dagger}  Y_\nu} \Big(2 \Big(-3 g_{L}^{2}  + 3 \mbox{Tr}\Big({Y_u  Y_{u}^{\dagger}}\Big)  + |b_c|^2 + \mbox{Tr}\Big({Y_\nu  Y_{\nu}^{\dagger}}\Big)\Big) + 3 |b|^2 \Big)\nonumber \\ 
 &+ F_c^* \Big(2 g_{R}^{2}  + 3 g_{BL}^{2}  - |a^1_c|^2  - \mbox{Tr}\Big({f^1_c  f_{c}^{1 \dagger}}\Big)  - \mbox{Tr}\Big({f_{c}^{1 \dagger}  f_{c}^{1 T}}\Big) \Big)\\ 
\gamma_{\hat{H}_d}^{(1)} & =  
3 \mbox{Tr}\Big({Y_d  Y_{d}^{\dagger}}\Big)  -\frac{1}{2} g_{R}^{2}  + \frac{3}{2} |b|^2  -\frac{3}{2} g_{L}^{2}  + |b_c|^2 + \mbox{Tr}\Big({Y_e  Y_{e}^{\dagger}}\Big)\\ 
\gamma_{\hat{H}_d}^{(2)} & =  
\frac{1}{4} \Big(27 g_{L}^{4} +6 g_{L}^{2} g_{R}^{2} +19 g_{R}^{4} +2 \Big(32 g_{3}^{2}  + g_{BL}^{2}\Big)\mbox{Tr}\Big({Y_d  Y_{d}^{\dagger}}\Big) +6 g_{BL}^{2} \mbox{Tr}\Big({Y_e  Y_{e}^{\dagger}}\Big) \nonumber \\ 
 &+|b|^2 \Big(-18 \mbox{Tr}\Big({Y_u  Y_{u}^{\dagger}}\Big) -15 |b|^2 -19 |b_c|^2  + 24 g_{L}^{2}  -6 \mbox{Tr}\Big({Y_\nu  Y_{\nu}^{\dagger}}\Big) \Big)\nonumber \\ 
 &-4 |b_c|^2 \Big(3 |b_c|^2  + 3 \mbox{Tr}\Big({Y_u  Y_{u}^{\dagger}}\Big)  + |a^1_c|^2  + \mbox{Tr}\Big({Y_\nu  Y_{\nu}^{\dagger}}\Big)\Big)-36 \mbox{Tr}\Big({Y_d  Y_{d}^{\dagger}  Y_d  Y_{d}^{\dagger}}\Big) \nonumber \\
& -12 \mbox{Tr}\Big({Y_d  Y_{d}^{\dagger}  Y_u  Y_{u}^{\dagger}}\Big) -12 \mbox{Tr}\Big({Y_e  Y_{e}^{\dagger}  Y_e  Y_{e}^{\dagger}}\Big) -4 \mbox{Tr}\Big({Y_e  Y_{e}^{\dagger}  Y_\nu  Y_{\nu}^{\dagger}}\Big) \Big)\\ 
\gamma_{\hat{H}_u}^{(1)} & =  
3 \mbox{Tr}\Big({Y_u  Y_{u}^{\dagger}}\Big)  -\frac{1}{2} g_{R}^{2}  + \frac{3}{2} |b|^2  -\frac{3}{2} g_{L}^{2}  + |b_c|^2 + \mbox{Tr}\Big({Y_\nu  Y_{\nu}^{\dagger}}\Big)\\ 
\gamma_{\hat{H}_u}^{(2)} & =  
\frac{1}{4} \Big(27 g_{L}^{4} +6 g_{L}^{2} g_{R}^{2} +19 g_{R}^{4} \nonumber \\
& + |b|^2 \Big(-18 \mbox{Tr}\Big({Y_d  Y_{d}^{\dagger}}\Big)  -15 |b|^2 -19 |b_c|^2  + 24 g_{L}^{2}  -6 \mbox{Tr}\Big({Y_e  Y_{e}^{\dagger}}\Big) \Big)\nonumber \\ 
 &-4 |b_c|^2 \Big(3 |b_c|^2  + 3 \mbox{Tr}\Big({Y_d  Y_{d}^{\dagger}}\Big)  + |a^1_c|^2   + \mbox{Tr}\Big({Y_e  Y_{e}^{\dagger}}\Big)\Big)+2 \Big(32 g_{3}^{2}  + g_{BL}^{2}\Big)\mbox{Tr}\Big({Y_u  Y_{u}^{\dagger}}\Big) \nonumber \\ 
 &+6 g_{BL}^{2} \mbox{Tr}\Big({Y_\nu  Y_{\nu}^{\dagger}}\Big) -4 \mbox{Tr}\Big({f^1_c  f_{c}^{1 \dagger}  Y_{\nu}^{T}  Y_\nu^*}\Big) -4 \mbox{Tr}\Big({f^1_c  Y_{\nu}^{\dagger}  Y_\nu  f_{c}^{1 \dagger}}\Big) -12 \mbox{Tr}\Big({Y_d  Y_{d}^{\dagger}  Y_u  Y_{u}^{\dagger}}\Big) \nonumber \\ 
 &-4 \mbox{Tr}\Big({Y_e  Y_{e}^{\dagger}  Y_\nu  Y_{\nu}^{\dagger}}\Big) -36 \mbox{Tr}\Big({Y_u  Y_{u}^{\dagger}  Y_u  Y_{u}^{\dagger}}\Big) -4 \mbox{Tr}\Big({Y_\nu  f_{c}^{1 \dagger}  f_{c}^{1 T}  Y_{\nu}^{\dagger}}\Big)  \nonumber \\ 
 &-12 \mbox{Tr}\Big({Y_\nu  Y_{\nu}^{\dagger}  Y_\nu  Y_{\nu}^{\dagger}}\Big) -4 \mbox{Tr}\Big({f_{c}^{1 \dagger}  Y_{\nu}^{T}  Y_\nu^*  f_{c}^{1 T}}\Big) \Big)\\ 
\gamma_{\hat{\Delta}^{c 0}}^{(1)} & =  
-2 g_{R}^{2}  -3 g_{BL}^{2}  + |a^1_c|^2 + \mbox{Tr}\Big({f^1_c  f_{c}^{1 \dagger}}\Big) + \mbox{Tr}\Big({f_{c}^{1 \dagger}  f_{c}^{1 T}}\Big)\\ 
\gamma_{\hat{\Delta}^{c 0}}^{(2)} & =  
\frac{1}{2} \Big(72 g_{BL}^{4} +24 g_{BL}^{2} g_{R}^{2} +44 g_{R}^{4} -4 |a^1_c|^2 \Big(|a^1_c|^2  +| b_c|^2 \Big)\nonumber \\ 
 &- \Big(2 g_{R}^{2}  + 3 g_{BL}^{2} \Big)\Big(\mbox{Tr}\Big({f^1_c  f_{c}^{1 \dagger}}\Big) + \mbox{Tr}\Big({f_{c}^{1 \dagger}  f_{c}^{1 T}}\Big)\Big)-4 \mbox{Tr}\Big({f^1_c  f_{c}^{1 \dagger}  f^1_c  f_{c}^{1 \dagger}}\Big) \nonumber \\
& -8 \mbox{Tr}\Big({f^1_c  f_{c}^{1 \dagger}  f_{c}^{1 T}  f_{c}^{1 \dagger}}\Big) -8 \mbox{Tr}\Big({f^1_c  f_{c}^{1 \dagger}  f_{c}^{1 T}  f^{1 *}_c}\Big) -4 \mbox{Tr}\Big({f^1_c  f_{c}^{1 \dagger}  Y_{\nu}^{T}  Y_\nu^*}\Big) \nonumber \\
& -4 \mbox{Tr}\Big({f^1_c  Y_{\nu}^{\dagger}  Y_\nu  f_{c}^{1 \dagger}}\Big) -4 \mbox{Tr}\Big({Y_\nu  f_{c}^{1 \dagger}  f_{c}^{1 T}  Y_{\nu}^{\dagger}}\Big) -4 \mbox{Tr}\Big({f_{c}^{1 \dagger}  f_{c}^{1 T}  f_{c}^{1 \dagger}  f_{c}^{1 T}}\Big) \nonumber \\
&-8 \mbox{Tr}\Big({f_{c}^{1 \dagger}  f_{c}^{1 T}  f^{1 *}_c  f_{c}^{1 T}}\Big) -4 \mbox{Tr}\Big({f_{c}^{1 \dagger}  Y_{\nu}^{T}  Y_\nu^*  f_{c}^{1 T}}\Big) \Big)\\ 
\gamma_{\hat{\bar{\Delta}}^{c 0}}^{(1)} & =  
-2 g_{R}^{2}  -3 g_{BL}^{2}  + |a^1_c|^2\\ 
\gamma_{\hat{\bar{\Delta}}^{c 0}}^{(2)} & =  
12 g_{BL}^{2} g_{R}^{2}  + 22 g_{R}^{4}  + 36 g_{BL}^{4}  \nonumber \\
& - |a^1_c|^2 \Big(2 |a^1_c|^2  + 2 |b_c|^2  + \mbox{Tr}\Big({f^1_c  f_{c}^{1 \dagger}}\Big) + \mbox{Tr}\Big({f_{c}^{1 \dagger}  f_{c}^{1 T}}\Big)\Big)\\ 
\gamma_{\hat{\Omega}}^{(1)} & =  
-4 g_{L}^{2}  + |b|^2\\ 
\gamma_{\hat{\Omega}}^{(2)} & =  
28 g_{L}^{4}  - |b|^2 \Big(2 |b_c|^2  + 3 |b|^2  + 3 \mbox{Tr}\Big({Y_d  Y_{d}^{\dagger}}\Big)  + 3 \mbox{Tr}\Big({Y_u  Y_{u}^{\dagger}}\Big)  \nonumber \\
& - g_{R}^{2}  + g_{L}^{2} + \mbox{Tr}\Big({Y_e  Y_{e}^{\dagger}}\Big) + \mbox{Tr}\Big({Y_\nu  Y_{\nu}^{\dagger}}\Big)\Big)\\ 
\gamma_{\hat{\Omega}^{c 0}}^{(1)} & =  
2 |b_c|^2  + |a^1_c|^2\\ 
\gamma_{\hat{\Omega}^{c 0}}^{(2)} & =  
-2 |a_{c}^{1}|^4 - |a^1_c|^2 \Big(-4 g_{R}^{2}  -6 g_{BL}^{2}  + \mbox{Tr}\Big({f^1_c  f_{c}^{1 \dagger}}\Big) + \mbox{Tr}\Big({f_{c}^{1 \dagger}  f_{c}^{1 T}}\Big)\Big) \nonumber \\ 
 &-2 |b_c|^2 \Big(2 |b_c|^2  + 3 |b|^2  -3 g_{L}^{2}  + 3 \mbox{Tr}\Big({Y_d  Y_{d}^{\dagger}}\Big)  + 3 \mbox{Tr}\Big({Y_u  Y_{u}^{\dagger}}\Big)  \nonumber \\
& - g_{R}^{2}  + \mbox{Tr}\Big({Y_e  Y_{e}^{\dagger}}\Big) + \mbox{Tr}\Big({Y_\nu  Y_{\nu}^{\dagger}}\Big)\Big)
\end{align} } 

In these expressions we have defined

\begin{equation}
F_c = f_{c}^{1}  f_{c}^{1 \dagger}  +  f_{c}^{1}  f^{1 *}_c  + f_{c}^{1 T}  f_{c}^{1 \dagger}  +  f_{c}^{1 T}  f^{1 *}_c 
\end{equation}

\subsubsection{Beta functions for soft breaking masses of sleptons}
Again, the results for the slepton soft SUSY breaking masses at 1-loop are shown. The beta functions read
\begin{eqnarray}
16 \pi^2 \frac{d}{dt} m_L^2 &=& 2 Y_e m_{e^c}^2 Y_e^\dagger + 2 m_{H_d}^2 Y_e Y_e^\dagger + 2 m_{H_u}^2 Y_\nu Y_\nu^\dagger + m_L^2 Y_e Y_e^\dagger \nonumber \\
&& + Y_e Y_e^\dagger m_L^2 + m_L^2 Y_\nu Y_\nu^\dagger + Y_\nu Y_\nu^\dagger m_L^2 + 2 Y_\nu m_{\nu^c}^2 Y_\nu^\dagger \nonumber \\
&& + 2 T_e T_e^\dagger + 2 T_\nu T_\nu^\dagger - (3 g_{BL}^2 |M_1|^2 + 6 g_L^2 |M_L|^2 + \frac{3}{4} g_{BL}^2 S_2)  {\bf 1} \\
16 \pi^2 \frac{d}{dt} m_{e^c}^2 &=& 2 Y_e^\dagger Y_e m_{e^c}^2 + 2 m_{e^c}^2 Y_e^\dagger Y_e + 4 m_{H_d}^2 Y_e^\dagger Y_e + 4 Y_e^\dagger m_L^2 Y_e \nonumber \\
&& + 4 T_e^\dagger T_e - (3 g_{BL}^2 |M_1|^2 + 2 g_R^2 |M_R|^2 - \frac{3}{4} g_{BL}^2 S_2 - \frac{1}{2} g_R^2 S_3 )  {\bf 1}
\end{eqnarray}
where
\begin{eqnarray}
S_2 &=& 2 ( m_{\Delta^{c \: 0}}^2 - m_{\bar{\Delta}^{c \: 0}}^2 ) + Tr \left[ m_{d^c}^2 - m_{e^c}^2 + m_{u^c}^2 - m_{\nu^c}^2 + 2 m_L^2 - 2 m_Q^2 \right] \\
S_3 &=& 2 ( m_{\Delta^{c \: 0}}^2 - m_{\bar{\Delta}^{c \: 0}}^2 - m_{H_d}^2 + m_{H_u}^2 ) + Tr \left[ 3 m_{d^c}^2 + m_{e^c}^2 - 3 m_{u^c}^2 - m_{\nu^c}^2 \right]
\end{eqnarray}

\subsubsection{Beta functions for gauge couplings}
\label{sec:beta2}
{\allowdisplaybreaks  \begin{align} 
\beta_{g_{BL}}^{(1)} & =  
9 g_{BL}^{3} \\ 
\beta_{g_{BL}}^{(2)} & =  
\frac{1}{2} g_{BL}^{3} \Big(16 g_{3}^{2} +50 g_{BL}^{2} +18 g_{L}^{2} +30 g_{R}^{2} -12 |{a_c^1}|^2 -9 \mbox{Tr}\Big({{f_c^1}  {f_c^1}^{\dagger}}\Big) -2 \mbox{Tr}\Big({Y_d  Y_d^{\dagger}}\Big) \nonumber \\ 
 &-6 \mbox{Tr}\Big({Y_e  Y_e^{\dagger}}\Big) -2 \mbox{Tr}\Big({Y_u  Y_u^{\dagger}}\Big) -6 \mbox{Tr}\Big({Y_v  Y_v^{\dagger}}\Big) -9 \mbox{Tr}\Big({{f_c^1}^{\dagger}  {f_c^1}^{T}}\Big) \Big)\\ 
\beta_{g_L}^{(1)} & =  
3 g_{L}^{3} \\ 
\beta_{g_L}^{(2)} & =  
\frac{1}{3} g_{L}^{3} \Big(-28 |b|^2  + 3 \Big(24 g_{3}^{2}  -2 |b_c|^2  -2 \mbox{Tr}\Big({Y_e  Y_e^{\dagger}}\Big)  -2 \mbox{Tr}\Big({Y_v  Y_v^{\dagger}}\Big)  + 3 g_{BL}^{2}  + 49 g_{L}^{2} \nonumber \\ 
 &  -6 \mbox{Tr}\Big({Y_d  Y_d^{\dagger}}\Big)  -6 \mbox{Tr}\Big({Y_u  Y_u^{\dagger}}\Big)  + g_{R}^{2}\Big)\Big)\\ 
\beta_{g_R}^{(1)} & =  
9 g_{R}^{3} \\ 
\beta_{g_R}^{(2)} & =  
g_{R}^{3} \Big(24 g_{3}^{2} +15 g_{BL}^{2} +3 g_{L}^{2} +15 g_{R}^{2} -4 |{a_c^1}|^2 -4 |b|^2 -2 |b_c|^2 -3 \mbox{Tr}\Big({{f_c^1}  {f_c^1}^{\dagger}}\Big) \nonumber \\
& -6 \mbox{Tr}\Big({Y_d  Y_d^{\dagger}}\Big) -2 \mbox{Tr}\Big({Y_e  Y_e^{\dagger}}\Big) -6 \mbox{Tr}\Big({Y_u  Y_u^{\dagger}}\Big) -2 \mbox{Tr}\Big({Y_v  Y_v^{\dagger}}\Big) -3 \mbox{Tr}\Big({{f_c^1}^{\dagger}  {f_c^1}^{T}}\Big) \Big)\\ 
\beta_{g_3}^{(1)} & =  
-3 g_{3}^{3} \\ 
\beta_{g_3}^{(2)} & =  
g_{3}^{3} \Big(14 g_{3}^{2}  + 3 g_{R}^{2}  -4 \mbox{Tr}\Big({Y_d  Y_d^{\dagger}}\Big)  -4 \mbox{Tr}\Big({Y_u  Y_u^{\dagger}}\Big)  + 9 g_{L}^{2}  + g_{BL}^{2}\Big)
\end{align}}

\end{document}